\documentclass[trackchanges,twocolumn]{aastex701}

\usepackage{booktabs} 
\usepackage{amsmath}
\usepackage{cancel}
\usepackage{subcaption}
\usepackage{bbold}

\newcommand{\fion}[2]{\mbox{[#1\,\uppercase\expandafter{\romannumeral #2\relax}]}}
\newcommand{\sfion}[2]{\mbox{#1\,\uppercase\expandafter{\romannumeral #2\relax}]}}

\newcommand{\appropto}
{\mathrel{\vcenter{\offinterlineskip\halign{\hfil$##$\cr\propto\cr\noalign{\kern2pt}\sim\cr\noalign{\kern-2pt}}}}}
\newcommand{\bagpipes}{{\tt{Bagpipes}}}
\newcommand{\prospector}{{\tt{Prospector}}}

\newcommand{\sextractor}{{\tt{SExtractor}}}

\newcommand{\eazy}{{\tt{EaZy}}}

\newcommand{\eazypy}{{\tt{EaZy-py}}}
\newcommand{\galfind}{{\tt{galfind}}}
\newcommand{\jwst}{\emph{JWST}}
\newcommand{\hst}{\emph{HST}}
\newcommand{\euclid}{\emph{Euclid}}

\newcommand{\stpsf}{{\tt{STPSF}}}
\newcommand{\Muv}{$M_{\mathrm{UV}}$}
\newcommand{\galfit}{{\tt{GALFIT}}}
\newcommand{\bpass}{{\tt{BPASS}}}

\newcommand{\cloudy}{{\tt{CLOUDY}}}

\newcommand{\beagle}{{\tt BEAGLE}}
\newcommand{\tweakreg}{{\tt tweakreg}}
\newcommand{\photutils}{{\tt photutils}}
\newcommand{\reproject}{{\tt reproject}}
\newcommand{\bdfinder}{{\tt BD-Finder}}

\newcommand{\aperpy}{{\tt aperpy}}
\newcommand{\pypher}{{\tt pypher}}
\newcommand{\pysersic}{{\tt pysersic}}
\newcommand{\epochsone}{{\tt EPOCHS-DR1}}
\newcommand{\epochstwo}{{\tt EPOCHS-DR2}}

\newcommand{\lya}{Ly$\alpha$}
\newcommand{\ha}{H$\alpha$}
\newcommand{\hbeta}{H$\beta$}

\newcommand{\xiion}{$\xi_{\rm ion}$}

\newcommand{\Ndotion}{$\dot{N}_{\rm ion}$}

\newcommand{\fesc}{$f_{\rm esc}^{\rm LyC}$}

\newcommand{\uvbeta}{$\beta_{\mathrm{UV}}$}
\newcommand{\oiiihbeta}{\fion{O}{3}$+${\hbeta}}

\usepackage{amsmath}
\usepackage{cancel}
\usepackage{subcaption}
\usepackage{rotating}
\usepackage{longtable}
\usepackage{placeins}
\usepackage{multirow}

\begin{document}

\title{EPOCHS-DR2 I: Expanded data release and properties of $6.5<z<16.5$ photometrically selected galaxies across NIRCam deep fields covering $\sim700\,\rm arcmin^2$ with the \galfind\ software}

\author[0000-0003-0519-9445]{Duncan Austin}
\affiliation{Jodrell Bank Centre for Astrophysics, Alan Turing Building, University of Manchester, Oxford Road, Manchester M13 9PL, UK}
\email[show]{duncan.austin@manchester.ac.uk}  

\author[0000-0003-1949-7638]{Christopher J. Conselice}
\affiliation{Jodrell Bank Centre for Astrophysics, Alan Turing Building, University of Manchester, Oxford Road, Manchester M13 9PL, UK}
\email[]{conselice@manchester.ac.uk}

\author[0000-0002-4130-636X]{Thomas Harvey}
\affiliation{Jodrell Bank Centre for Astrophysics, Alan Turing Building, University of Manchester, Oxford Road, Manchester M13 9PL, UK}
\email[]{thomas.harvey-3@manchester.ac.uk}

\author[0000-0003-4875-6272]{Nathan J. Adams}
\affiliation{Jodrell Bank Centre for Astrophysics, Alan Turing Building, University of Manchester, Oxford Road, Manchester M13 9PL, UK}
\email[]{nathan@theadamshouse.co.uk}

\author[0009-0008-8375-8605]{Louis Quilley}
\affiliation{Jodrell Bank Centre for Astrophysics, Alan Turing Building, University of Manchester, Oxford Road, Manchester M13 9PL, UK}
\email[]{louis.quilley@manchester.ac.uk}

\author[0000-0002-9816-1931]{Jordan C. J. D'Silva}
\affiliation{Jodrell Bank Centre for Astrophysics, Alan Turing Building, University of Manchester, Oxford Road, Manchester M13 9PL, UK}
\email[]{jordan.dsilva@manchester.ac.uk}

\author[0000-0001-7633-3985]{Vadim Rusakov}
\affiliation{Jodrell Bank Centre for Astrophysics, Alan Turing Building, University of Manchester, Oxford Road, Manchester M13 9PL, UK}
\email[]{vadim.rusakov@manchester.ac.uk}

\author[0000-0002-3119-9003]{Qiong Li}
\affiliation{Jodrell Bank Centre for Astrophysics, Alan Turing Building, University of Manchester, Oxford Road, Manchester M13 9PL, UK}
\email[]{qiong.li@manchester.ac.uk}

\author[0009-0008-8642-5275]{Lewi Westcott}
\affiliation{Jodrell Bank Centre for Astrophysics, Alan Turing Building, University of Manchester, Oxford Road, Manchester M13 9PL, UK}
\email[]{lewi.westcott@manchester.ac.uk}

\author[0009-0006-4799-4497]{Caio Goolsby}
\affiliation{Jodrell Bank Centre for Astrophysics, Alan Turing Building, University of Manchester, Oxford Road, Manchester M13 9PL, UK}
\email[]{caio.goolsby@postgrad.manchester.ac.uk}

\author[0009-0003-7198-6591]{James Arcidiacono}
\affiliation{Jodrell Bank Centre for Astrophysics, Alan Turing Building, University of Manchester, Oxford Road, Manchester M13 9PL, UK}
\email[]{james.arcidiacono@manchester.ac.uk}

\author[0009-0006-4327-7618]{Kai Madgwick}
\affiliation{Jodrell Bank Centre for Astrophysics, Alan Turing Building, University of Manchester, Oxford Road, Manchester M13 9PL, UK}
\email[]{kyle.madgwick@postgrad.manchester.ac.uk}

\author[0000-0002-3257-8806]{William J. Roper}
\affiliation{Oxford Research Software Engineering Group, MPLS Doctoral Training Centre, University of Oxford, 1-4 Keble Road, Oxford OX1 3NP, UK}
\affiliation{Astronomy Centre, University of Sussex, Falmer, Brighton BN1 9QH, UK}
\email[]{william.roper@dtc.ox.ac.uk}


\author[0000-0003-1625-8009]{Brenda Frye}
\email[]{brendafrye@gmail.com}
\affiliation{Department of Astronomy/Steward Observatory, University of Arizona, 933 N Cherry Ave, Tucson, AZ, 85721-0009, USA}

\author[0000-0001-9440-8872]{Norman A. Grogin}
\email[]{nagrogin@stsci.edu}
\affiliation{Space Telescope Science Institute, 3700 San Martin Drive, Baltimore, MD 21218, USA}

\author[0000-0002-6150-833X]{Rafael Ortiz~III}
\affiliation{School of Earth and Space Exploration, Arizona State University,
Tempe, AZ 85287-1404, USA}
\email{rortizii@asu.edu}

\author[0000-0001-8156-6281]{Rogier A. Windhorst}
\affiliation{School of Earth and Space Exploration, Arizona State University,
Tempe, AZ 85287-6004, USA}
\email{Rogier.Windhorst@asu.edu}

\begin{abstract}

We present the \epochstwo\ sample of $2452$ star forming galaxies at $6.5<z_{\rm phot}<16.5$ produced using the \galfind\ photometric pipeline across $\sim650-740~\mathrm{arcmin}^2$ in six of the most widely used deep, blank field \jwst/NIRCam imaging surveys. $293$ galaxies are spectroscopically confirmed and a spectroscopic--photometric redshift comparison yields a normalized median absolute deviation, $\sigma_{\rm NMAD}=0.017$, an $\eta=4.3\%$ outlier rate, and a mean offset $\langle\Delta z\rangle=-0.022$. We find $34$ little red dots (LRDs) with ``V--shaped'' SEDs, as well as $69$ L--type, $36$ T--type, and $2$ compact Y--type brown dwarf candidates, which we remove from our sample. The fiducial sample yields a flattening $\beta_{\rm UV}-M_{\rm UV}$ relation with only moderately blue median $\beta_{\rm UV}(M_{\rm UV}=-19)=-2.32^{+0.08}_{-0.07}$ at $z\simeq12.5$, implying dust-polluted stellar populations at the highest redshifts. A ``gold'' sample of $62$ sources are identified at $z_{\rm phot}>10.5$, $\sim85\%$ of which are without NIRSpec follow-up, including $15$ candidates at $z_{\rm phot}>13.5$ across a diverse range of proper sizes $30\,\rm pc\lesssim r_{\rm c}\lesssim1.5\,\rm kpc$ and UV slopes $-4.0\lesssim\beta_{\rm UV}\lesssim-1.5$. Two very promising candidates stand out, GS-W-4608 and GN-Med-1881, with $\beta_{\rm UV}\lesssim-3$, $\gtrsim1.6\,\rm mag$ \lya\ break colors, and F335M and F410M detections which improve their high-redshift solution robustness. Complementary to our high-redshift sample, this public data release provides the community with optical/NIR catalogs, empirical PSFs, and \eazypy, \bagpipes, and \bdfinder\ SED fitting results, enhancing the legacy value of photometric JWST data in the upcoming era of wide-field imaging datasets with Euclid and Roman.
\end{abstract}

\keywords{ }

\section{Introduction}

Over the course of the first four years since the launch of the \textit{James Webb Space Telescope} (\jwst), the unprecedented sensitivity of Near-InfraRed Camera \citep[NIRCam;][]{Rieke2005, Rieke2023} imaging at $0.7-4.8\,\mu \rm m$ has pushed the high-redshift frontier out to $z\simeq14-16$, $\sim250\,\rm Myr$ after the Big Bang. Substantial efforts have been made to produce large samples of star forming galaxies at $z\gtrsim10$ \citep[e.g.][]{Adams2023, Austin2023, Donnan2023, Finkelstein2023, Harikane2023, Hainline2024a, 
Adams2024,
Conselice2025, Hainline2026}, with Cycle 1 and 2 programs generating the bulk of these candidates across sky areas covered by complementary photometric legacy data, most notably from the Hubble Space Telescope (\hst) Cosmic Assembly Near-IR Deep Extragalactic Legacy Survey \citep[CANDELS;][]{Koekemoer2011-CANDELS, Grogin2011-CANDELS} and the Hubble Legacy Fields \citep[HLF;][]{Illingworth2016, Whitaker2019-HLF}.

UV luminosity functions (UVLFs) from these samples have revealed an abundance of luminous $M_{\rm UV}\lesssim-20$ sources at $z>9$ \citep{Bouwens2023, Donnan2023, Harikane2023, Leung2023, Adams2024, Donnan2024, Finkelstein2024, McLeod2024, Willott2024, Franco2025, Morishita2025, Weibel2025, Whitler2025}, breaking the tension between \hst\ derived number density estimates \citep{McLeod2016, Oesch2018}. Many theories exist to explain the observed number densities, including the possibility of an enhanced star formation efficiency \citep[SFE;][]{Dekel2023, Ceverino2024, Li2024, Feldmann2025} or increasingly bursty star formation \citep{Furlanetto2022, Pallottini2023, Sun2023a, CarvajalBohorquez2025} producing an inflated UV luminosity--halo mass scatter \citep[$\sigma_{\rm UV}$;][]{Mason2023, Mirocha2023, Shen2023, Ciesla2024, Gelli2024, Kravtsov2024} in the early Universe. The degeneracy between these can be split by galaxy clustering measurements at the UVLF bright end \citep{Munoz2023}, however the small NIRCam FOV makes this challenging even across the largest COSMOS-Web areas \citep{Paquereau2025} in the pre-Roman era. Another possibility is that compact ``blue monsters'' with UV continuum slopes $\beta_{\rm UV}<-2$ \citep{Ferrara2023, Ziparo2023, Fiore2023, Ferrara2024} produce super-Eddington radiation driven outflows \citep{Ferrara2024} which expel dust from the sites of UV emission in these sources; this can help explain the reducing dust content inferred from \jwst\ $\beta_{\rm UV}$ studies \citep{Topping2022, Cullen2023, Cullen2024, Austin2025a}.

An abundance of ``overmassive'' galaxies, including so-called ``little red dots'' \citep[LRDs; e.g.][]{Labbe2023, Greene2024, Kokorev2024, Carranza-Escudero2025} observed in early NIRCam data hinted at potential inconsistencies with the standard $\Lambda\rm CDM$ model of cosmology \citep{BoylanKolchin2023, Lovell2023}. Multi-object spectroscopy with the Near-InfraRed Spectrograph \citep[NIRSpec;][]{Jakobsen2022, Ferruit2022, Boker2023}, however, has shown that at least a subsample of these LRDs are instead active galactic nuclei \citep[AGN;][]{Greene2024, Matthee2024}.

The redshift frontier has proved particularly challenging to constrain on account of limited spectroscopic confirmations and catastrophic photo-z failures, and many $z>15$ photometric candidates \citep{Austin2023, Donnan2023, PerezGonzalez2023, Yan2023, Conselice2025, PerezGonzalez2025} have now been confirmed as lower redshift interlopers \citep[e.g.][]{ArrabalHaro2023b} by NIRSpec. Thus far, $4$ sources have been spectroscopically confirmed at $z>13$ \citep{CurtisLake2023, Carniani2024, Donnan2026}, with the most distant (MoM-z14 in the COSMOS field) residing at $z=14.44\pm 0.02$ \citep{Naidu2026}. 

Wide area samples produced from pure-parallel NIRCam imaging suggest a dearth of photometrically selected sources at $z>14.5$ \citep{McLeod2026, Weibel2026}, with only three candidates selected by \citet{McLeod2026} across $\gtrsim0.6~\rm deg^2$. One such candidate, PAN-z14-1 at $z_{\rm phot}\simeq15$, has been subsequently detected at the slightly lower redshift $z_{\rm spec}=13.53^{+0.05}_{-0.06}$ \citep{Donnan2026} in JWST Cycle 4 (PID 6954; PIs C. Donnan, D. McLeod) as a consequence of line-of-sight absorption by a damped \lya\ system (DLA) which are known to bias photometric redshifts high \citep[i.e. $z_{\rm phot}>z_{\rm spec}$;][]{Asada2025, Heintz2026}. These are expected to become increasingly common and more prominent at high redshift as the time reduces between the onset of accretion from cold gas reservoirs to form early stars and galaxies.

Star formation rate density (SFRD; $\rho_{\rm SFR}$) estimates at $z\gtrsim13$ inferred from \jwst\ photometry, however, remain inconsistent between studies due to differing selection methodologies, especially towards the sensitivity limit of NIRCam surveys. It is therefore of the utmost importance to construct datasets in a homogeneous way such that biases and uncertainties can be fully defined and measured, enabling more reliable and reproducible results. We started this with the \epochsone\ paper series \citep[e.g.][]{Conselice2025} of $>10$ papers examining the properties of $1165$ galaxies at $6.5<z<16.5$ found in $\sim180\,\rm arcmin^2$ deep imaging from the Cosmic Evolution Early Release Science \citep[CEERS;][]{Bagley2023-CEERS}, the North Ecliptic Pole Time Domain Field \citep[NEP-TDF;][]{Jansen2018}, the Next Generation Deep Extragalactic Public \citep[NGDEEP;][]{Bagley2024-NGDEEP}, the Grism Lens-Amplified Survey from Space \citep[GLASS;][]{Treu2022-GLASS}, the first epoch of the JWST Advanced Deep Extragalactic Survey \citep[JADES;][]{Bunker2020, Rieke2020, Rieke2023b, Eisenstein2026} imaging in the Great Observatories Origins Deep Survey \citep[GOODS;][]{Giavalisco2004} South field, and behind the SMACS-0723, MACS-0416, and El-Gordo strong-lensing clusters. 

In this new \epochstwo\ paper series, we greatly enlarge our sample of distant $6.5<z<16.5$ galaxies to $2452$ over $\sim650-740\,\rm arcmin^2$ in Ultra Deep Survey (UDS) and COSMOS Public Release IMaging for Extragalactic Research (PRIMER) imaging, as well as CEERS, NEP-TDF, NGDEEP, and the full JADES GOODS-South/North surveys. This sample is entirely constructed using functionality available in the introduce the public \galfind\ python software package, which we formally introduce in this paper. The highly modular \galfind\ ``photometric toolbox'' rapidly produces galaxy candidates and properties from reduced imaging and allows for a plethora of catalogs to be constructed using, for example, different aperture sizes, deblending methodologies, detection bands, SED fitting templates/tools, selection functions, etc. Functionality is highlighted whenever used throughout this paper, making results in this \epochstwo\ series almost entirely reproducible. In this data release paper, we present properties of the full photometric sample in \autoref{sec:properties} and take a closer look at $13.5<z<16.5$ candidates in \autoref{sec:highz_sources}. In future papers, UVLFs, global stellar mass functions, and the total ionizing output will be produced from this sample with greater precision than previous \jwst\ studies, all done using the \galfind\ package. 

As usual, we assume a flat $\Lambda\rm CDM$ cosmology with $H_0 = 70\,\rm km\,s^{-1}\,Mpc^{-1}$, $\Omega_{\Lambda}=0.7$, and $\Omega_{\rm m}=0.3$ in this work. The AB magnitude system is used throughout \citep{Oke1974, Oke1983}.

\section{Data products} 
\label{sec:data_products}

\subsection{Fiducial data reduction procedure}


We use our own modified version of the official \jwst\ pipeline for data reduction, adopting either Calibration Reference Data System (CRDS) v1210, v1293, or v1364 depending on the survey in question (our internal v12, v13, and v14 versions respectively). Custom wisp templates from \citet{Adams2025} are used in between stage 1 and 2, derived from a stack of all \epochsone\ reduced imaging for the F150W, F200W widebands and, when used, the F182M and F210M medium band filters. After stage 2, we apply the $1/f$ noise correction derived by Chris Willott\footnote{\url{https://github.com/chriswillott/jwst}}. A custom 2D sky background is performed with \photutils\ \citep{Bradley2022} instead of the default stage 3 sky background subtraction. Our WCS is derived from the alignment of the F444W imaging to Gaia DR3 \citep{GAIA-DR3} before all other images (both from NIRCam and \hst's Advanced Camera for Surveys (ACS) Wide Field Channel (WFC)) are matched to the same WCS using \tweakreg\ and XY-pixel aligned using \reproject\ \citep{Robitaille2020} on a $0.03$\arcsec\ pixel scale. 

\begin{figure*}
    \centering
    \includegraphics[width=1.0\linewidth]{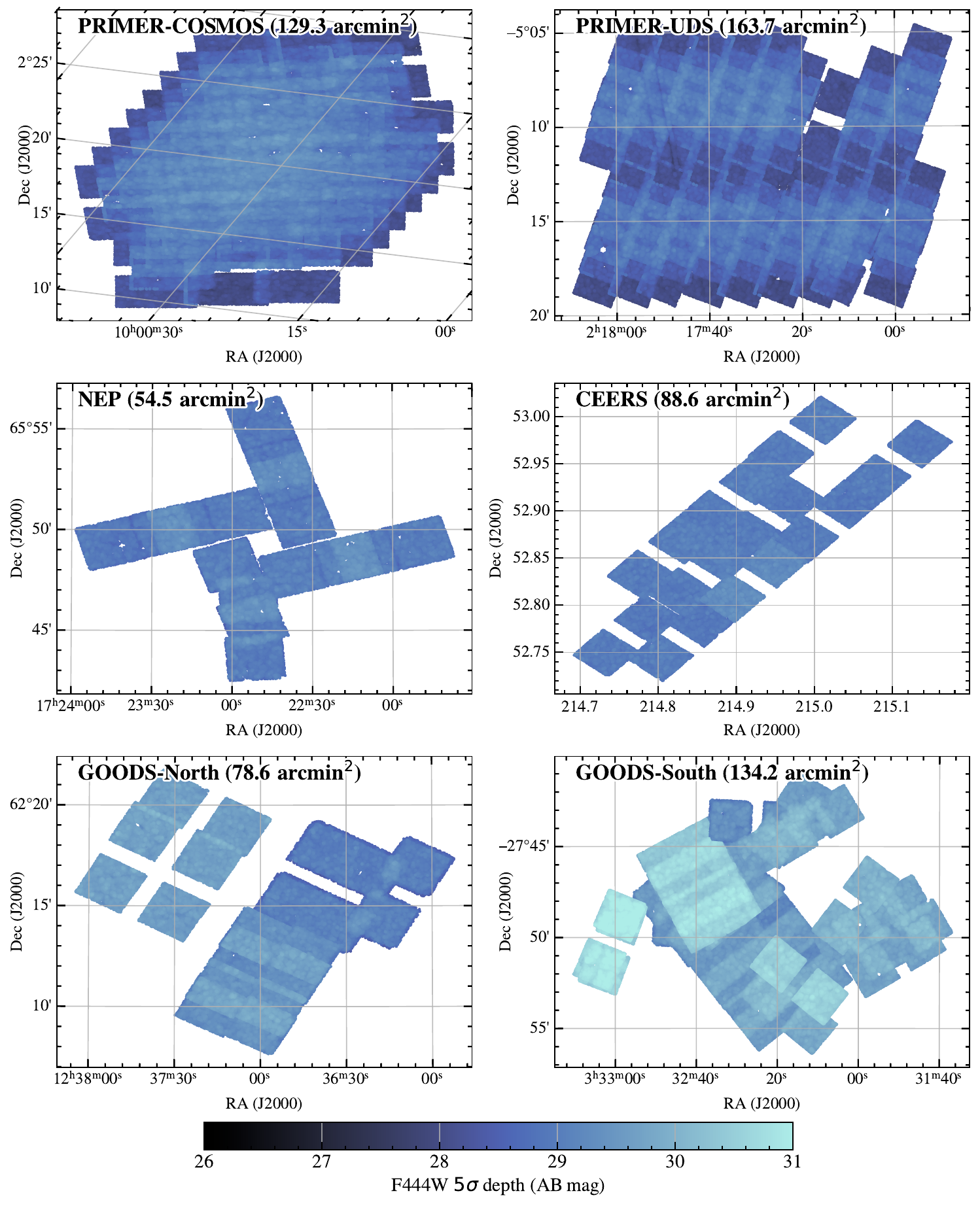}
    \caption{Sky maps of the 6 deep and wide area surveys used in the \epochstwo\ sample, including PRIMER, NEP-TDF, CEERS, NGDEEP, and JADES across the UDS, COSMOS, NEP, EGS, and GOODS-North/South fields. The unmasked areas of each survey used for sample selection are given in the upper left hand side of each plot, with maps colored by F444W local depth calculated via the method explained in \autoref{sec:EPOCHS_v2_cataloguing_and_depths}.}
    \label{fig:EPOCHS_v2_data}
\end{figure*}

In the subsections below, we outline data reduction specifics for each of the fields used in this work. These surveys are displayed in \autoref{fig:EPOCHS_v2_data}, colored by F444W local depth (see \autoref{sec:EPOCHS_v2_cataloguing_and_depths} for a discussion as to how these are calculated).

\subsubsection{JADES (GOODS-North/South)}

The deepest data used in this paper series comes from the JWST Advanced Deep Extragalactic Survey \citep[JADES;][see also \citealt{Bunker2024} and \citealt{Hainline2024a}]{Eisenstein2026}. This includes the F090W, F115W, F150W, F200W, F277W, F335M, F356W, F410M, and F444W NIRCam data presented in \citealt{Rieke2023b} (DR1) and additional area from PIDs 1180 (PI: D. Eisenstein), 1286 (PI: L{\"u}tzgendorf), and 1287 (PI: Isaak). As well as data in the Great Observatories Origins Deep Survey (GOODS)-South, the JADES program includes slightly shallower data ($5\sigma$ depths $m_{\rm F277W}\simeq29.2-29.8$) in GOODS-North from program 1181 (PI: D. Eisenstein) using F090W, F115W, F150W, F200W, F277W, F335M, F356W, F444W. F070W is additionally used across a three NIRCam pointings worth of sky area. We divide the full JADES-GS area into 4 and JADES-GN into 3 sky regions and use CRDS v1293 to reduce the data following the prescription described above.

The JADES-DR3 data includes the JADES Origins Field (JOF; \citealt{Eisenstein2023_JOF}; PIDs 1210, 3215), which is a combined JWST Cycle 1 and 2 survey located in the flanking portion of the GOODS-South field. This field contains 14 JWST/NIRCam bands, including both SW (F162M, F182M, F210M) and LW (F250M, F300M, F335M, F410M) medium band imaging for more precisely determined photometric properties and redshifts. Also included in GOODS-South is the JWST Extragalactic Medium-band Survey (JEMS; \citealt{Williams2023}; PID: 1963; PIs: C. C. Williams, S. Tacchella \& M. V. Maseda), which contains a single pointing of the less sensitive F070W, F162M, F182M, F210M, F250M, and F300M bands, which are also used sporadically throughout the rest of the GOODS-South area.

HST ACS-WFC legacy data for the F435W, F606W, F775W, F814W, and F850LP filters is taken from v2.5 of the HLF \citep[][]{Illingworth2016, Whitaker2019-HLF}. The total usable (i.e. unmasked) area for robust photometric candidate selection is $\simeq134.2\,\rm arcmin^2$ in GOODS-South and $\simeq78.6\,\rm arcmin^2$ in GOODS-North.

\subsubsection{NGDEEP}

As part of the GOODS-South dataset we include imaging from the Next Generation Deep Extragalactic Exploratory Public \citep[NGDEEP, PID: 2079, PIs: S. Finkelstein, C. Papovich, N. Pirzkal;][see also \citealt{Leung2023}]{Bagley2024-NGDEEP} survey. Our initial data reduction using the first imaging epoch lead to the discovery of a single $z_{\rm phot}=15.6^{+0.4}_{-0.3}$ candidate \citep[NGD-z16a from][also selected as NGD-1156 in this work]{Austin2023}. The updated LW in-flight flat-fields used alongside CRDS v1364, combined with the second epoch of NIRCam imaging, has brought about a vast $\Delta m_{\rm F444W}\sim1.0-1.4$ improvement in median image depth. The NGDEEP NIRCam filters are identical to those used in the CEERS survey, and F435W, F606W, F775W, F814W, and F850LP broadband HST/ACS-WFC imaging is collated from the Hubble Ultra Deep Field (HUDF) second parallel field in v2.5 of the HLF archive \citep{Illingworth2016, Whitaker2019-HLF}. 

\subsubsection{PRIMER (UDS/COSMOS)}
\label{sec:primer_data}

We reduce F090W, F115W, F150W, F200W, F277W, F356W, F410M, and F444W imaging from the PRIMER survey (PI: J. Dunlop, PID: 1837) in the Ultra Deep Survey (UDS; $163.7\,\rm arcmin^2$) and Cosmological Evolution Survey (COSMOS; $129.3\,\rm arcmin^2$) using CRDS v1210. Since these surveys have tiered exposure times, we split both fields into ``Wide'' and ``Deep'' subregions using a K-means clustering \citep{MacQueen1967} algorithm applied to the weight map to reduce systematics in the local depth measurements.

For the PRIMER-UDS field, an additional step before $1/f$ noise removal in an attempt to reduce the ``claw-like'' artefact, the cause of which is not well established in the literature; the image was rotated $100.25^{\circ}$ to make the artefact horizontal along the x-axis, before removing the median for each row along that axis and rotating the image back to its original orientation. We note that this did not completely remove the artefact, so we also take extreme care when handling sources in this sky area. In addition to this, our UDS mosaic has 2/3 missing NIRCam exposures, reducing the usable sky area by $\sim5-10~\mathrm{arcmin}^2$.

In addition to the NIRCam data, we use F606W and F814W imaging from CANDELS \citep[][]{Koekemoer2011-CANDELS, Grogin2011-CANDELS} to provide non-detections for our $z\simeq6$, crucial for the robust identification of these sources at these redshifts.

\subsubsection{North Ecliptic Pole}

As well as the JADES and PRIMER surveys, we additionally include data across $\sim54.5\,\rm arcmin^2$ in the North Ecliptic Pole Time Domain Field \citep[NEP-TDF;][]{Jansen2018}. This data is taken from the Prime Extragalactic Areas for Reionization and Lensing Science \citep[PEARLS; PIDs: 1176, 2738; PIs: R. Windhorst \& H. Hammel;][]{Windhorst2023} and was reduced using CRDS v1364. We also include bluer ACS-WFC F435W and F606W imaging in the NEP-TDF from the GO-15278 (PI: R.~Jansen) and GO-16252/16793 \citep[PIs: R.~Jansen \&
N.~Grogin, see][]{OBrien2024} HST programs.

\subsubsection{CEERS}

The Early Release Science (ERS) Cosmic Evolution Early Release Science Survey \citep[CEERS, PID: 1345, PI: S. Finkelstein;][]{Bagley2023-CEERS} is a 10 NIRCam pointing program covering $\sim 88.6\,\rm arcmin^2$ in the Extended Groth Strip \citep[EGS;][]{Groth1994}. This comprises the bulk of early JWST UVLF \citep[e.g.][]{Bouwens2023, Donnan2023, Harikane2023, Adams2024, Finkelstein2024} and morphological studies \citep[e.g.][]{Ferreira2022, Kartaltepe2023, Ferreira2023, LeConte2024, Ormerod2024} and is also included in the \epochsone\ series. We reduce the NIRCam data using CRDS v1364 in 6 broadband filters (F115W, F150W, F200W, F277W, F356W, F444W) as well as the F410M medium band filter. Notably this excludes F090W which is crucial for narrow redshift PDF solutions at $z\simeq6$; this dataset is therefore not included in the 
\ha\ study by \citet{Austin2025b}. Similarly to the UDS and COSMOS fields, we additionally include deep F606W and F814W HST/ACS-WFC data from CANDELS \citep{Koekemoer2011-CANDELS, Grogin2011-CANDELS}.

\subsection{PSF homogenization}
\label{sec:PSF_homogenization}

We use the functionality in {\tt galfind.imaging.PSF} (derived from a modified version of \aperpy\footnote{Available on GitHub (\url{https://github.com/astrowhit/aperpy}) and Zenodo \citep{Weaver2023}}) to produce empirical point spread functions (PSFs) in each filter for each \epochstwo\ field. This methodology follows the prescription by \citet{Skelton2014} and \citet{Whitaker2019-HLF} and is implemented in \jwst/NIRCam data by \citet{Weaver2024} and \citet{Harvey2025b}, which is summarized briefly below. We first identify stellar point sources in each survey using {\tt PSF.find\_stars()} which are known to occupy the so-called ``elephant's trunk'' in size-magnitude parameter space. 

\begin{figure}
    \centering
    \includegraphics[width=0.98\linewidth]{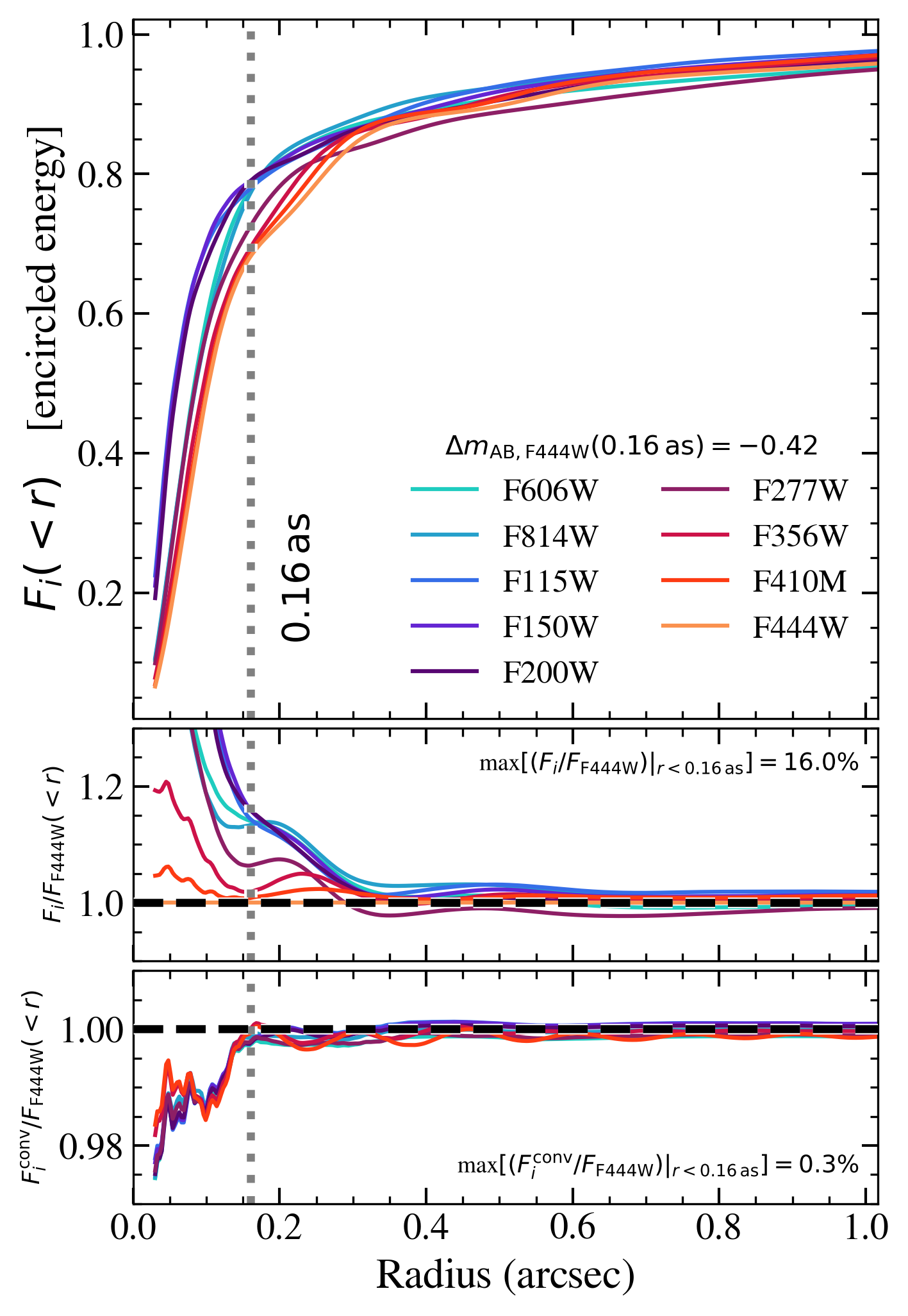}
    \caption{\textbf{(Upper panel)} Encircled energies for our empirical PSFs generated from stacked stars detected across all \epochstwo\ surveys. \textbf{(Central panel)} PSF growth curves relative to F444W. 
    \textbf{(Lower panel)} PSF growth curves compared to F444W post convolution with \pypher\ derived homogenization kernels.}
    \label{fig:empirical_psf_eecs}
\end{figure}

In this method, we use the concentration, $C=f_{\nu}(r<0.16'')/f_{\nu}(r<0.32'')$, as a size proxy and our $0.32''$ diameter aperture fluxes to compute the magnitudes. To ensure robust measurements, source centering was performed with the {\tt centroid\_com} function from {\tt photutils.centroids}. A simple power law was fitted to sources with $m_{\rm AB}<25$ and $1.5<C<5.0$, where sources beyond $3.5\sigma$ from the best-fit line are sigma clipped out. These candidate stars are then stacked using {\tt PSF.stack\_stars()}; $\mathrm{SNR}>1000$ sources are centered and stacked (sigma clipping at $<2.8\sigma$ on a pixel-by-pixel basis), and those with centers of mass $>3.5\,\rm pix$ from the flux peak removed before repeating the procedure. The PSFs were then normalized to STScI tabulated encircled energies within a $4.0\,\rm arcsec$ radius\footnote{\hst\ ACS/WFC: \url{https://www.stsci.edu/hst/instrumentation/acs/data-analysis/aperture-corrections}; \jwst\ NIRCam: \url{https://jwst-docs.stsci.edu/jwst-near-infrared-camera/nircam-performance/nircam-point-spread-functions}}. The approximate number density of stars used in the final PSF construction range from $\sim0.05\,/\,\rm arcmin^{-2}$ (F814W) to $\sim0.30\,/\,\rm arcmin^{-2}$ (F090W), coinciding with between $35$ and $200$ across the entire \epochstwo\ sky area. The full \galfind\ implementation of this procedure can be found in {\tt{galfind.imaging.PSF\_Cutout.from\_empirical\_psf()}}. All stars were then stacked to produce global empirical PSF models, the encircled energy (EE) curves for which are displayed in the upper panel of \autoref{fig:empirical_psf_eecs}. 

Once we have produced our PSF cutouts, we compute homogenization kernels for each filter using \pypher\ \citep{Boucaud2016} with a regularization parameter $10^{-4}$ to the broadest wideband empirical F444W PSF run with the {\tt{psf\_homogenize("F444W")}} method from the {\tt galfind.imaging.Data} class. Convolution kernels are run with $3\times$ oversampling and later rescaled to our $30\,\rm mas$ pixel scale. This method additionally uses these kernels to PSF homogenize the science, error, and weight maps to the reference filter, in this case F444W, for each field. The central panel of \autoref{fig:empirical_psf_eecs} displays the curve of growth of each PSF relative to F444W; the lower two panels show the filter curve of growth relative to F444W pre and post PSF homogenization, demonstrating an improvement from $16.0\%$ to $0.3\%$ at $r<0.16\,\rm \arcsec$.


While we do not study the NIRCam PSFs in detail in this paper series, we compare our results to \stpsf\ \citep{Perrin2012-WebbPSF, Perrin2014-WebbPSF} models extracted on the approximate date in question in the same sky area. This minimizes the impact of wavefront variation stemming from minor misalignment between the $18$ primary mirror segments. As found in many previous studies \citep[e.g.][]{Ono2023, Ito2024}, we find that our empirical PSFs are between $10-20\%$ wider at $r=0.1\,\rm arcsec$; this can be explained by adding telescope jitter \citep{Morishita2024} which broadens the \stpsf\ models.

\subsection{Photometric cataloguing pipeline}
\label{sec:EPOCHS_v2_cataloguing_and_depths}

High-quality and reproducible photometric catalogs from these six \jwst\ blank fields are produced using methods from the {\tt galfind.imaging.Data} class. We use {\tt segment()} and {\tt perform\_forced\_phot()} to perform image segmentation, deblending, and extraction in a inverse variance weighted stack of the JWST/NIRCam LW widebands (F277W+F356W+F444W) in $0.32$\arcsec\ diameter circular apertures using \sextractor\ {\tt v2.25.0} \citep{Bertin1996-SExtractor}. A similar setup to \citet{Adams2023} is used with a couple of notable differences to aid deblending of compact, faint sources, which are detailed below. \sextractor's {\tt DEBLEND\_NTHRESH} (the number of deblending sub-thresholds) are varied from $32$ to $64$ and {\tt DEBLEND\_MINCONT} (the minimum contrast parameter for deblending) from $5\times10^{-3}$ to $10^{-4}$ in a trial and error method until successful deblending of all spectroscopic sources presented in \autoref{sec:spec_confirmed_sources} is achieved. We note that the increased ``shredding'' of clumpy and extended galaxies at lower redshifts into multiple sources comes as a natural consequence. 

Following this, we produce fits masks for each filter in each field to identify and remove image edges and stellar diffraction spikes using the {\tt mask()} method. By default this runs an automated masking procedure, masking pixels within $50~\rm pix$ of image edges and uses appropriately scaled stellar templates to mask bright stars and their associated diffraction spikes in the image using the position of known stars from the GAIA-DR3 catalogue. For the NEP-TDF and GOODS-South (excluding NGDEEP) surveys with more complex geometries and multiple PAs, we use {\tt mask(method="manual")} to use our manually created masks produced from ds9 ``.reg'' files.

Depths are calculated with {\tt run\_depths()}, which inserts $0.32$\arcsec\ diameter apertures into blank regions as defined by the \sextractor\ segmentation map and the image mask, with the local depth calculated as the NMAD (normalized median absolute deviation) of the closest $200$ apertures to each source. Appropriate flux errors with a minimum $5\%$ error floor are computed from the local depths, and the empirical PSFs outlined in \autoref{sec:PSF_homogenization} are used to aperture correct the raw flux measurements. Image depths and areas are presented in \autoref{tab:EPOCHS_v2_ACS+NIRCam_SW_depths}, \autoref{tab:EPOCHS_v2_NIRCam_LW_depths}, and \autoref{tab:EPOCHS_v2_areas} in Appendix~\ref{sec:areas_depths}, where further details regarding the splitting of fields into regions of differing depth/filter coverage are provided.

\section{Sample}
\label{sec:sample}

\subsection{Sample selection criteria}
\label{sec:EPOCHS_v2_sample_selection}

To robustly select a sample of $z>6.5$ galaxies, we use a slightly modified \epochsone\ selection criteria from \citet{Conselice2025} to account for the reduction in minimum flux percentage error from $10\%$ to $5\%$. We run \eazypy\ \citep{Brammer2008-EAZY} using $4$ different template sets to calculate photometric redshifts, including the ``fsps\_larson'' \citep{Larson2023a} and ``fsps\_jades'' \citep{Hainline2024a} template sets part-built using the Flexible Stellar Population Synthesis (FSPS) code \citep{Conroy2010_software}, as well as the ``sfhz'' \citep{Brammer2008-EAZY} and ``sfhz\_blue\_agn'' sets, the latter of which includes a model replicating the $\mathrm{ID}=4590$ extreme emission line galaxy in the SMACS-0723 cluster at $z\simeq8.5$ from \citet{Carnall2023}. For each template set we include the damped \lya\ (DLA) photo-$z$ correction procedure using the average $N_{\rm HI}$ column densities from \citet{Asada2025}, removing the systematic photo-z overestimation at the highest redshifts. The sample selection criteria using the photo-$z$ PDFs from \eazypy\ are as follows:

\begin{enumerate}
    \item All bands entirely bluewards of \lya\ should be $<2.0\sigma$ detected ($\mathrm{SNR}_{\rm blue\,Ly\alpha}<2.0$).
    \item The first two widebands entirely redwards of \lya\ should be $>5\sigma$ detected ($\mathrm{SNR}_{\rm red\,Ly\alpha}>\{5.0, 5.0\}$).
    \item All widebands redwards of \lya\ should be $>2\sigma$ detected ($\mathrm{SNR}_{\rm red\,Ly\alpha}>2.0$).
    \item The galaxy \eazypy\ SED should be reasonably well fitting ($\chi_{\mathrm{red, gal}}^2<6.0$).
    \item The $\chi^2$ from the \eazypy\ run limited to $z_{\rm max}=6.0$ should be worse fitting with $2\sigma$ significance: 
    \begin{equation*}
        \Delta\chi^2_{\rm low-z}=\chi^2_{z_{\mathrm{max}}=6.0}-\chi^2_{z_{\mathrm{free}}}>4.0.
    \end{equation*}
    \item The best-fitting \eazypy\ redshift PDF should be tightly constrained such that $>60\%$ of the PDF lies within $10\%$ of the best-fitting redshift, $z_{\rm best}$:
    \begin{equation*}
        \int^{1.1\times z_{\rm best}}_{0.9\times z_{\rm best}} P(z)\mathrm{d}z>0.6.
    \end{equation*}
    \item The \sextractor\ half light radii should exceed $R_{\rm e}>45.0\,\rm mas$ ($1.5~{\mathrm{pix}}$) in every selection band (F277W, F356W, F444W) to remove any potential hot pixels.
    \item Unmasked in at least 2 ACS/WFC bands and 6 NIRCam bands, which must include F277W, F356W, and F444W, as well as the first two wideband filters redwards of \lya\ and at least one bluewards of \lya.
\end{enumerate}

These criteria are computed using the {\tt galfind.selection.EPOCHS\_Selector} called on a {\tt galfind.catalogue.Catalogue} object which loops through an array of {\tt galfind.catalogue.Galaxy} objects outputting the boolean results and keyword arguments to a FITS formatted table. In addition to the above selection criteria, $38$ ($1.5\%$) objects impacted by unmasked diffraction spikes and close to NIRCam SW image edges, primarily in the PRIMER-UDS field, are removed by eye. For selection criteria \#$(1,2,3)$, the band order (blue $\rightarrow$ red) is determined using the central filter wavelengths excluding F070W (due to its sporadic use and the common depth differential with deep F606W and F814W ACS/WFC imaging) and F850LP (due to its near complete overlap with F090W). The wavelength limits are taken to be those at which there is $50\%$ filter transmission.

The above criteria are used to select a sample of $2,511$ candidates at $6.5<z<13.5$. To account for the increasing contamination observed at $13.5<z<16.5$ (see \autoref{sec:specz_photoz_comparison}), we tighten selection criteria \#2 to $\mathrm{SNR}_{\rm red\,Ly\alpha}>\{8.0, 8.0\}$ in this redshift regime. $127$ and $227$ dithered NIRCam LW hot pixels in the PRIMER-COSMOS and PRIMER-UDS fields respectively pass criteria \#7 and masquerade as dropout candidates at $z\simeq 16.1-16.5$ are removed from the sample by visual inspection, producing a fiducial sample of $73$ $13.5<z<16.5$ candidates.

\begin{figure*}
    \centering
    \includegraphics[width=1.0\linewidth]{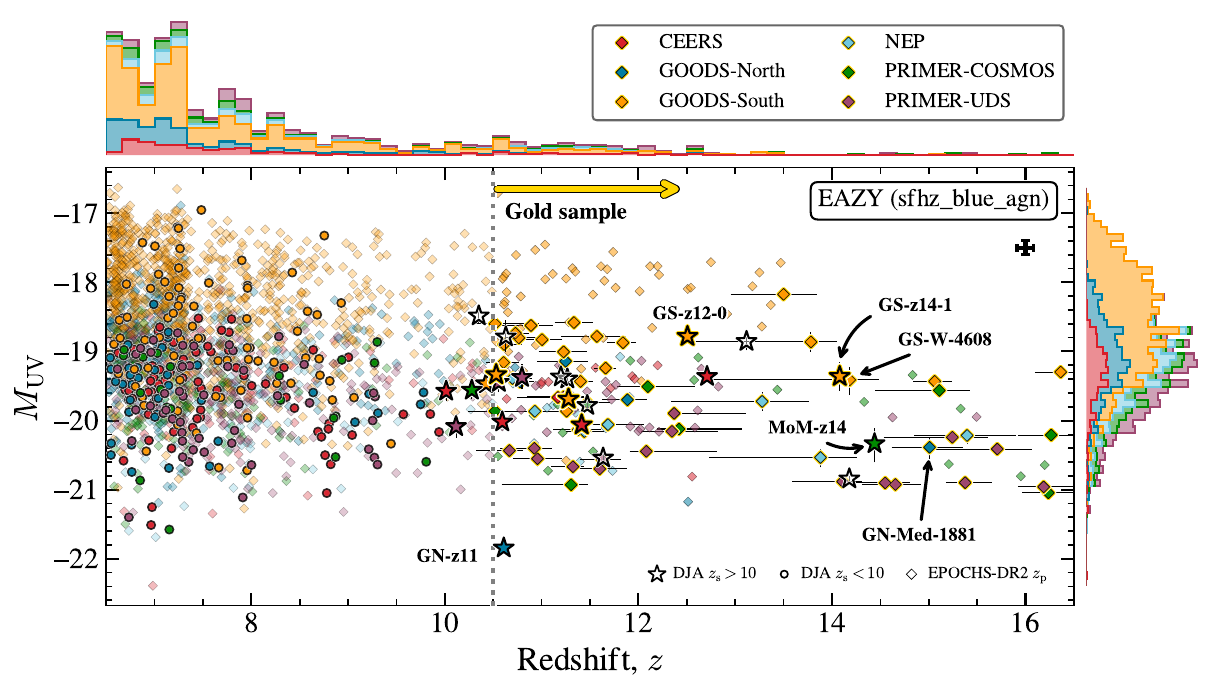}
    \caption{Overview \Muv$-z$ parameter space for photometric galaxy candidates (diamonds) and spectroscopically confirmed sources from the DJA (circles at $z_{\rm spec}<10$, stars at $z_{\rm spec}>10$) across the \epochstwo\ fields. Properties of the high-redshift spectroscopic sample are given in \autoref{sec:spec_confirmed_sources}. Sources are colored by the fields from which they are identified, which includes the NEP, CEERS, GOODS North and South, and PRIMER. Objects that are selected by our ``gold'' photometric selection criteria at $z>10.5$ are appropriately outlined in gold, which are plotted with errorbars included. The median redshift and \Muv\ error and axes histograms are shown for the fiducial photometric sample.}
    \label{fig:Muv_z}
\end{figure*}

In addition we introduce a ``gold sample'' of sources at $z_{\rm phot}>10.5$ discussed further in \autoref{sec:highz_candidates}. To produce this sample, criteria $\#2$ is tightened to $\mathrm{SNR}_{\rm red\,Ly\alpha}>\{8.0, 8.0\}$ at $10.5<z<13.5$ and $\mathrm{SNR}_{\rm red\,Ly\alpha}>\{10.0, 8.0\}$ at $13.5<z<16.5$ to further reduce the contamination. An additional stacked $\mathrm{SNR^{stack}_{blue\,Ly\alpha}}<2.0\sigma$ for all filters bluewards of \lya\ is required, excluding a further $4$ sources. Finally, we remove $5$ source by eye at $10.5<z<13.5$ ($10\%$) and $25$ at $13.5<z<16.5$ ($60\%$) that are in the diffuse outskirts of brighter sources, leaving a final ``gold'' sample size of $62$ at $z_{\rm phot}>10.5$. A summary of these fiducial and ``gold'' samples is presented in \autoref{fig:Muv_z}, where the UV magnitudes are computed directly from the rest-frame photometric fluxes (see \autoref{sec:UV_properties}) and spectroscopic cross-matches from the DJA (see \autoref{sec:specz_photoz_comparison}) are shown for comparison.

\subsection{Brown dwarf removal}
\label{sec:brown_dwarfs}

The \epochstwo\ selection criteria outlined in \autoref{sec:EPOCHS_v2_sample_selection} produces a sample of galaxy candidates without consideration of contamination from compact foreground Milky Way brown dwarfs. These point-source brown dwarf populations masquerade as dropout galaxies and, while the number density at $m_{\rm AB}\gtrsim26$ is not well constrained, they have been found to comprise a sizable proportion of high-redshift samples at $z\gtrsim6$ produced from \jwst/NIRCam \citep[e.g.][]{Langeroodi2023, Hainline2024b} and \euclid/NISP \citep[e.g.][]{Weaver2025} photometry respectively.

In an attempt to alleviate this issue, we fit L, T, and Y--type brown dwarf templates from the {\tt sonora bobcat} \citep{Marley2021}, {\tt cholla} \citep{Karalidi2021}, {\tt diamondback} \citep{Morley2024}, {\tt elf owl} \citep{Mukherjee2024}, and {\tt LOWZ} \citep{Meisner2021} libraries to our ACS-WFC+NIRCam photometry using the \bdfinder\ python package\footnote{\url{https://github.com/tHarvey303/BD-Finder}}. Brown dwarf candidates are selected on the basis of $\chi^2_{\rm red, BD}<\chi^2_{\rm red, gal}$. Our brown dwarf candidates must additionally fulfill our compactness criterion, $C/C_{\rm PSF}<1.1$, where $C=f_{\nu}(r<0.32'')/f_{\nu}(r<0.16'')$ is the source compactness and $C_{\rm PSF}$ is the compactness measured from the empirical F444W PSF (see \autoref{sec:PSF_homogenization} for details of its computation) which is approximately $C_{\rm PSF}\simeq1.3$ and fairly consistent across our surveys.

\begin{figure*}
    \centering
    \includegraphics[width=0.95\linewidth]{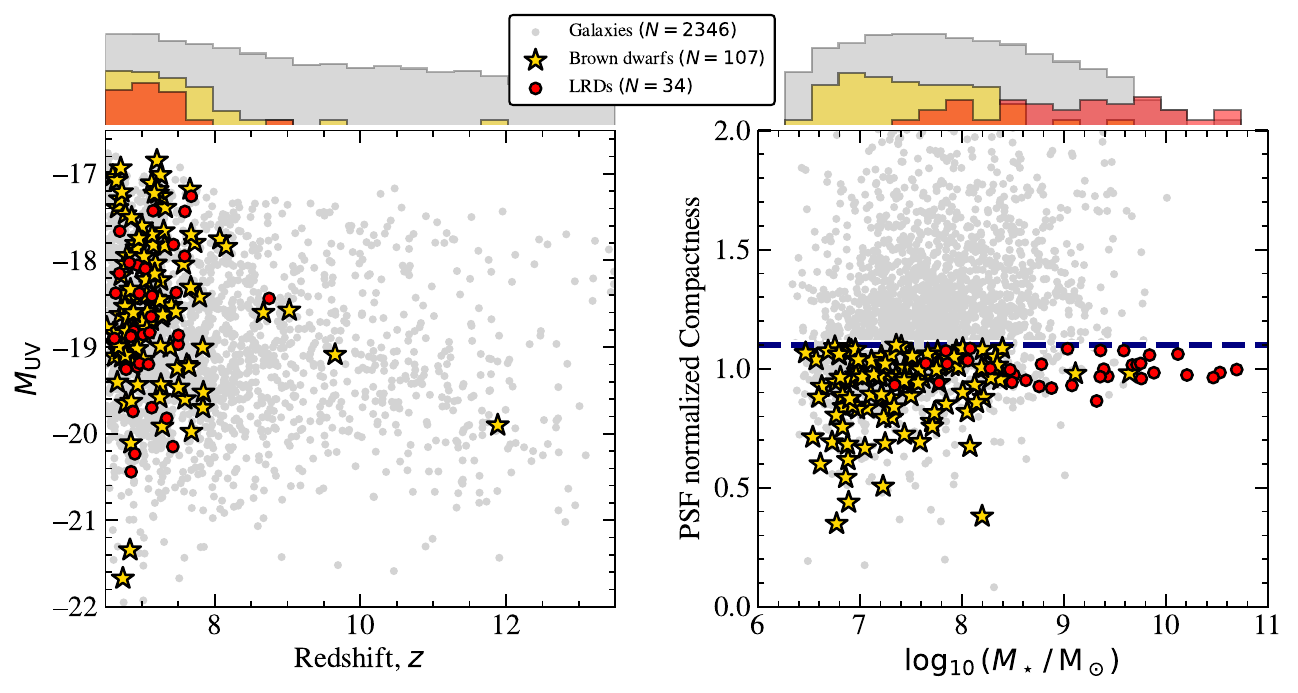}
    \caption{Brown dwarf/LRD diagnostic plot showing the inferred $z_{\rm phot}-M_{\rm UV}$ parameter space (\textbf{left panel}) and PSF normalized compactness, $C/C_{\rm PSF}$, as a function of fiducial \bagpipes\ stellar mass (\textbf{right panel}). The $107$ brown dwarfs and $34$ LRDs are shown as gold stars and red dots respectively, with the x-axis histograms shown in log-space against the fiducial \epochstwo\ sample in gray.  Note that the redshifts of the brown dwarfs are for illustrative purposes to show were they would be found assuming they were galaxies.}
    \label{fig:bd_lrd_schematic}
\end{figure*}

In total, $107$ brown dwarf candidates are selected by the fiducial $6.5<z<13.5$ selection criteria and thus removed from our \epochstwo\ galaxy sample. The vast majority ($69$) are L--type with surface temperatures $1300<T\,/\, \rm K<2400$, while $36$ are T--type with $600<T\,/\,\rm K<1300$, and $2$ are Y--dwarf candidates with $T=600\,\rm K$. \autoref{fig:bd_lrd_schematic} shows that these primarily exist in the $6.5<z<7.5$ redshift bin, $\sim 1-2\,\rm mag$ below the knee of the UVLF, and with stellar masses $\log_{10}(M_{\star}\,/\,\rm M_{\odot})\sim7.0-8.5$ (see \autoref{sec:bagpipes} for the details of the SED fitting).
A single source within our sample ({\tt SURVEY\_ID=14287} in the Southern tile of GOODS-South) is classified as both a brown dwarf and LRD (as outlined in \autoref{sec:LRDs}). It is best-fit by the ``t1300g31f3\_m0.0\_co1.0\_resample'' brown dwarf template from the {\tt sonora\_diamondback} library with temperature $T=1300\,\rm K$, making it spectral type L9--T2.

Since our \epochstwo\ sample is already selected based on a well fitting \eazy\ SED with $\chi^2_{\rm red, gal}<6.0$ (criteria $\#4$), it follows that this brown dwarf sample is not complete and only includes ``galaxy-like'' brown dwarfs. As a consequence of the larger number density of compact $z\simeq7-7.5$ galaxies than Milky Way L-- and early T--type brown dwarfs, we note that many of these brown dwarf candidates are likely compact galaxies with broadband photometry scattering into the $\chi^2_{\rm red, BD}<\chi^2_{\rm red, gal}$ parameter space. This template degeneracy can be split either with spectroscopic data \citep[e.g.][]{Burgasser2024, Tu2025} or proper motion searches at NIR wavelengths \citep[see e.g.][]{Hainline2024b, Hainline2026b, Liu2026}, which we leave for a future study.

\subsection{LRD identification}
\label{sec:LRDs}

As well as the $107$ brown dwarfs, we additionally identify $34$ compact ``little red dots'' (LRDs) with blue rest-UV $\rm F150W - F200W<0.8$, and red rest-optical $\rm F277W - F444W>0.7$ and $\rm F356W - F444W>0.6$ colors, corresponding to the {\tt red2} selection criteria for $z>6$ galaxies from \citet{Kokorev2024}. A wide variety of alternative photometric selection techniques exist \citep[e.g.][]{Greene2024, Hainline2025, Kocevski2025}, however we do not provide LRD samples using these criteria in this work.

As expected, many of these sources occupy the high mass regime, as seen in the right hand panel of \autoref{fig:bd_lrd_schematic}. In fact, almost all sources with $M_{\star}>10^{10}\,\rm M_{\odot}$ are classified as LRDs when fit with stellar only templates, and thus the impact of these sources on the high mass end of the global stellar mass function is clear. While much work has been spent to determine whether these sources are of stellar or AGN origin \citep[e.g.][]{Matthee2024, Carranza-Escudero2025, Rusakov2026}, the question is still mostly unanswered and hence this remains a very active area of research. Since any robust conclusions on this topic require high SNR spectroscopic data, this photometric study simply provides a high-redshift LRD database for future spectroscopic follow-up and legacy science usage.

\subsection{Spectroscopic redshift comparison}
\label{sec:specz_comparison}

\subsubsection{Spectroscopic sample}
\label{sec:spec_sample}

To determine the quality of our photometric redshifts, we compare to a spectroscopic redshift catalog comprising NIRSpec PRISM/CLEAR sources collated from v4.4 of the DAWN JWST Archive (DJA) and reduced using the \texttt{msaexp}\footnote{\doi{10.5281/zenodo.7299500}} pipeline \citep[see][]{Heintz2024b, deGraaff2025}. We cross-match our catalogs within a $0.3$\arcsec\ tolerance finding $12,141$ with unmasked ``zgrade'' spectroscopic redshifts and DJA grade 3 classifications (i.e. the most robust candidates determined by eye) from a total of $18,843$. The spectroscopic programs that contribute sources to this catalog include JADES (PI: N. L\"{u}etzgendorf; PIDs: GTOs 1210, 1212, 1286 (GOODS-South), 1213 (CEERS), 1214 (COSMOS), 1215 (UDS) + PI: K. Isaak; PID: GTO 1211 \citep[GOODS-North;][]{Maseda2024} and GTO 1287 (GOODS-South) + PI: D. Eisenstein; PIDs: GTO 1180 \citep[GOODS-South;][]{D'Eugenio2025}, GTO 1181 (GOODS-North), and GO 3215 \citep{Eisenstein2026}), UDS CAPERS (PI: M. Dickinson; PID: GO 6368), CEERS ERS \citep[PI: S. Finkelstein, PID: 1345][]{Finkelstein2023}, EGS DDT 2750 (PI: P. Arrabal-Haro), GO 4106 (PI: E. Nelson), RUBIES \citep[PID: GO 4233; PI: A. de Graaff;][]{deGraaff2025}, COSMOS DDT 6585 (PI: D. Coulter), GO 2565 \citep[PI: K. Glazebrook;][]{Nanayakkara2025} and Miracle or Mirage (MoM; PID: 5224; PI: P. Oesch) in COSMOS and UDS, as well as DDT 6541 (PI: E. Egami) and GO 2198 \citep[PI: L. Barrufet;][]{Barrufet2025} in GOODS-South. No spectroscopic data from NIRSpec/PRISM is currently available in the NEP-TDF.

To produce a 1-to-1 relation between spectroscopic and photometric sources we retain the candidate with the brightest F444W flux for each unique spectrum as well as the highest rest-frame UV SNR spectrum for each unique photometric source, reducing our catalog size to $11,038$. Of these uniquely cross-matched spectra, $672$ are identified with $z_{\rm spec}>6.5$ ($142$ in GOODS-South, $77$ in GOODS-North, $203$ in the EGS, $173$ in the UDS and $77$ in the COSMOS field), $564$ ($84\%$) of these are unmasked by the \galfind\ masking procedure and we select $293$ ($52\%$ of the unmasked spectroscopic sample). We additionally include three sources spectroscopically confirmed at $z_{\rm spec}>10$ with non-grade 3 DJA spectroscopy at their published redshifts, namely GS-20030902 and CAPERS-EGS-43539 from \citet{RobertsBorsani2025}, and MoM-z14 \citep{Naidu2026}. Details of these sources can be found in \autoref{tab:highz_spec} in \autoref{sec:spec_confirmed_sources}.

\subsubsection{Photometric-spectroscopic redshift comparison}
\label{sec:specz_photoz_comparison}

\begin{figure*}
    \centering
    \includegraphics[width=1.0\linewidth]{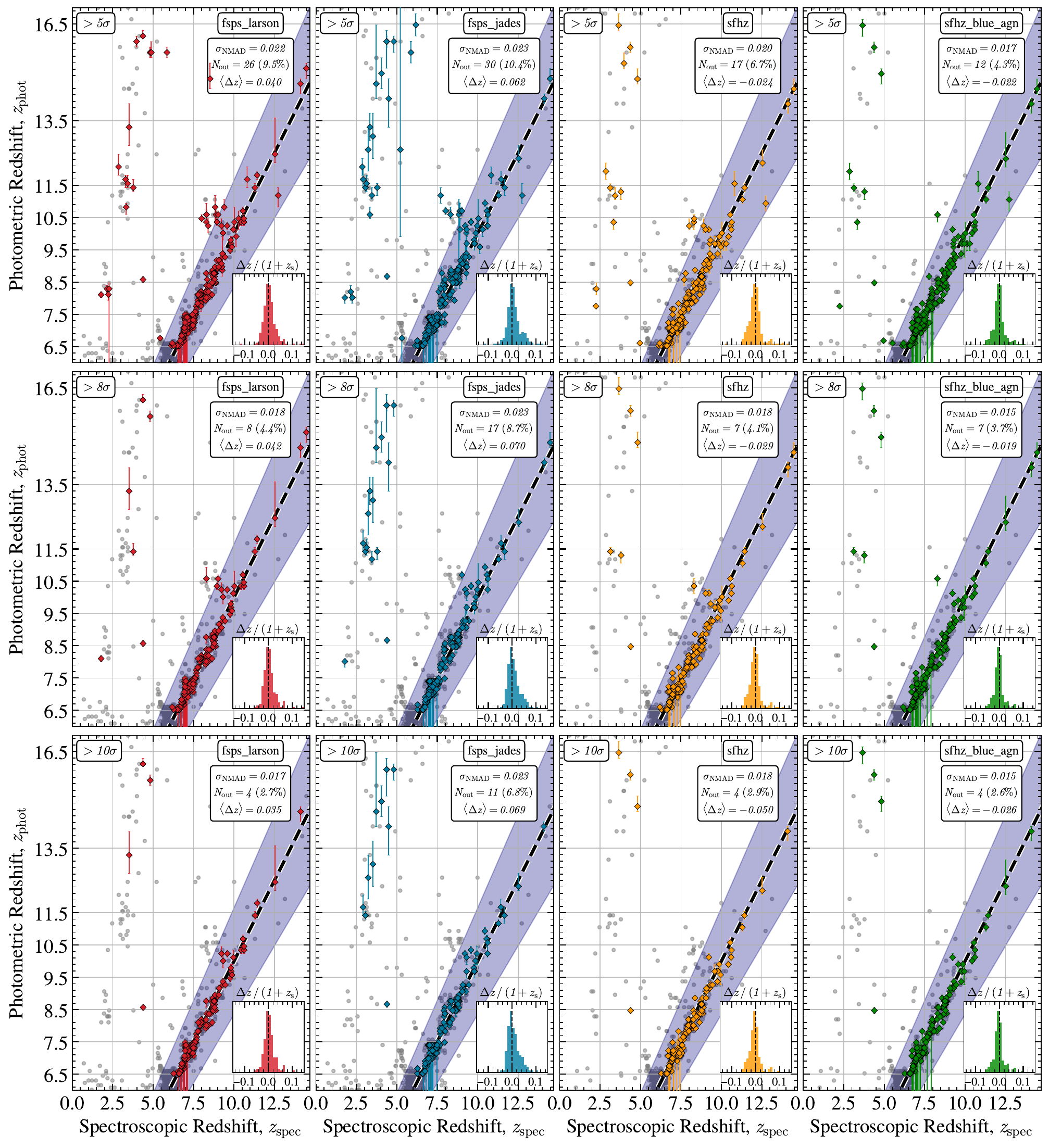}
    \caption{Photometric-spectroscopic redshift comparison for the $4$ template sets used in our \eazypy\ SED fitting procedure, where grey/colored points highlight unselected/selected cross-matches respectively. The top row shows our fiducial \epochstwo\ sample selected using the criteria in \autoref{sec:EPOCHS_v2_sample_selection}, with subsequent rows showing the impact of tightening criteria \#2 to $\mathrm{SNR}_{\rm red\,Ly\alpha}>\{8.0, 8.0\}$ (middle) and $\mathrm{SNR}_{\rm red\,Ly\alpha}>\{10.0, 10.0\}$ (bottom). Spectroscopic redshifts are collated from the DJA v4.4 grade 3 sample (using the ``zgrade'' column) outlined in \autoref{sec:spec_sample} combined with $3$ confirmed sources at $z_{\rm spec}>10$ from \citealt{Naidu2026} (MoM-z14) and \citealt{RobertsBorsani2025} (GS-20030902 and CAPERS-EGS-43539)  not classified as grade 3 in the DJA. The upper right corner highlights $\sigma_{\rm NMAD}$ (from \autoref{eq:nmad}), the number of extreme outliers outside the $15\%$ region in navy, $N_{\rm out}$ (from \autoref{eq:noutliers}), and the mean offset from the 1-to-1 line in dashed black for the non-outlier sample, $\langle\Delta z\rangle$. We adopt the ``sfhz\_blue\_agn'' results as our fiducial photo-z's since they consistently perform the best across all metrics for all SNR selection limits.}
    \label{fig:specz_vs_photoz_comparison}
\end{figure*}

The redshift comparison for the entire sample is shown as grey dots in \autoref{fig:specz_vs_photoz_comparison}. Each panel highlights the photometrically selected spectra in our sample from the different \eazypy\ template sets used in this study at different $\mathrm{SNR_{red\,Ly\alpha}}$ thresholds. From these samples we determine the best performing template set by comparing the mean offset from the 1-to-1 line, $\langle\Delta z\rangle$ (where $\Delta z = z_{\rm phot} - z_{\rm spec}$), as well as the normalized median absolute deviation,
\begin{equation}
    \sigma_{\rm NMAD}=1.48 \times\mathrm{median}\bigg(\frac{\lvert\Delta z\rvert}{1+z_{\rm spec}}\bigg),
    \label{eq:nmad}
\end{equation}
and the fraction of extreme outliers, $\eta = N_{\rm out}/N_{\rm tot}$, where the number of outliers is given by
\begin{equation}
N_{\rm out} = \sum_{i=0}^{N_{\rm tot}}\,\mathbb{1}\big[\,\lvert\Delta z\rvert_i>0.15\times(1+z_{\mathrm{spec}, i})\,\big]
\label{eq:noutliers}
\end{equation}
and $\mathbb{1}$ is the indicator function that yields $1$ if true and $0$ if false. It is worth noting that only non photo-z outliers are used in the computation of $\langle\Delta z\rangle$.

Since the ``sfhz\_blue\_agn'' template set has the smallest $\sigma_{\rm NMAD}=0.017$, the fewest percentage of extreme outliers $\eta=4.3\%$, and the smallest mean offset $\langle\Delta z\rangle=-0.022$ at $\mathrm{SNR}_{\rm red\,Ly\alpha}>\{5.0,5.0\}\sigma$, we decide to use the photometric redshifts and sample derived using this template set throughout the rest of this paper series. This includes $293$ spectroscopically confirmed and photometrically selected sources, with $88$ across GOODS-South, $45$ in GOODS-North, $82$ in CEERS, $53$ in the PRIMER-UDS footprint, and $25$ in PRIMER-COSMOS. A number of these high redshift sources are explored in \autoref{sec:highz_sources}.

The number of extreme outliers for our fiducial $\mathrm{SNR_{red \, Ly\alpha}}>\{5.0, 5.0\}\sigma$ selected sample is a mere $\eta=4.3\%$, dropping to $\eta=2.6\%$ at $\mathrm{SNR_{red \, Ly\alpha}}>\{10.0, 10.0\}$, across the full $6.5<z_{\rm phot}<16.5$ redshift range. Due to our small number statistics, we prefer to quote $68\%$ ($1\sigma$) Clopper-Pearson binomial intervals \citep{ClopperPearson1934} for these $\eta$ rather than the simpler point estimates, yielding $\eta=4.3_{-1.2}^{+1.6}\,\%$ and $\eta=2.6_{-1.2}^{+2.0}\,\%$ respectively. 
It is clear from \autoref{fig:specz_vs_photoz_comparison} that the number of catastrophic photo-z failures is redshift dependent, and thus we choose to also compute these outlier rates as a function of redshift. We find a low $\eta=1.9_{-0.8}^{+1.3}\,\%$ at $6.5<z_{\rm phot}<10.5$, rising to $\eta=40_{-18}^{+21}\,\%$ at $10.5<z_{\rm phot}<13.5$ and $\eta=60_{-30}^{+25}\,\%$ at $13.5<z_{\rm phot}<16.5$ where only two objects (GS-z14-1 from \citealt{Carniani2024} and MoM-z14 from \citealt{Naidu2026}) are correctly spectroscopically confirmed at high redshift.

\subsubsection{Comparison to other large photometric studies}
\label{sec:photometric_comparison}

To determine the quality of our \epochstwo\ sample, we cross-match our photometric catalog to several other wide-area NIRCam studies from the literature within a $0.3$\arcsec\ radius.

\textbf{EPOCHS-DR1:} The \epochsone\ sample from \citet{Conselice2025} contains $1011$ $z>6.5$ sources across $214\,\rm arcmin^2$ primarily in the NEP, CEERS, and JADES-GS regions \citep[see also][]{Adams2024,Austin2025a,Harvey2025a}. Across this area, the \epochstwo\ catalog contains $1082$ selected sources at $6.5<z<13.5$ compared to their $956$, with $489$ in common ($198$ in GOODS-South which includes the East tile and NGDEEP, $138$ in the NEP-TDF, and $153$ in the EGS). This gives an overlap of approximately $\sim 45-50\%$, rising to $\sim 55-60\%$ at $\mathrm{SNR}_{\rm F444W}>10$. Our \epochstwo\ sample has a slighter tighter NMAD and outlier fraction than the $\sigma_{\rm NMAD}=0.021$ and $\eta\sim7\%$ ($6/83$) reported for the $z_{\rm phot}>6.5$ \epochsone\ sample. This stems as a result of the switch from the ``fsps\_larson'' template set as well as the larger spectroscopic sample which is now available; $\eta\rightarrow8.4\%$ when including the single source within $\Delta z\sim0.2$ of the $15\%$ limit in their Fig.~2.


\textbf{JADES:} The JADES photometric data release by \citet{Hainline2024a} contains $717$ candidates at $z_{\rm phot}>8$ in GOODS-North and GOODS-South and $700$ at $8.0<z_{\rm phot}<13.5$. $370$ sources are selected in \epochstwo\ across the same redshift range and sky area, with $204$ ($\sim55\%$) in common, increasing to $\simeq60\%$ at $\mathrm{SNR}_{\rm F444W}>10$. This yields $\sigma_{\rm NMAD}=0.02$, $\langle\Delta z\rangle=0.16$, and no outliers at $z_{\rm spec}>7$ from $123$ spectroscopically confirmed sources for the fiducial template set used in their work. The low number of outliers may be due to their $z_{\rm spec}>7$ cut (as seen in their Figs. 10 \& 11), and the \prospector\ fits from the \citet{Hainline2026} value-added catalog made by \citet{Duan2026} shows a handful of outliers at $z_{\rm phot}>8$. In total \citet{Duan2026} find $\sigma_{\rm NMAD}=0.026$ and $\eta=7.1\%$ across all redshifts $z_{\rm phot}<15$; this increase compared to our study is likely caused by a reduced photo-z precision and accuracy at $z<6.5$ where fewer bands lie bluewards of the Lyman break.

\subsubsection{Summary of reasons for photometric selection failure}
\label{sec:photz_selection_failure_reasons}

To identify the primary reasons for \epochstwo\ non-selection, we present the number of spectra that fail each selection criteria from \autoref{sec:EPOCHS_v2_sample_selection} in \autoref{tab:spec_fail_reasons} as a function of redshift.

It is clear from \autoref{tab:spec_fail_reasons} that the $>5\sigma$ SNR criteria for the first two widebands entirely redwards of \lya\ (criteria \#2) is the primary reason for the non-selection of spectra failed by $\simeq62\%$ of the unmasked and unselected spectra and $\simeq30\%$ of all $z_{\rm spec}>6.5$ unmasked spectra. More minor failure contributions come from the $\Delta\chi^2_{\rm low-z}>4.0$ requirement (criteria \#5) and $2\sigma$ SNR detection/non-detection criteria (\#1 and \#3). We note that while we observe a low failure rate, loosening these \epochstwo\ selection  criteria to increase the sample completeness will also increase the contamination from lower-redshift sources.

Our $6.5<z<7.5$ PRIMER-COSMOS and PRIMER-UDS samples have the lowest success rate ($12/44$ ($27\%$) and $28/84$ ($33\%$) respectively) which likely emerges as a direct consequence of the inhomogeneous spectroscopic selection function and the reduced depth and wideband coverage in these fields. The PRIMER fields host the largest number of spectra from rest-frame optical selected HST dark galaxies \citep[e.g.][]{Barrufet2023, Nelson2023, Perez-Gonzalez2023} and LRDs \citep[e.g.][]{Akins2025}, which are also more prevalent at $z\simeq7$ than at $z\gtrsim12-14$ due to both hierarchical structure formation and an increase in dust reddening over time post the onset of AGB dust formation \citep[e.g.][]{Finkelstein2012, Schneider2024}. Many of these sources fail rest-frame UV SNR selection criteria and are more degenerate with Balmer break galaxies at lower redshift and thus removed by criteria \#5.

\begin{table*}
    \centering
    \setlength{\tabcolsep}{2pt}
    \fontsize{8}{10}\selectfont
    \renewcommand{\arraystretch}{1.}
    \caption{Summary of the photometric-spectroscopic cross-matched sample. Outlined are the number of NIRSpec PRISM/CLEAR ``grade 3'' spectra from the DJA cross-matched to our \epochstwo\ photometry, the number of these that remain unmasked in our photometry, and the total number (and percentage compared to the number of unmasked spectra) of these that are fiducially \epochstwo\ selected, as a function of \textit{spectroscopic redshift}. The number and percentage of sources that fail photometric selection are given in terms of the specific criteria given in \autoref{sec:EPOCHS_v2_sample_selection}.}
    \vspace{-0.8em}
    \label{tab:spec_fail_reasons}
    \begin{tabular}{l|c|c|c|c|c|c|c}
    \toprule
    & $6.5<z<7.5$ & $7.5<z<8.5$ & $8.5<z<9.5$ & $9.5<z<10.5$ & $10.5<z<11.5$ & $11.5<z<13.5$ & $13.5<z<16.5$ \\
    \midrule
    \textbf{Total spectra...} & $447$ & $125$ & $69$ & $15$ & $10$ & $4$ & $3$ \\
    \quad \textbf{(of which unmasked)} & $379$ & $98$ & $59$ & $14$ & $9$ & $4$ & $3$ \\
    \textbf{Selected spectra} & $180~(47\%)$ & $54~(55\%)$ & $39~(66\%)$ & $9~(64\%)$ & $7~(78\%)$ & $2~(50\%)$ & $2~(67\%)$ \\
    \midrule
    \multicolumn{8}{c}{\textbf{Selection failure reasons}} \\
    \midrule
    $\mathbf{(\#1)}\,\,\mathrm{SNR}_{\mathrm{blue\,Ly\alpha}}<2.0$ & $35~(9\%)$ & $11~(11\%)$ & $6~(10\%)$ & $2~(14\%)$ & $1~(11\%)$ & $1~(25\%)$ & $0~(0\%)$ \\
    $\mathbf{(\#2)}\,\,\mathrm{SNR}_{\mathrm{red\,Ly\alpha}}>\{5.0,5.0\}$ & $127~(34\%)$ & $26~(27\%)$ & $12~(20\%)$ & $2~(14\%)$ & $1~(11\%)$ & $0~(0\%)$ & $1~(34\%)$ \\ 
    $\mathbf{(\#3)}\,\,\mathrm{SNR}_{\mathrm{red\,Ly\alpha}}>2.0$ & $35~(9\%)$ & $10~(10\%)$ & $2~(3\%)$ & $0~(0\%)$ & $0~(0\%)$ & $0~(0\%)$ & $1~(34\%)$ \\ 
    
    $\mathbf{(\#4)}\,\,\chi^2_{\rm red, gal}<6.0$ & $22~(6\%)$ & $7~(7\%)$ & $2~(3\%)$ & $1~(7\%)$ & $0~(0\%)$ & $0~(0\%)$ & $1~(34\%)$ \\
    $\mathbf{(\#5)}\,\,\Delta\chi^2_{\rm low-z}>4.0$ & $65~(17\%)$ & $11~(11\%)$ & $5~(8\%)$ & $0~(0\%)$ & $0~(0\%)$ & $1~(25\%)$ & $1~(34\%)$ \\

    $\mathbf{(\#6)}\,\,R_{\rm e}>45.0\,\mathrm{mas}$ & $3~(1\%)$ & $0~(0\%)$ & $0~(0\%)$ & $0~(0\%)$ & $0~(0\%)$ & $0~(0\%)$ & $0~(0\%)$ \\
    $\mathbf{(\#7)}\,\,\int^{1.1\times z_{\rm best}}_{0.9\times z_{\rm best}} P(z)\mathrm{d}z>0.6$ & $2~(1\%)$ & $1~(1\%)$ & $0~(0\%)$ & $0~(0\%)$ & $0~(0\%)$ & $0~(0\%)$ & $0~(0\%)$ \\
    \bottomrule
    \end{tabular}
\end{table*}

\subsection{Summary of the \texorpdfstring{$6.5<z<16.5$}{6.5<z<16.5} sample}
\label{sec:sample_summary}

Our final \epochstwo\ sample comprises $2452$ galaxies at $6.5<z<13.5$, $33$ of which are classified as LRDs, and a further $73$ less robust candidates at $13.5<z<16.5$. We detect $107$ brown dwarf candidates (including a single LRD), the majority of which are $1300<T\,/\,\rm K<2400$ L-dwarfs with best-fitting ``sfhz\_blue\_agn'' \eazypy\ photo-z solutions at $z\simeq7$. ``Gold'' samples of $45$ $10.5<z<13.5$ and $17$ $13.5<z<16.5$ sources are produced after increasing the detection SNR threshold redwards of \lya. The photometric redshift accuracy is quantified from $673$ cross-matched spectroscopically confirmed sources from DJA v4.4; $566$ of which are unmasked and $293$ ($52\%$) are \epochstwo\ selected, yielding $\sigma_{\rm NMAD}=0.017$, $\eta=4.3\%$, and $\langle\Delta z\rangle=-0.022$.

A summary of the properties of the fiducial \epochstwo\ photometric sample is presented in \autoref{tab:sample_properties}, and properties for individual candidates in the ``gold'' sample at $10.5<z<16.5$ are shown in \autoref{tab:gold_sample} in \autoref{sec:gold_properties}. Our final public catalog contains boolean flags denoting successfully cross-matched sources to a number of photometric studies, from which we find the overlap between \epochstwo\ candidates and those in other large photometric studies across the same sky area to be $\simeq 60\%$ at $\rm SNR_{\rm F444W}>10$.


\begin{deluxetable*}{l|cccccc}
    \tablewidth{0pt}
    \tablecaption{Sample summary showing the number of fiducially \epochstwo\ selected galaxy candidates as a function of redshift. In addition, key photometrically-derived properties computed from power-law fits to the rest-frame UV fluxes (\Muv\ and \uvbeta) and the \bagpipes\ posteriors setup (all other properties) are provided. The properties of extended sources are appropriately corrected for following the procedures outlined in \autoref{sec:UV_properties} and \autoref{sec:bagpipes}.\label{tab:sample_properties}}
    \tablehead{
    \colhead{Property} & \colhead{$6.5<z<7.5$} & \colhead{$7.5<z<8.5$} & \colhead{$8.5<z<9.5$} & \colhead{$9.5<z<10.5$} & \colhead{$10.5<z<11.5$} & \colhead{$11.5<z<13.5$}
    }
    \startdata
    $N_{\rm gals}$ & $1382$ & $558$ & $224$ & $136$ & $102$ & $84$ \\
    $M_{\rm UV}$ & $-18.83_{-0.88}^{+1.00}$ & $-19.16_{-0.77}^{+1.02}$ & $-19.07_{-0.89}^{+0.84}$ & $-19.19_{-0.76}^{+1.08}$ & $-19.46_{-0.80}^{+0.96}$ & $-19.48_{-0.77}^{+1.36}$ \\
    $\beta_{\rm UV}$ & $-2.31_{-0.55}^{+0.50}$ & $-2.43_{-0.69}^{+0.63}$ & $-2.34_{-0.66}^{+0.66}$ & $-2.43_{-0.65}^{+0.60}$ & $-2.39_{-0.67}^{+0.64}$ & $-2.34_{-0.73}^{+0.68}$ \\
    $\log_{10}(M_\star\,/\,\mathrm{M}_\odot)$ & $7.56_{-0.60}^{+0.74}$ & $7.59_{-0.50}^{+0.73}$ & $7.66_{-0.56}^{+0.60}$ & $7.88_{-0.66}^{+0.69}$ & $8.15_{-0.83}^{+0.79}$ & $8.37_{-0.88}^{+0.57}$ \\
    $\log_{10}(\mathrm{SFR}_{10}\,/\,\mathrm{M}_\odot\,\mathrm{yr}^{-1})$ & $0.16_{-0.53}^{+0.55}$ & $0.22_{-0.56}^{+0.45}$ & $0.20_{-0.63}^{+0.49}$ & $-0.12_{~\,-1.42}^{\dagger\,+0.59}$ & $-0.25_{~\,-1.74}^{\dagger\,+0.65}$ & $-0.46_{~\,~\,-2.19}^{\dagger\dagger\,+0.87}$ \\
    $\log_{10}(\mathrm{SFR}_{100}\,/\,\mathrm{M}_\odot\,\mathrm{yr}^{-1})$ & $-0.50_{-0.56}^{+0.61}$ & $-0.44_{-0.48}^{+0.59}$ & $-0.37_{-0.54}^{+0.49}$ & $-0.19_{-0.64}^{+0.59}$ & $-0.06_{-0.71}^{+0.70}$ & $0.21_{-0.72}^{+0.49}$ \\
    $\log_{10}(\mathrm{\Phi_{\rm SF}})$ & $0.90_{~\,~\,-0.69}^{\dagger\dagger\,+0.10}$ & $0.90_{~\,~\,-0.87}^{\dagger\dagger\,+0.10}$ & $0.89_{~\,~\,-1.13}^{\dagger\dagger\,+0.11}$ & $0.24_{~\,~\,-1.93}^{\dagger\dagger\,+0.74}$ & $-0.24_{~\,~\,-1.99}^{\dagger\dagger\,+1.17}$ & $-0.82_{~\,~\,-2.29}^{\dagger\dagger\,+1.69}$ \\
    $\log_{10}(\mathrm{sSFR}_{10}\,/\,\mathrm{yr}^{-1})$ & $-7.11_{~\,-1.03}^{\dagger\,+0.11}$ & $-7.11_{~\,-1.18}^{\dagger\,+0.11}$ & $-7.11_{~\,-1.34}^{\dagger\,+0.12}$ & $-7.82_{~\,~\,-2.15}^{\dagger\dagger\,+0.80}$ & $-8.37_{~\,~\,-2.44}^{\dagger\dagger\,+1.31}$ & $-8.94_{~\,~\,-2.47}^{\dagger\dagger\,+1.83}$ \\
    $\log_{10}(\mathrm{sSFR}_{100}\,/\,\mathrm{yr}^{-1})$ & $-8.00_{~\,-0.18}^{\dagger\,+0.03}$ & $-8.00_{~\,-0.15}^{\dagger\,+0.03}$ & $-8.00_{~\,-0.11}^{\dagger\,+0.03}$ & $-7.99_{~\,~\,-0.17}^{\dagger\dagger\,+0.05}$ & $-7.99_{~\,~\,-0.44}^{\dagger\dagger\,+0.05}$ & $-7.99_{~\,~\,-0.42}^{\dagger\dagger\,+0.06}$ \\
    $\log_{10}(\xi_{\rm ion, 0}\,/\,\mathrm{Hz}\,\mathrm{erg}^{-1})$ & $25.47_{~\,-0.33}^{\dagger\,+0.17}$ & $25.49_{~\,-0.43}^{\dagger\,+0.17}$ & $25.48_{~\,~\,-0.65}^{\dagger\dagger\,+0.19}$ & $25.09_{~\,~\,-1.17}^{\dagger\dagger\,+0.52}$ & $24.88_{~\,~\,-1.59}^{\dagger\dagger\,+0.74}$ & $24.61_{~\,~\,-1.91}^{\dagger\dagger\,+0.92}$ \\
    $\log_{10}(\dot{N}_{\rm ion, 0}\,/\,\mathrm{Hz})$ & $53.69_{-0.61}^{+0.59}$ & $53.79_{-0.68}^{+0.49}$ & $53.73_{-0.77}^{+0.60}$ & $53.26_{~\,-1.18}^{\dagger\,+0.75}$ & $53.18_{~\,-1.50}^{\dagger\,+0.79}$ & $52.96_{~\,~\,-1.99}^{\dagger\dagger\,+0.93}$ \\
    \enddata
    \tablecomments{``$\dagger$'': Somewhat prior-driven posterior ($\mathcal{D}_{\rm KL}<1$ in $>16\%$ of the sample).\\``$\dagger\dagger$'': Predominantly prior-driven posterior ($\mathcal{D}_{\rm KL}<1$ in $>50\%$ of the sample).}
\end{deluxetable*}

\section{Photometric properties}
\label{sec:properties}

\subsection{UV properties}
\label{sec:UV_properties}

The rest-frame UV properties, including UV continuum slopes, $\beta_{\rm UV}$, and absolute magnitudes, \Muv, for the entire catalog are calculated directly from the rest-frame UV photometry where available rather than the best-fitting SED from either \eazypy\ or \bagpipes. While this makes only a minor difference for measurements of \Muv, the SED fitting $\beta_{\rm UV}$ measurements are strongly impacted by the parameter space covered by the SED template set being fitted, excluding the extreme blue $\beta_{\rm UV}$ results observed with \jwst\ in both NIRCam imaging \citep[e.g.][]{Topping2022, Cullen2023, Cullen2024, Austin2025a} and in NIRSpec PRISM spectroscopy \citep[e.g.][]{Nanayakkara2023, Napolitano2025, Morales2025, Yanagisawa2025, Saxena2026}.

To calculate $\beta_{\rm UV}$ for our sample, we adopt the methodology from \citet{Austin2025a}, which fits a $f_{\lambda}\propto\lambda^{\beta_{\rm UV}}$ power law to the photometric fluxes lying entirely within $1250<\lambda_{\mathrm{rest}}\,/\,\rm \AA<3000$, as determined by wavelengths corresponding to their upper and lower $50\%$ filter throughput. We note that this technique is susceptible to unavoidable biases from the $2175\,\rm \AA$ dust bump, strong rest-UV emission lines, and redshift biases induced by DLAs and strong \lya\ emission. \Muv\ posteriors are produced from these power law fits using a tophat with $100$\,\AA\ width centered at $\lambda_{\mathrm{rest}}=1500\,\rm \AA$ and corrected for extended sources when necessary by the ratio of \sextractor\ \texttt{FLUX\_AUTO} to \texttt{FLUX\_APER} in the closest filter to $\lambda_{\rm rest}=1500\,\rm \AA$ clipped to a maximum of $10$. 



\subsection{A moderate evolution towards blue UV continua at the highest redshifts}
\label{sec:UV_slopes}

\begin{figure}
    \centering
    \includegraphics[width=0.95\linewidth]{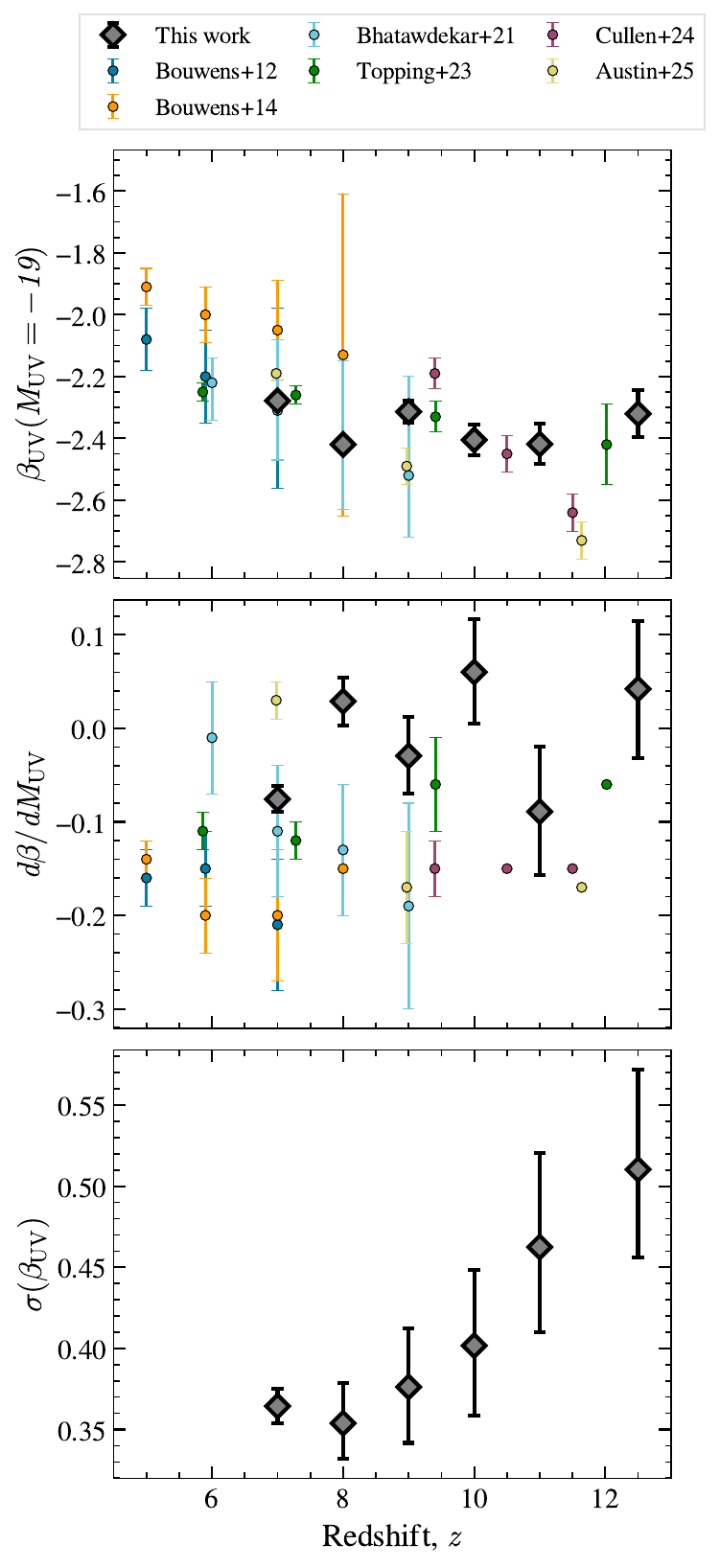}
    \caption{Evolution of the best fitting $\beta_{\rm UV}-M_{\rm UV}$ power law parameters, $\beta_{\rm UV}(M_{\rm UV}=-19)$ amplitude (\textbf{upper panel}), $\mathrm{d}\beta_{\rm UV}/\mathrm{d}M_{\rm UV}$ slope (\textbf{central panel}), and $\sigma_{\beta_{\rm UV}}$ scatter (\textbf{lower panel}) for our fiducial \epochstwo\ sample at $6.5<z<13.5$. Results are compared to photometric studies both from \hst\ \citep{Bouwens2012, Bouwens2014, Bhatawdekar2021} and \jwst\ \citep{Topping2024a, Cullen2024, Austin2025a}.}
    \label{fig:beta_Muv_evo}
\end{figure}

One of the key insights from early NIRCam studies has highlighted a transition from moderately dust-reddened systems at the end of the EoR to dust-poor or even dust-free stellar populations at the highest redshifts \citep[e.g.][]{Topping2022, Cullen2023, Cullen2024, Topping2024a, Austin2025a}. In this work we determine the redshift evolution of $\beta_{\rm UV}-M_{\rm UV}$ using the UV properties calculated using the power law method outlined in \autoref{sec:UV_properties}. This avoids both \uvbeta\ systematics induced by SED fitting \citep[see e.g.][]{Dunlop2013, Austin2025a, Morales2025} and the fact that $\sim 32-60\%$ of our sources at $z>9$ are dominated by the SED fitting hyperprior on \uvbeta\ as highlighted in \autoref{sec:KL_divergences}. \autoref{fig:beta_Muv_evo} shows the power law fitting parameters to each $\beta_{\rm UV}-M_{\rm UV}$ relation in six redshift bins from $z\simeq6.5$ to $z\simeq13.5$ for our fiducial \epochstwo\ sample, where we include the scatter, $\sigma_{\beta_{\rm UV}}$, as a free parameter. 

Our results suggest a moderate evolution of $\beta_{\rm UV}(M_{\rm UV}=-19)$ from $\beta_{\rm UV}=-2.28\pm0.01$ at $z\simeq7$ to $\beta_{\rm UV}=-2.42^{+0.07}_{-0.06}$ at $z\simeq11$ and $\beta_{\rm UV}=-2.32^{+0.08}_{-0.07}$ at $z\simeq12.5$. This is consistent with the JADES study by \citet{Topping2024a}, although both the \citet{Cullen2024} and \epochsone\ \citep{Austin2025a} samples predict $\Delta\beta_{\rm UV}\sim0.3-0.4$ bluer than this work at $M_{\rm UV}=-19$ and $z\simeq11$. The \citet{Cullen2024} sample also predicts a much more rapid build up of dust between $z\sim11.5$ and $z\sim9.5$ than this work, albeit with consistent dust content at $z\simeq10$, most likely due to their inclusion of wider-area COSMOS/UltraVISTA data from the ground \citep{McCracken2012}. It is not likely that there are age or metallicity differences between these two samples, which nevertheless have a subdominant impact on the UV slope reddening compared to attenuation by dust.

Since galaxies formed at $z\simeq18-20$ can be no more than $\sim150-200\,\rm Myr$ in age at this epoch, AGB stars which dominate dust production over $\gtrsim300\,\rm Myr$ timescales \citep{Schneider2024} are likely not the major contributor to the budget. Instead our results favor the scenario of rapid enrichment  by either Type II SNe with minimal destruction in the reverse shock \citep[e.g.][]{Nozawa2007,Bianchi2009,Micelotta2018}, large populations of Red Super Giant (RSG) or Wolf Rayet (WR) stars from a more top-heavy IMF \citep[e.g.][]{Jermyn2018, Steinhardt2023, Cueto2024, Hutter2025} which have been observed to produce copious amounts of dust in the local Universe \citep{Lau2022} on $\sim3-10\,\rm Myr$ timescales, or more likely a combination of the two. The latter scenario is backed up by observations of the carbonaceous $2175\,\rm \AA$ dust bump produced by carbon-rich Wolf Rayet stars \citep[WCs; e.g.][]{Witstok2023, Fisher2025, Markov2025, Ormerod2025} at $z\simeq7$. 

Our sample transitions towards a flatter relation ($\mathrm{d}\beta/\mathrm{d}M_{\rm UV}=0.04\pm 0.07$) at the highest redshifts than previous \hst\ \citep{Bouwens2012, Bouwens2014, Bhatawdekar2021} and \jwst\ studies \citep{Topping2024a, Cullen2024, Austin2025a}. 
An evolution to  $\mathrm{d}\beta/\mathrm{d}M_{\rm UV}=-0.08\pm 0.01$ at $z\simeq7$ suggests that AGB dust production, potentially with assistance by grain-grain growth in the ISM \citep{Dwek1998, Draine2009, Asano2013, Galliano2018}, has started to dominate in the more luminous systems between these two epochs. 
The scatter is observed to rapidly increase from $\sigma_{\beta_{\rm UV}}=0.36\pm0.01$ at $z\simeq7$ to $\sigma_{\beta_{\rm UV}}=0.51^{+0.06}_{-0.05}$ at $z\simeq12.5$, highlighting the increasing heterogeneity of ages, metallicities, and dust content in SFGs at these epochs. An increasingly turbulent and volatile ISM/CGM \citep[e.g.][]{Sun2026} induced by clumpy and stochastic star formation \citep[e.g.][]{McClymont2025b} also likely increases the diversity of SNe driven dust outflows from high $\Sigma_{\rm SFR}$ regions \citep{Ziparo2023, Ferrara2024}, offsetting the dust from UV regions \citep{Bowler2022, Punyasheel2026}. 
In addition, since free-free nebular continuum emission acts to redden the slope \citep[e.g.][]{Saxena2026}, an increasing number of nebular dominated systems, which have already been observed by NIRSpec \citep{Cameron2024-nebular, Katz2025, Trussler2026}, could help explain this $\sigma_{\beta_{\rm UV}}$ evolution.

\subsection{Bagpipes SED fitting}
\label{sec:bagpipes}

We perform Bayesian SED fitting using a modified version of \bagpipes\ \footnote{Available at \url{https://github.com/tHarvey303/bagpipes}} \citep{Carnall2018} with \bpass\ v2.2.1 stellar population synthesis models \citep{Eldridge2017-BPASS, Stanway2018-BPASS} and an assumed \citet{Kroupa2001} IMF. We run SED fitting for our fiducial \epochstwo\ sample at $6.5<z<13.5$ with the \galfind\ python wrapper using the {\tt galfind.sed\_fitting.Bagpipes(Catalogue)} called with an initialized {\tt galfind.catalogues.Catalogue} object for each survey. A gaussian redshift prior centered on the ``sfhz\_blue\_agn'' \eazypy\ redshift with $3\sigma_{\rm z, EAZY}$ width is used, where $\sigma_{\rm z, EAZY}$ is the standard deviation of the \eazypy\ redshift PDF. We adopt the standard \citet{Calzetti2000} dust attenuation curve with no additional contribution from natal birth clouds and the ``continuity bursty'' star formation history (SFH) parameterization from \citet{Leja2019} and \citet{Tacchella2022} with additional $0-3~\rm Myr$ and $3-10~\rm Myr$ low-age bins included. We follow the prescription of \citet{Inoue2014} for the attenuation of Lyman-series photons by the IGM. 

Three different setups are run in this paper, namely ``fiducial'', ``fesc'', and ``uniform''; we first outline the ``fiducial'' below. Lognormal priors are set for the dust attenuation, $A_V$, and the metallicity, $Z_{\star}$, and a flat uniform prior on the logarithm of the ionization parameter ($-4<\log U<-1$) is assumed, the upper limit of which has been increased from the standard \bagpipes\ {\tt BC03} setup with \cloudy\ \citep{Ferland2017}. 

In the ``fesc'' setup, we include \fesc\ as a free parameter with uniform prior $f_{\rm esc}^{\rm LyC}\in[0, 1]$. The ``uniform'' setup instead switches the lognormal priors on $A_V$ and $Z_{\star}$ with uniform priors across the same wide prior range. As with the \Muv\ values computed from the photometry, we correct the \Muv, SFR, \Ndotion, and stellar masses output by \bagpipes\ using the same technique in the relevant rest-frame UV (for \Muv, SFR, and \Ndotion) or F444W (for the stellar masses) filter. 



\subsection{Testing the reliability of Bayesian SED fitting parameter posteriors}
\label{sec:KL_divergences}

One particular concern which propagates throughout photometric studies of the galaxy population is the reliability of properties obtained via SED fitting. Modern Bayesian SED fitting codes such as \bagpipes\ \citep{Carnall2018}, \beagle\ \citep{Chevallard2016}, and \prospector\ \citep{Leja2017, Johnson2021} have regularly been used to determine galaxy properties from photometry, however, there are often a greater number of free parameters than available fluxes indicating that perhaps not all of these are constrained during the fitting procedure. 

To test this we determine the information gained for the most widely used parameters left free in the \bagpipes\ fitting of our \epochstwo\ sample using the fiducial setup described in \autoref{sec:bagpipes}. We compare each posterior distribution, $P(x)$, to its corresponding (hyper-) prior, $Q(x)$, and compute the Kullback-Leibler divergence \citep[$\mathcal{D}_{\rm KL}$;][]{KullbackLeibler1951} as 
\begin{equation}
    \mathcal{D}_{\rm KL}(P\,||\,Q)
    \simeq\frac{1}{N}\sum_{i=0}^N\ln\bigg(\frac{P(x_i)}{Q(x_i)}\bigg),
\label{eq:kl_divergence}
\end{equation}
where $x_i$ are a set of $N$ Monte-Carlo posterior draws from $P(x)$. This $\mathcal{D}_{\rm KL}(P\,||\,Q)$ is computed using {\tt scipy.stats.entropy} and is a measure of the relative entropy between the two distributions. We note that the KL divergence from \autoref{eq:kl_divergence} is positive definite and values $\mathcal{D}_{\rm KL}(P\,||\,Q)\gtrsim1$ indicate a parameter that is strongly constrained by the data. Note that this does not mean that the parameter is accurately calculated and without SED fitting biases.

\begin{figure*}
    \centering
    \includegraphics[width=0.99\linewidth]{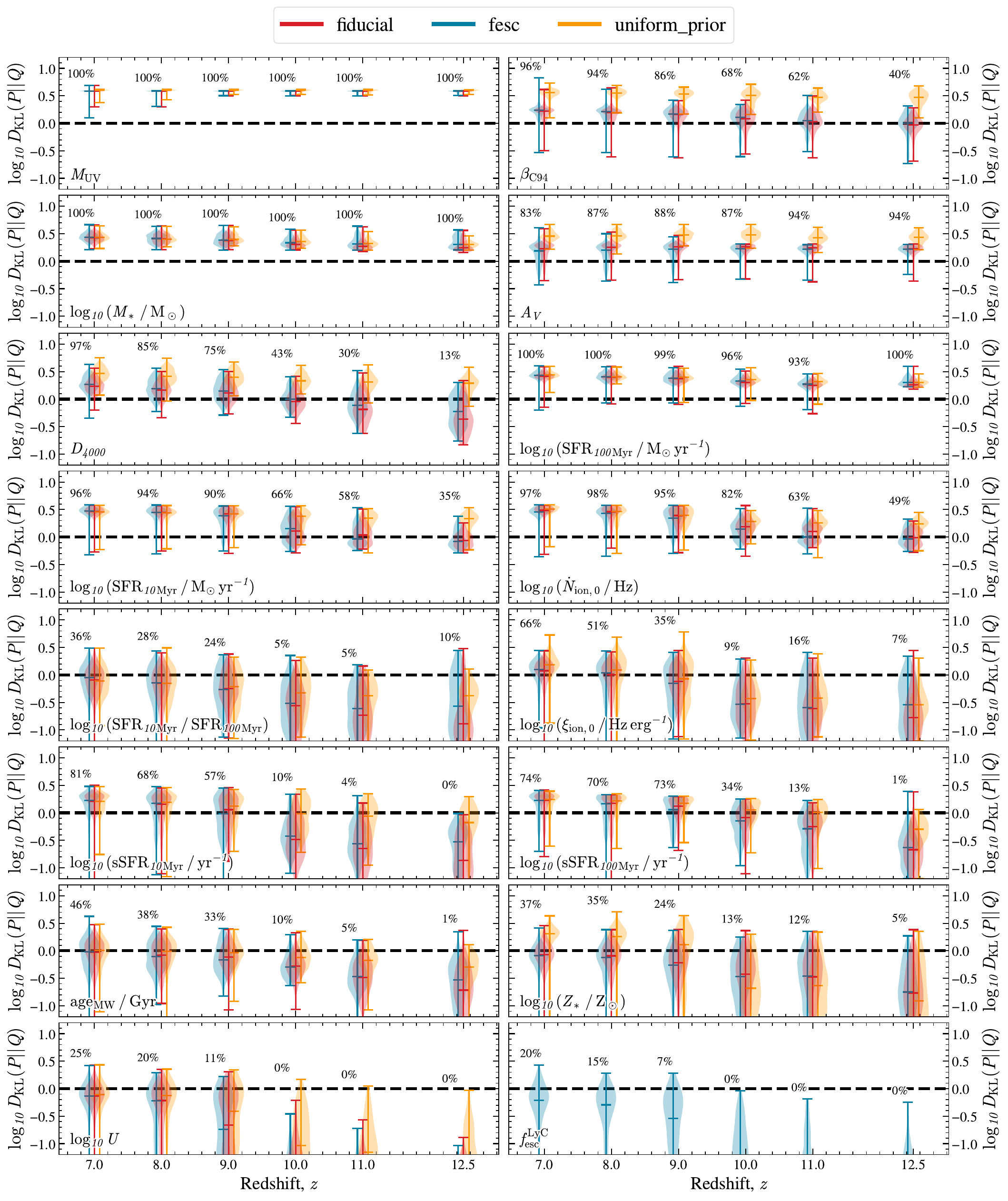}
    \caption{Evolution of KL divergence distributions computed as per \autoref{eq:kl_divergence} from the fiducial \epochstwo\ sample at $6.5<z<13.5$. $5$ free and $11$ hyper-parameters are shown from our ``fiducial'' (red), ``fesc'' (blue), and ``uniform'' (yellow) \bagpipes\ fitting procedures; details regarding these setups are given in \autoref{sec:bagpipes}. Violins show the distribution in each redshift bin, with caps showing the median and extrema. The percentage of sources that exceed the $\mathcal{D}_{\rm KL}(P\,||\,Q)>1$ threshold to be considered ``strongly constrained'' for the fiducial \bagpipes\ setup (and for ``fesc'' for $f_{\rm esc}^{\rm LyC}$) are shown to the upper left of each distribution. Only in cases when the majority of the distribution (i.e. $84\%$ or so) falls above the threshold is the data constraining enough to draw meaningful conclusions from each observable. Similar results are obtained for all 3 setups, with the exception of a moderately better constrained $\beta_{\rm UV}$, $A_V$, $Z_{\star}$, and $D_{\rm Balmer}$ with the ``uniform'' \bagpipes\ run.}
    \label{fig:kl_divergences}
\end{figure*}

The redshift evolution of KL-divergences for $16$ free/hyper-parameters from our ``fiducial'', ``fesc'', and ``uniform'' \bagpipes\ setups is presented in \autoref{fig:kl_divergences}. It is clear that \Muv\ is the most well constrained parameter with all $\mathcal{D}_{\rm KL}(P\,||\,Q)>3$, which is perhaps to be expected given that our sample is UV selected. Encouragingly the stellar masses are also well constrained across all redshift bins even though the Balmer break, produced by A-type stars \citep[e.g.][]{Trussler2024} in the absence of AGN \citep[see e.g.][]{Inayoshi2025}, is no longer observed in NIRCam at $z\gtrsim11.5$.

\subsubsection{The UV continuum slope and dust constraints}

The rest-frame UV slope ($\beta_{\rm C94}$; measured in the 10 \citealt{Calzetti1994} windows) is strongly constrained for $96\%$ of sources in the $6.5<z<7.5$ bin compared to only $40\%$ at $11.5<z<13.5$. While we retain complete rest-frame UV wavelength coverage across redshifts, this trend is dominated by the number of ultra-blue $\beta_{\rm C94}\lesssim-2.8$ candidates increasing with redshift which SED fitting templates fail to reproduce \citep[see e.g.][]{Cullen2024, Austin2025a, Morales2025}. 
Closely linked to $\beta_{\rm C94}$ is the V-band dust attenuation, $A_V$, which is well constrained in $83-94\%$ of sources across redshifts in our one-parameter dust model. The dust measurements, however, come with the caveat that a fixed \citet{Calzetti2000} dust slope may not match the intrinsic value, leading to large systematic errors.

\subsubsection{Star formation history constraints}

The KL divergences on the SFR, SF burstiness (defined as $\Phi_{\rm SF}=\mathrm{SFR}_{\rm 10}/\mathrm{SFR}_{\rm 100}$), ionizing photon production rate ($\dot{N}_{\rm ion}=7.28\times10^{11}L_{{\mathrm{H}\alpha}}(1-f_{\rm esc}^{\rm LyC})^{-1}\,\,[\rm Hz]$), ionizing photon production efficiency ($\xi_{\rm ion}=\dot{N}_{\mathrm{ion}}/L_{\nu, \rm UV}\,\,[\rm Hz\,erg^{-1}]$), and mass-weighted age ($\mathrm{age}_{\rm MW}$) hyper-parameters directly pertain to the continuity bursty SFH \citep{Leja2019, Tacchella2022} adopted in our \bagpipes\ SED fitting. Since $\mathrm{SFR}_{100\rm Myr}$ is predominantly derived from the well-constrained rest-frame UV continuum emission of massive O/B-type main-sequence stars, it follows that this is well constrained for $\sim95-100\%\%$ of the sample. In contrast, $\mathrm{SFR}_{10\rm Myr}$ and \Ndotion\ stem from the rest-frame optical emission line fluxes, and thus become poorly constrained at $z\gtrsim9$ where the \oiiihbeta\ complex lies redwards of F444W. Both $\Phi_{\rm SF}$ and \xiion\ are poorly constrained at $z\lesssim9$ ($36\%$ and $28\%$ respectively at $6.5<z<7.5$ and $6.5<z<7.5$) and become even worse ($\sim 5-10\%$ constrained) at the highest redshifts as a result of the aforementioned loss of rest-optical emission lines from NIRCam. One exception occurs when \oiiihbeta\ (or equivalently \ha\ at lower redshift) falls within a medium-width filter, for instance F410M at $z\simeq7.5$. Lastly, the mass-weighted age becomes more constrained with redshift, presumably as a consequence of the reduction in SFH diversity in our \epochstwo\ sample \citep[i.e. smouldering galaxies with a recent downturn in SFR, e.g.][which are more selectable at $z\simeq7$ than at $z\simeq12.5$]{Looser2024, Trussler2025, Austin2025b}.

\subsubsection{Unconstrained parameters and induced systematics}

While a small subset of sources at $z\simeq7$ have constrainable ionization parameters ($\log_{10}U$; $25\%$) and stellar metallicities ($\log_{10}Z_{\star}$; $37\%$), these are among the least well constrained parameters in \autoref{fig:kl_divergences} and thus we do not draw strong conclusions using these posteriors throughout this paper series. It is perhaps not a sensible idea either to make inferences regarding the Lyman continuum escape fraction, \fesc, from photometric data alone. This is only constrained for a maximum $20\%$ of sources at $z\simeq7$ and not at all at $z>9$. A further investigation finds that these $20\%$ are young, low mass, blue starburst systems that \textit{require} small $f_{\rm esc}^{\rm LyC}\lesssim10\%$ to somewhat relieve the dust--\xiion\ degeneracy outlined by \citet{Austin2025b} whereby the stellar emission cannot match the observed spectroscopic \xiion\ values from \citet{Atek2024} even with fixed $f_{\rm esc}^{\rm LyC}=0$.

A similar study regarding the constrainability of SED fitting parameters was done on the JADES-DR5 galaxy stellar population catalog \citep{Duan2026} using using \prospector\ adopting a more complex dust attenuation model (including a \citet{Salim2018} modified \citet{Calzetti2000} slope and \citet{CharlotFall2000} natal birth cloud attenuation free parameters) as well as MIR/FIR dust \citep{DraineLi2007} and AGN emission prescriptions. All of these additional free parameters were found to be more prior-dominated than the stellar metallicity even when including deep MIRI SMILES photometry \citep{Alberts2024, Alberts2026}. As with all SED fitting results, we urge extreme caution when interpreting prior-driven posteriors with $\mathcal{D}_{\rm KL}(P\,||\,Q)<1$ which can induce further biases from mismatching prior-intrinsic property distributions.







\section{High redshift sources}
\label{sec:highz_sources}

\subsection{Spectroscopically confirmed sources at \texorpdfstring{$z\gtrsim10$}{z>10}}
\label{sec:spec_confirmed_sources}

Since early \jwst\ spectroscopic studies with NIRSpec \citep[e.g.][]{Nakajima2023, Fujimoto2023}, a number of $z\gtrsim10$ sources have been identified in these major \jwst\ NIRCam fields. Below we outline $22$ objects at $z_{\rm spec}>10$ from the literature, all of which exist in our cross matched DJA v4.4 catalogue.

\begin{figure*}
    \centering
    \includegraphics[width=0.95\linewidth]{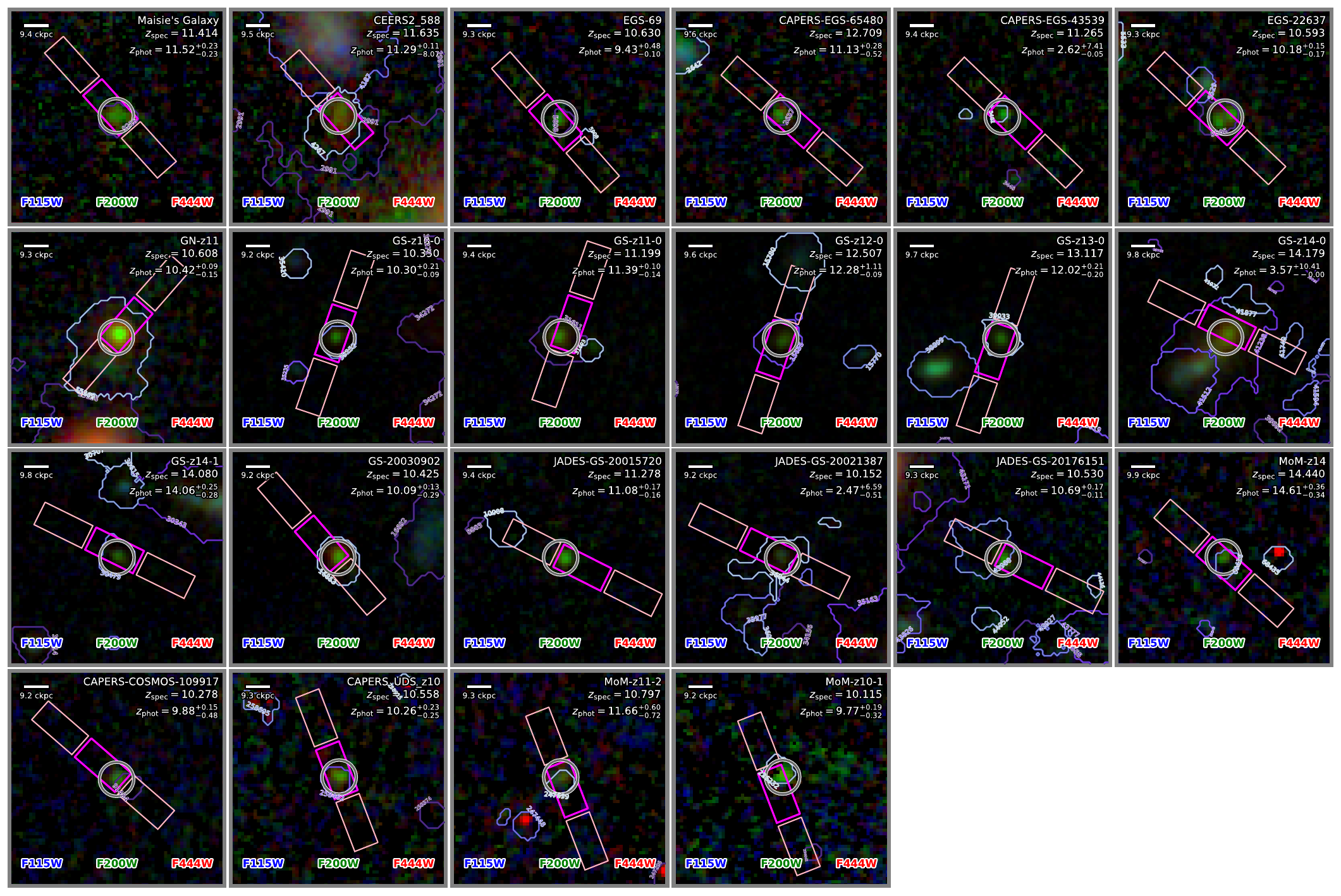}
    \caption{$2''\times 2''$ RGB cutouts (native non-PSF homogenized $\rm B=F115W$, $\rm G=F200W$, $\rm R=F444W$) from the \epochstwo\ catalogs for the $22$ $z_{\rm spec}>10$ sources highlighted in \autoref{tab:highz_spec}. The best-matching NIRSpec slitlet is plotted in magenta (on source) and hot pink (nods) using {\tt galfind.Spectrum.plot\_slitlet()}. Individual sources from the F277W+F356W+F444W segmentation map are outlined and labelled with their {\tt SURVEY\_ID}. The scaling is set by a modified $\operatorname{asinh}$ stretch from \citet{Lupton2004} with $\mathrm{stretch}=0.001$ and softening $Q=10.0$ in $\mu\rm Jy$ flux units.}
    \label{fig:highz_spec_cutouts}
\end{figure*}


The majority of these sources are identified in GOODS-South JADES imaging, which hosts the deepest NIR imaging and greatest NIRSpec spectroscopic coverage. One of the earliest spectroscopically confirmed samples of four galaxies at $z\gtrsim10$ came from \citet{CurtisLake2023} in JADES-DR1 (our GOODS-South ``East'' tile) and include the record-breaking GS-z13-0 at $z_{\rm spec}=13.117$, although only GS-z12-0 is selected in our \epochstwo\ sample. In May 2024, \citet{Carniani2024} published confirmation of two new spectroscopic redshift record holders at $z_{\rm spec}\sim14$ from our ``South'' tile of GOODS-South imaging. GS-z14-0 at $z_{\rm spec}=14.32^{+0.08}_{-0.20}$ is the 2024 redshift record holder, detected with the Mid-InfraRed Instrument \citep[MIRI;][]{Bouchet2015, Rieke2015, Wright2023}  at $7.7\,\mu\rm m$ \citep{Helton2025} and in an Atacama Large Millimeter/sub-millimeter Array (ALMA) Band 6 line scan at $6.6\sigma$ corresponding to \fion{O}{3}$_{88\mu\mathrm{m}}$ at $z=14.1793\pm0.0007$ \citep{Schouws2025b}, eclipsing the previous MACS-1149 JD1 detection record holder at $z=9.1096$ \citep{Hashimoto2018} and providing state-of-the-art cold gas content and kinematics in the early Universe. It has an \fion{O}{3}$_{88\mu\mathrm{m}}$ derived dynamical mass $\log_{10}(M_{\rm dyn}\,/\,\rm M_\odot)=9.0\pm0.2$ FWHM \citep{Carniani2025} and is a damped \lya\ (DLA) system hosting a large proximate neutral gas reservoir with \ion{H}{1} column density $\log_{10}(N_{\rm HI}\,/\,\rm cm^{-2})=22.27^{+0.08}_{-0.09}$ \citep{Heintz2025b}. GS-z14-1, in contrast, is compact ($r_{\rm UV}<160\,\rm pc$) and blue ($\beta_{\rm UV}=-2.71\pm0.19$) at $z_{\rm spec}=13.90\pm0.17$ with an interesting $\sim2\sigma$ \ion{C}{2}$-\lambda1335$ doublet absorption feature. 

The next record-breaking source was identified by \citet{Naidu2026} in the PRIMER-COSMOS field by the ``Miracle or Mirage'' survey (MoM; PID: 5224; PI: P. Oesch), known as MoM-z14 at $z_{\rm spec}=14.44\pm0.02$. This source is compact ($r_{\rm eff}=74^{+15}_{-12}\,\rm pc$) with strong rest-frame UV lines and enhanced $[\rm N/C]>1$ symbolic of local globular clusters, reinforcing the bimodality in high-z sources in \sfion{N}{4}--$r_{\rm eff}$ parameter space identified by \citet{Harikane2025}.

JADES GOODS-North imaging hosts another large spectroscopic database, albeit with fewer spectra and shallower imaging than in GOODS-South. Few interesting sources are detected except the remarkably bright GN-z11 at $z_{\rm spec}=10.6034\pm 0.0013$ and $M_{\rm UV}=-21.5$ \citep{Bunker2023} in our ``Medium'' tile, first discovered in \hst\ imaging \citep{Bouwens2010b} and subsequently by HST/WFC3 slitless grism spectroscopy \citep{Oesch2016} and Keck/MOSFIRE \citep{McLean2012} by \citet{Jiang2021}. NIRSpec Multi-Object Spectroscopy (MOS) and Integral Field Spectroscopy (IFS) has subsequently lead to detections of broad-line \ha\ indicative of AGN activity \citep{Maiolino2024a, Maiolino2024c},  \ion{He}{2}$-\lambda1640$ emission revealing possible halo Pop. III signatures \citep{Maiolino2024b}, extended \lya\ emission \citep{Bunker2023, Scholtz2024}, and nitrogen enhancement \citep[e.g.][]{Belokurov2023, Kobayashi2024, Watanabe2024} as a potential signature of globular cluster precursors \citep{Cameron2023-GNz11, Charbonnel2023, MarquesChaves2024, Senchyna2024}. Photometric imaging from JADES is also presented in \citet{Tacchella2023-GNz11}.

Early NIRSpec spectroscopy became available as early as March 2023 in the EGS field. Perhaps the most notable study is DDT 2750 \citep[PI: P. Arrabal-Haro;][]{ArrabalHaro2023b}, which includes targets independently discovered in a number of early NIRCam studies \citep{Finkelstein2022-CEERSKP1, Bouwens2023, Donnan2023, Harikane2023}. Three notable galaxies at $z_{\rm phot}>10$ (in our ``CEERSP2'' imaging) are highlighted, including Maisie's galaxy (CEERS2\_5429) at $z_{\rm spec}=11.416\pm0.005$ originally found in CEERS imaging by \citet{Finkelstein2022-z12} and CEERS2\_588 at $z_{\rm spec}=11.043\pm0.003$. CEERS-93316 Initially detected by \citet{Donnan2023}, this remarkably bright $M_{\rm UV}=-21.66$ candidate at $z_{\rm phot}\simeq15.8$ (our {\tt SURVEY\_ID=5613}) is photometrically selected by our \epochstwo\ criteria. Spectroscopy unfortunately places this as an $A_{\rm V}=2.3\pm 0.2$ dusty interloper in a foreground overdensity at $z_{\rm spec}=4.912\pm 0.001$, as proposed by \citet{Naidu2022b}, and therefore not included in our high-redshift sample.

In mid-2025, the first results by \citet{Kokorev2025b} from the CANDELS--Area Prism Epoch of Reionization Survey (CAPERS)
program (PID: 6368, PI: M. Dickinson) in the UDS field identified $2$ sources at $z>10$. This includes CAPERS\_UDS\_z10 and CAPERS\_UDS\_z11, the latter of which falls in the PRIMER-UDS area that is not included in our analysis. A number of other high-redshift sources are also identified in multi-field studies by \citet{RobertsBorsani2025} and \citet{Tang2025}.


An RGB mosaic for all of these sources is plotted in \autoref{fig:highz_spec_cutouts}, where the slit positions and outlines of the \sextractor\ segmentation maps are shown. We note that the successful deblending of these sources was used to tweak our \sextractor\ setup as explained in \autoref{sec:EPOCHS_v2_cataloguing_and_depths}. \autoref{tab:highz_spec} summarizes the \epochstwo\ photometric selection successes and failures of these sources. We find that $4$ ($18\%$) are best-fit with low-redshift $z_{\rm phot}<4$ solutions (GS-z14-0, JADES-GS-20021387, CEERS2\_588, and CAPERS-EGS-43539), while GS-z10-0 falls within our CEERSP2 image mask, EGS-69 is not selected at $>5\sigma$ in the first two widebands redwards of \lya, and both GS-z13-0 and GS-z11-0 are detected bluewards of \lya\ at $\mathrm{SNR}>2$. This leaves us with an acceptable $64\%$ success rate rising to $68\%$ when considering sources within our quoted \epochstwo\ area. 

\begin{deluxetable*}{l|cccccccc}
    \tablewidth{\textwidth}
    \tabletypesize{\fontsize{8}{10}\selectfont}
    \setlength{\tabcolsep}{3pt}
    \tablecaption{NIRSpec PRISM spectra identified at $z_{\rm spec}>10$ by \citealt{Carniani2024} (C+24), \citealt{CurtisLake2023} (CL+23), \citealt{ArrabalHaro2023b} (AH+23b), \citealt{Finkelstein2022-z12} (F+22), \citealt{Bunker2023} (B+23), \citealt{Kokorev2025b} (K+25b), \citealt{Naidu2026} (N+26), \citealt{RobertsBorsani2025} (RB+25), and \citealt{Tang2025} (T+25). These spectra were collated from a cross-match of our photometry to the full DJA v4.4 catalog. The columns ``ID'' and ``$z_{\rm spec}$'' refer to the photometric {\tt SURVEY\_ID} and the ``zgrade'' spectroscopic redshift respectively. The failure reasons make reference to the numbered list of selection criteria in \autoref{sec:EPOCHS_v2_sample_selection}.\label{tab:highz_spec}}
    \tablehead{
    \colhead{Source name} & \colhead{Survey} & \colhead{ID} & \colhead{$z_{\rm spec}$} & \colhead{$z_{\rm phot, best}$} & \colhead{$z_{\rm phot}$} & \colhead{grade} & \colhead{Failure criteria} & \colhead{Reference(s)}
    }
    \startdata
    GS-z14-0 & GS-South & 41238 & $14.179$ & $3.57$ & $3.53_{-0.00}^{+0.05}$ & 3 & \#(2,3,4,5) & (C+24) \\
    GS-z14-1$^{\dagger\dagger}$ & GS-South & 30479 & $14.080$ & $14.06$ & $14.03_{-0.30}^{+0.45}$ & 3 & $-$ & (C+24) \\
    GS-z13-0 & GS-East & 39033 & $13.117$ & $12.02$ & $11.94_{-0.13}^{+0.26}$ & 3 & \#1 & (CL+23) \\
    GS-z12-0$^{\dagger\dagger}$ & GS-East & 15655 & $12.521$ & $12.28$ & $12.33_{-0.26}^{+1.09}$ & 3 & $-$ & (CL+23) \\
    GS-z11-0 & GS-East & 31611 & $11.594$ & $11.39$ & $11.30_{-0.12}^{+0.25}$ & 3 & \#1 & (CL+23) \\
    JADES-GS-20015720$^{\dagger\dagger}$ & GS-South & 10008 & $11.278$ & $11.08$ & $11.06_{-0.12}^{+0.24}$ & 3 & $-$ & (T+25) \\
    JADES-GS-20176151$^{\dagger\dagger}$ & GS-South & 43988 & $10.488$ & $10.69$ & $10.59_{-0.00}^{+0.12}$ & 3 & $-$ & (T+25) \\
    GS-z10-0 & GS-East & 35252 & $10.350$ & $10.30$ & $10.25_{-0.11}^{+0.22}$ & 3 & \#8 & (CL+23) \\
    JADES-GS-20021387 & GS-South & 36044 & $10.152$ & $2.47$ & $2.42_{-0.20}^{+0.37}$ & 2 & \#(1,2,3,5,8) & (T+25) \\
    GS-20030902$^{\S\ddagger\dagger}$ & GS-South & 16853 & $10.425$ & $10.09$ & $10.02_{-0.11}^{+0.22}$ & $-$ & $-$ & (RB+25) \\
    \hline
    GN-z11$^{\dagger}$ & GN-Medium & 15369 & $10.606$ & $10.42$ & $10.36_{-0.11}^{+0.23}$ & 3 & $-$ & (B+23) \\
    \hline
    CAPERS-EGS-65480$^{\dagger}$ & CEERS-P4 & 2637 & $12.709$ & $11.13$ & $11.06_{-0.36}^{+0.60}$ & 3 & $-$ & (RB+25) \\
    CEERS2\_588 & CEERS-P2 & 4267 & $11.635$ & $11.29$ & $3.39_{-0.21}^{+8.13}$ & 3 & \#5 & (AH+23b) \\
    Maisie's Galaxy$^{\dagger\dagger}$ & CEERS-P2 & 9216 & $11.414$ & $11.52$ & $11.43_{-0.12}^{+0.25}$ & 3 & $-$ & (F+22; AH+23b) \\
    CAPERS-EGS-43539 & CEERS-P9 & 3481 & $11.265$ & $2.62$ & $2.60_{-0.04}^{+0.07}$ & 2 & \#(1,2,3,5,8) & (RB+25) \\
    EGS-69 & CEERS-P2 & 5898 & $10.630$ & $9.43$ & $9.49_{-0.21}^{+0.42}$ & 3 & \#2 & (RB+25) \\
    EGS-22637$^{\dagger}$ & CEERS-P9 & 5505 & $10.593$ & $10.18$ & $10.13_{-0.11}^{+0.22}$ & 3 & $-$ & (RB+25) \\
    \hline
    CAPERS-COSMOS-109917$^{\dagger}$ & PRIMER-COSMOS & 91484 & $10.278$ & $9.88$ & $9.70_{-0.21}^{+0.43}$ & 3 & $-$ & (T+25) \\
    MoM-z14$^{\S\dagger}$ & PRIMER-COSMOS & 68416 & $14.44$ & $14.61$ & $14.49_{-0.15}^{+0.47}$ & 1 & $-$ & (N+26) \\ 
    \hline
    MoM-z11-2$^{\dagger}$ & PRIMER-UDS & 247699 & $10.797$ & $11.66$ & $11.55_{-0.25}^{+0.63}$ & 3 & $-$ & (RB+25) \\
    CAPERS\_UDS\_z10$^{\dagger}$ & PRIMER-UDS & 258462 & $10.558$ & $10.26$ & $10.25_{-0.11}^{+0.22}$ & 3 & $-$ & (K+25b) \\
    MoM-z10-1$^{\dagger}$ & PRIMER-UDS & 226232 & $10.115$ & $9.77$ & $9.70_{-0.21}^{+0.32}$ & 3 & $-$ & (RB+25) \\
    \enddata
    \tablecomments{``$\S$'': Incorrect DJA $z_{\rm spec}$ replaced with the published value; 
    ``$\ddagger$'': No assigned DJA grade; 
    ``$\dagger$'': Passes \epochstwo\ fiducial selection; 
    ``$\dagger\dagger$'': Passes \epochstwo\ ``gold'' $z_{\rm phot}>10.5$ selection}
\end{deluxetable*}

\subsection{Photometric candidates at \texorpdfstring{$z>10.5$}{z>10.5} with spectroscopic confirmation}
\label{sec:highz_candidates}


Our fiducial \epochstwo\ sample comprises $186$ candidates at $10.5<z_{\rm phot}<13.5$ and $26$ at $13.5<z_{\rm phot}<16.5$ with the loosened selection criteria \#2 from \autoref{sec:EPOCHS_v2_sample_selection}. The number of selected candidates passing criteria \#$\{1,3-8\}$ and \#2 with $\mathrm{SNR}_{\rm red\,Ly\alpha}=\{5.0,5.0\}$, $\mathrm{SNR}_{\rm red\,Ly\alpha}=\{8.0,8.0\}$, and $\mathrm{SNR}_{\rm red\,Ly\alpha}=\{10.0,8.0\}$ thresholds in each survey is outlined in \autoref{tab:highz_phot_selected}. Note that we additionally remove objects $\mathrm{SNR^{\rm stack}_{\rm blue\,Ly\alpha}}<2.0$ for wideband filters bluewards of \lya.

As discussed in \autoref{sec:EPOCHS_v2_sample_selection}, we produce a ``gold'' sample in each redshift bin by additionally removing sources that are too diffuse/lie within the low SNR outskirts of a nearby bright source and tightening criteria \#2 to $\mathrm{SNR}_{\rm red\,Ly\alpha}=\{8.0,8.0\}$ at $10.5<z<13.5$ and $\mathrm{SNR}_{\rm red\,Ly\alpha}=\{10.0,8.0\}$ at $13.5<z<16.5$. We additionally run \bagpipes\ with redshift free for this sample, finding that all redshifts fall within $\pm0.15\times(1+z_{\rm spec})$ of those computed by \eazypy. From \autoref{tab:highz_phot_selected} we see that $45$ and $17$ ``gold'' candidates are selected at $10.5<z_{\rm phot}<13.5$ and $13.5<z_{\rm phot}<16.5$ respectively.

\begin{figure*}
	\centering
	\begin{subfigure}{0.49\linewidth}
		\centering
		\includegraphics[width=\linewidth]{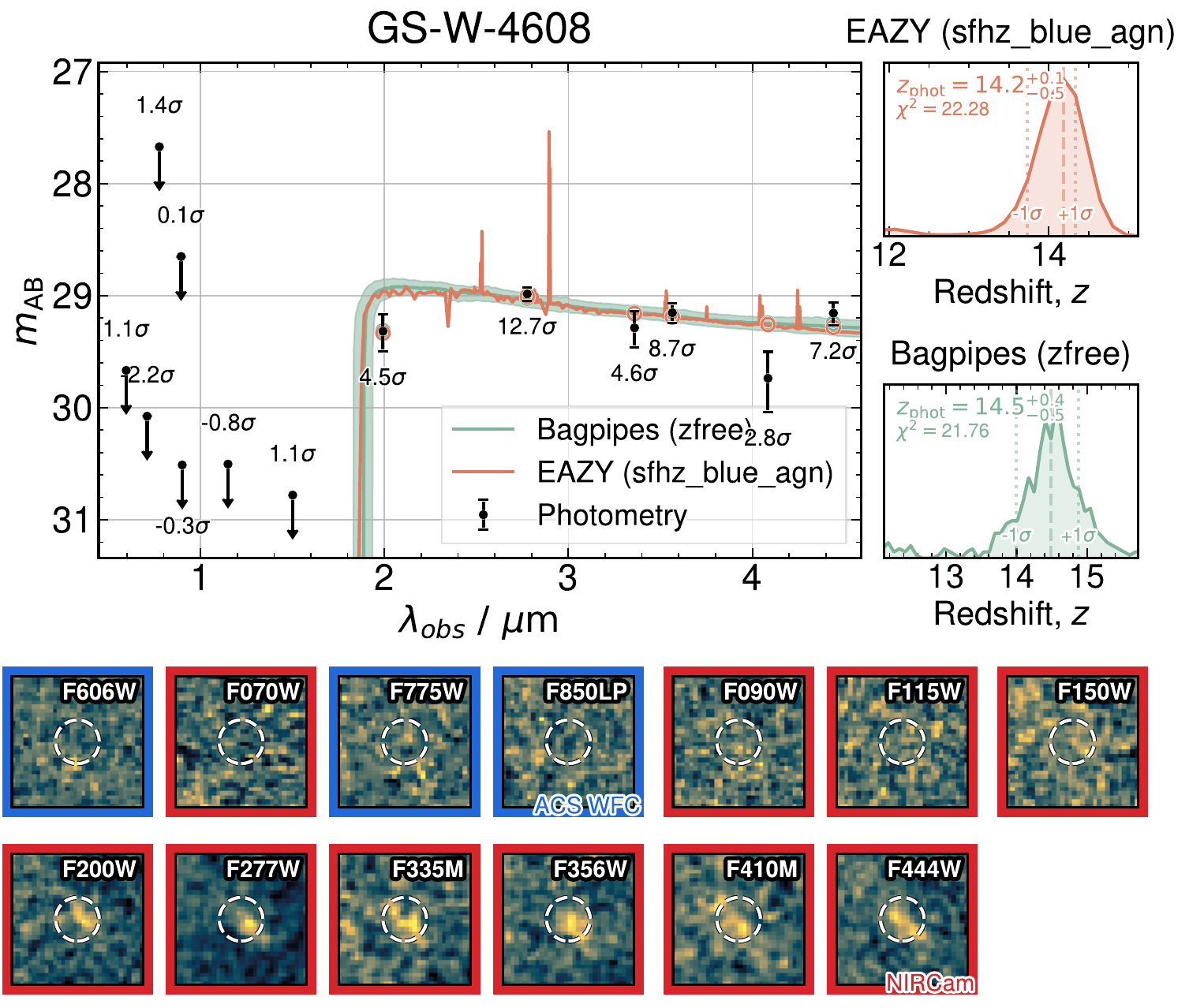}
		\label{fig:sed_4608}
	\end{subfigure}
    \hfill
    \begin{subfigure}{0.49\linewidth}
		\centering
		\includegraphics[width=\linewidth]{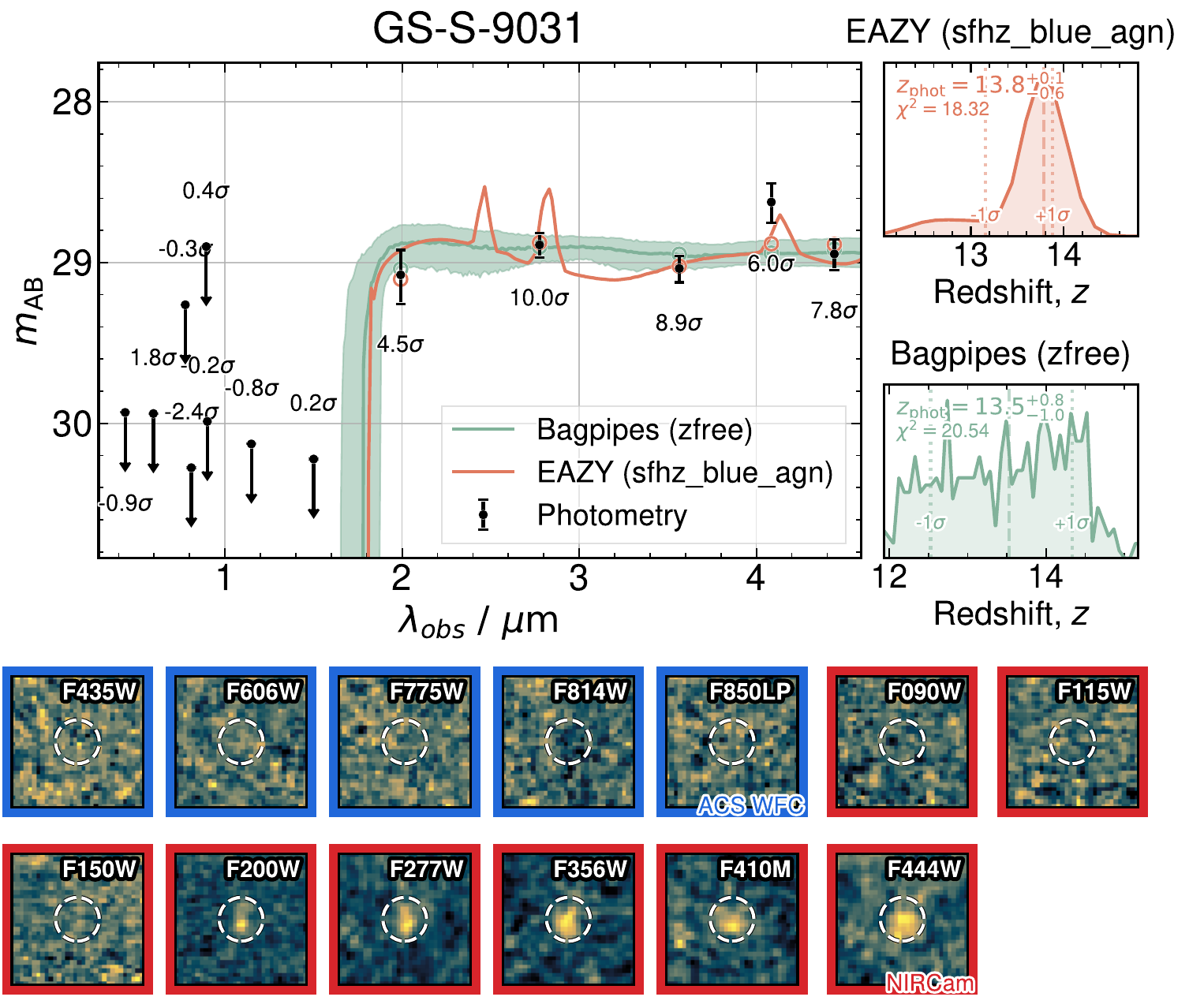}
		\label{fig:sed_9031}
	\end{subfigure}
    \\
    \begin{subfigure}{0.49\linewidth}
		\centering
		\includegraphics[width=\linewidth]{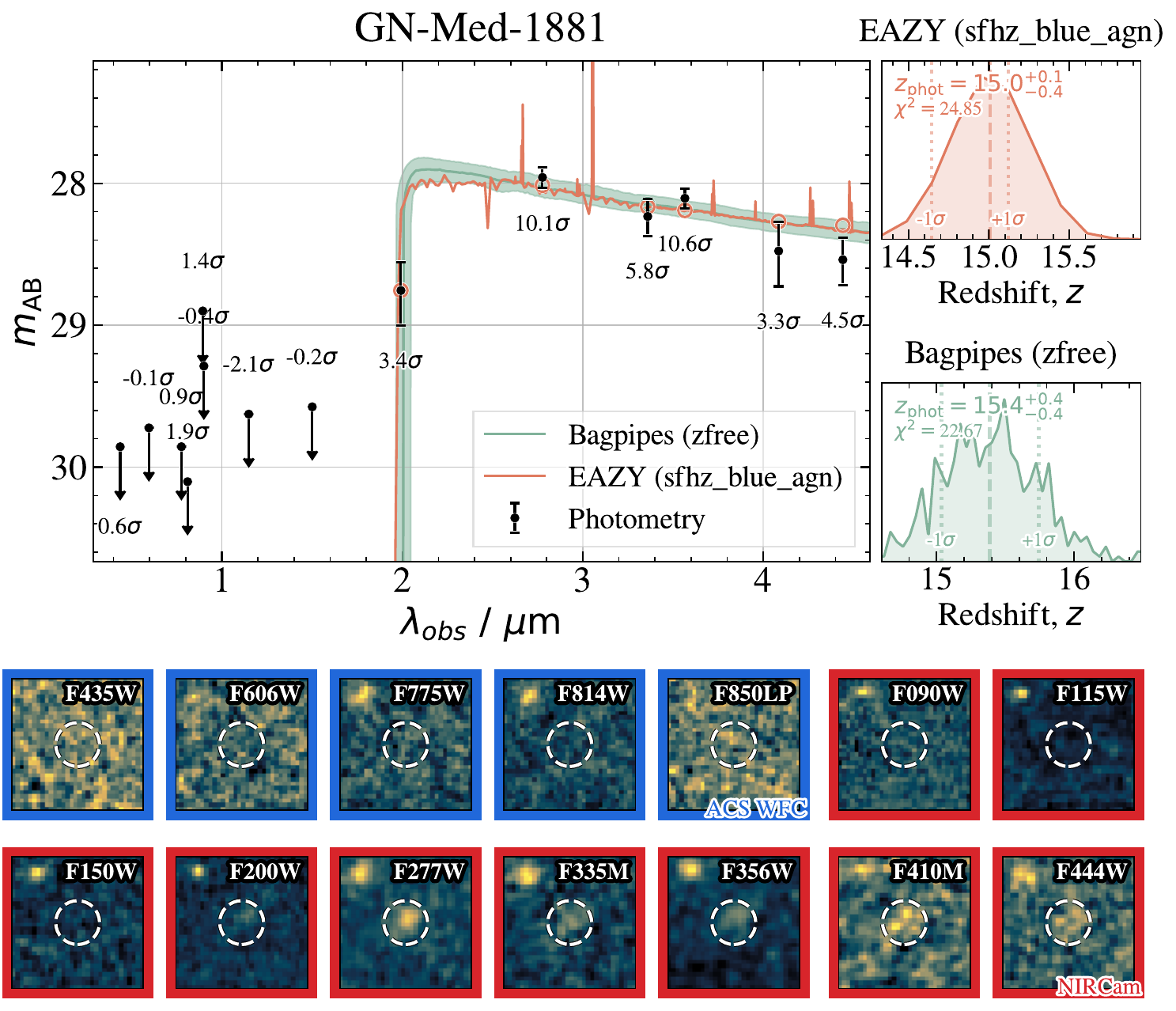}
		\label{fig:sed_1881}
	\end{subfigure}
    \hfill
    \begin{subfigure}{0.49\linewidth}
		\centering
		\includegraphics[width=\linewidth]{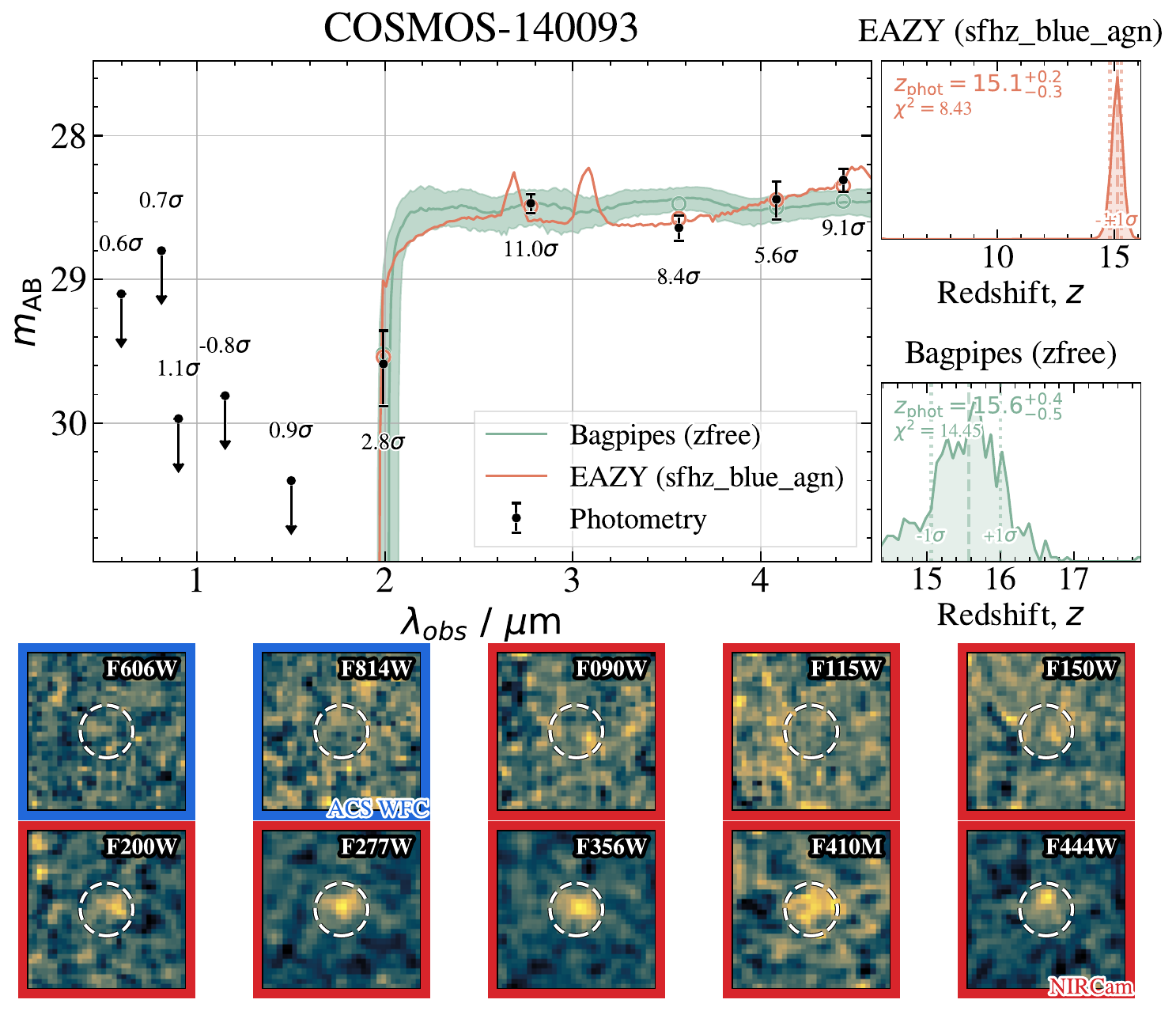}
		\label{fig:sed_140093}
	\end{subfigure}
	\caption{Photometric SED diagnostic plots for $4$ of our most promising ``gold'' candidates at $13.5<z<16.5$ without NIRSpec coverage produced with \texttt{galfind.Galaxy.plot\_phot\_diagnostic} and highlighted in \autoref{tab:gold_sample}. We show the best-fitting \eazypy\ and \bagpipes\ ``zfree'' Bayesian posterior SED in orange and green respectively, with photometric data shown in black alongside their corresponding SNR values. Non-detections are plotted as upper limits at the $2\sigma$ local depth. $0.96''\times0.96''$ ACS/WFC (blue outline) and NIRCam (red outline) cutouts are shown, with $0.32''$ diameter circular apertures outlined.}
	\label{fig:phot_seds}
\end{figure*}

We check our cross-matched NIRSpec/PRISM catalog to determine if any of these been previously targeted by NIRSpec MSA observations, finding $7$ DJA ``grade 3'' sources at $10.5<z<13.5$ and $2$ at $13.5<z<16.5$, with $4/7$ and $1/2$ correctly identified at high-redshift. 
The successful candidates are presented in \autoref{tab:highz_spec} and comprise entirely of previously published high-redshift sources, GS-z14-1 \citep{Carniani2024}, JADES-GS-20015720 and JADES-GS-20176151 \citep{Tang2025}, Maisie's galaxy \citep{Finkelstein2022-z12, ArrabalHaro2023b}, and GS-z12-0 \citep{CurtisLake2023}. Three candidates at $10.5<z_{\rm phot}<13.5$, namely COSMOS-171869 (\texttt{capers-cos13-v4 183727} at $z_{\rm spec}=8.288$), COSMOS-173946 (\texttt{capers-cos13-v4 11409} at $z_{\rm spec}=3.136$), UDS-302436 (\texttt{capers-udsp3-v4 123090} at $z_{\rm spec}=3.785$), and a single source at $z_{\rm phot}>13.5$, COSMOS-74973 (\texttt{capers-cos04-v4 39810} at $z_{\rm spec}=3.656$), are spectroscopically confirmed at a lower redshift by the CAPERS program. A table of these ``gold'' candidates is presented in \autoref{tab:gold_sample} and the diagnostic photometry plots for our $4$ most promising high-redshift candidates without NIRSpec spectroscopy, namely GS-W-4608, GS-S-9031, GN-Med-1881, and COSMOS-140093, made using {\tt Galaxy.plot\_phot\_diagnostic()} from the {\tt galfind.galaxy} module, are presented in \autoref{fig:phot_seds}. 

Our \autoref{fig:specz_vs_photoz_comparison} suggests contamination at $13.5<z_{\rm phot}<16.5$ is potentially as large as $\eta=75_{-37}^{+21}\,\%$ ($3/4$), although the limited number of confirmed sources at $z_{\rm spec}>13.5$ \citep[see][]{Carniani2024, Donnan2026, Naidu2026} makes this challenging to quantify. Assuming $\sim25\%$ of our ``gold'' photometric candidates are real high-redshift sources, we estimate there to be $\sim 4$ across the \epochstwo\ unmasked sky area. If we include MoM-z14 in our contamination estimates, which marginally fails our ``gold'' SNR criteria redwards of \lya, we reduce our contamination to $\sim60\%$ and the expected number of high-z sources jumps to $\sim6-7$. This is in slight tension with recent results from \jwst/NIRCam wide area \citep{McLeod2026} and pure-parallel \citep{Weibel2026} studies and highlights the fragility of number density estimates towards the sensitivity limit of photometric imaging datasets. Historically this tension has also been seen with \hst\ at $z\sim10$ in the mid-late 2010s \citep[e.g.][]{McLeod2016,Oesch2018}; \jwst\ has since observed an abundance of galaxies at this epoch, many of which have been spectroscopically confirmed as far as $z_{\rm spec}=14.44$ \citep[][see \autoref{sec:spec_confirmed_sources}]{Naidu2026}.

\begin{table}
    \setlength{\tabcolsep}{3pt}
    \centering
    \caption{Number counts for high redshift \epochstwo\ candidates in each field selected for different $\mathrm{SNR_{\rm red\,Ly\alpha}}$ thresholds. These include the $\mathrm{SNR^{\rm stack}_{\rm blue\,Ly\alpha}}<2.0$ criteria and exclude any sources removed by eye. Spectroscopic redshift successes and failures in each redshift range are highlighted. The ``$\dagger$'' indicates our ``gold'' sample; individual candidates are outlined in greater detail in \autoref{tab:gold_sample} in \autoref{sec:gold_properties}.}
    \label{tab:highz_phot_selected}
    \begin{tabular}{l|ccc}
    \toprule
    \multirow{2}{*}{\textbf{Survey}} & \multicolumn{3}{c}{\# selected} \\
    & $\{5.0, 5.0\}\sigma$ & $\{8.0, 8.0\}\sigma$ & $\{10.0, 8.0\}\sigma$ \\
    \midrule
    \multicolumn{4}{c}{$\mathbf{10.5 < z < 11.5}$} \\
    \midrule
    CEERS & 11 & 2 & 0 \\
    GOODS-North & 4 & 2 & 1 \\
    GOODS-South & 40 & 13 & 9 \\
    NEP & 10 & 4 & 2 \\
    PRIMER-COSMOS & 11 & 5 & 1 \\
    PRIMER-UDS & 16 & 4 & 2 \\
    \hline
    \textbf{Total} & $\mathbf{92}$ & $\mathbf{30^{\dagger}}$ & $\mathbf{15}$ \\
    NIRSpec (high-z) & 3 (3\%) & 2 (7\%) & 2 (15\%) \\
    NIRSpec (low-z) & 3 (3\%) & 3 (10\%) & 0 (0\%) \\
    \midrule
    \multicolumn{4}{c}{$\mathbf{11.5 < z < 13.5}$} \\
    \midrule
    CEERS & 7 & 1 & 1 \\
    GOODS-North & 2 & 1 & 0 \\
    GOODS-South & 27 & 5 & 4 \\
    NEP & 6 & 2 & 0 \\
    PRIMER-COSMOS & 6 & 1 & 0 \\
    PRIMER-UDS & 21 & 5 & 3 \\
    \hline
    \textbf{Total} & $\mathbf{69}$ & $\mathbf{15^{\dagger}}$ & $\mathbf{8}$ \\
    NIRSpec (high-z) & 3 (4\%) & 2 (13\%) & 2 (25\%) \\
    NIRSpec (low-z) & 1 (1\%) & 0 (0\%) & 0 (0\%) \\
    \midrule
    \multicolumn{4}{c}{$\mathbf{13.5 < z < 16.5}$} \\
    \midrule
    CEERS & -- & 0 & 0 \\
    GOODS-North & -- & 1 & 1 \\
    GOODS-South & -- & 5 & 5 \\
    NEP & -- & 2 & 2 \\
    PRIMER-COSMOS & -- & 11 & 4 \\
    PRIMER-UDS & -- & 7 & 5 \\
    \hline
    \textbf{Total} & -- & $\mathbf{26}$ & $\mathbf{17^{\dagger}}$ \\
    \textbf{NIRSpec (high-z)} & $-$ & 1 (4\%) & 1 (6\%) \\
    \textbf{NIRSpec (low-z)} & $-$ & 1 (4\%) & 1 (6\%) \\
    \botrule
    \end{tabular}
\end{table}

\subsection{A closer look at \texorpdfstring{$z>10$}{z>10} galaxy sizes}
\label{sec:highz_sizes}

Since one might expect more distant sources observed earlier in the Universe's history to be younger and thus smaller with reduced surface brightness ($\mu\propto(1+z)^{-4}$), we now take a closer look at the sizes of the $z_{\rm spec}>10$ spectroscopic sample detailed in \autoref{sec:spec_confirmed_sources}. We fit the light profile of each source using \pysersic\ adopting a simple single S{\'e}rsic model, 
\begin{equation}
    I(r) = I(r_{\mathrm{e}})\exp\Bigg\{-b_n\bigg[\bigg(\frac{r}{r_{\mathrm{e}}}\bigg)^{1/n} - 1\bigg]\Bigg\},
    \label{eq:single_sersic}
\end{equation}
where $I$ is the intensity at radius $r$, $n$ is the S{\'e}rsic index commonly assumed as $n\simeq4$ for ellipticals (giving the \citet{deVaucouleurs1948} profile) and $n\simeq1$ for disk galaxies \citep{Sersic1963, Ciotti1991, Caon1993}, and $b_n\simeq2n-\frac{1}{3}+ \frac{4}{405n} - \frac{46}{25515n^2}$ \citep{Ciotti1999}. The F277W filter is chosen for the $5''\times5''$ cutouts since it traces the rest-frame UV at these redshifts and often hosts the deepest depths across our dataset owing to its impressive sensitivity. 

We run our Bayesian fitting procedure using the empirical \epochstwo\ PSFs outlined in \autoref{sec:PSF_homogenization}, masking out neighbours identified by our \sextractor\ deblending procedure. In our fiducial run, the S{\'e}rsic index kept free across the default \pysersic\ $n\in [0.65, 8.0]$ prior. We notice that approximately $10-15\%$ of our sources exhibit bimodality in $n$ peaking towards the lower/upper prior limits, indicating that a multi-component point-source + single S{\'e}rsic or double S{\'e}rsic model may be more appropriate, potentially highlighting hidden AGN or early bulges in these sources. In addition, $\simeq70-80\%$ converge to $n\gtrsim6$, which is likely a natural consequence of fitting mostly low SNR unresolved sources. For this reason we choose to fix $n=1$ in the fitting procedure.

\begin{figure}
    \centering
    \includegraphics[width=0.99\linewidth]{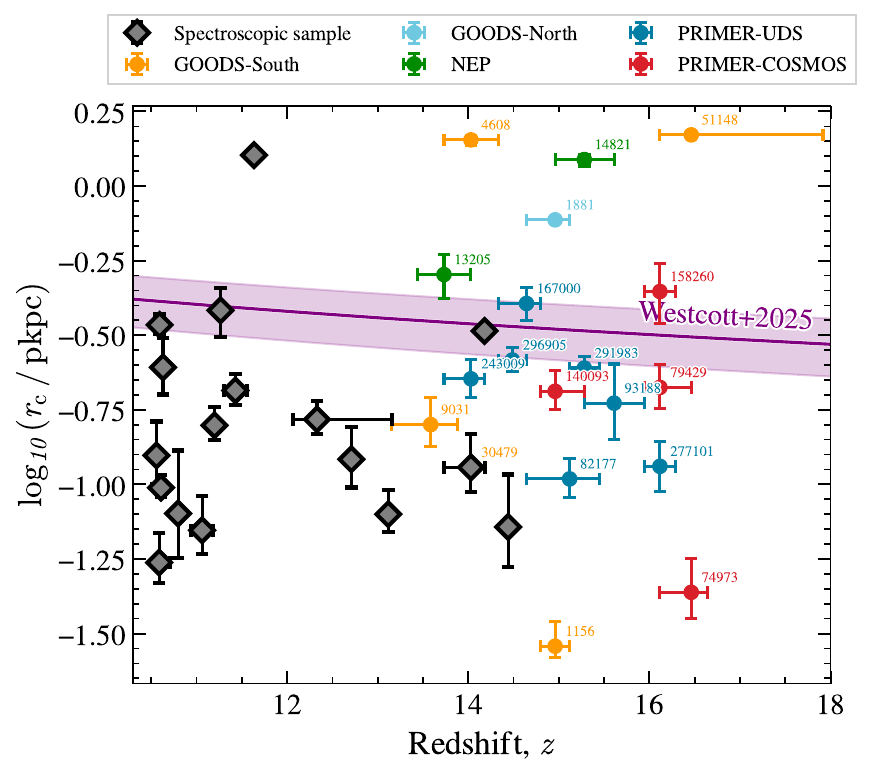}
    \caption{Redshift evolution of the circularized radius for $z_{\rm spec}>10$ spectroscopically confirmed galaxies and our ``gold'' $13.5<z_{\rm phot}<16.5$ photometric sample computed using fixed $n=1$ single S{\'e}rsic F277W Bayesian fits with \pysersic. Our spectroscopic sample yields median sizes $\sim0.3-0.5$ smaller than expected from the extrapolated \citet{Westcott2025} relation for star-forming galaxies, shaded in purple. The colored photometric sample highlights a large diversity in $z_{\rm phot}>13.5$ sizes including a number of extended $r_{\rm c}\gtrsim0.5-1\,\rm kpc$ sources.}
    \label{fig:sizes}
\end{figure}

Our circularized galaxy sizes, $r_{\rm c}=r_{\rm e}\sqrt{1-e}$ (where $e=1-b/a$ is the galaxy ellipticity and $r_{\rm e}$ is the semi-major axis) are shown in \autoref{fig:sizes} as a function of redshift. We compare our $r_{\rm c}$ sizes against the extrapolated \citet{Westcott2025} fit for the $r_{\rm e}$ size evolution of typical star-forming galaxies, finding an $\sim0.5\,\rm dex$ offset towards smaller sizes at $z_{\rm spec}>10.5$. This can be partly explained by a combination of differences in the \galfit\ vs the fully Bayesian \pysersic\ fitting procedures (for instance by fixing $n=1$, which reduces the median circularized radius of the spectroscopically confirmed sample, $\langle r_{\rm c}\rangle$, by $\sim0.4\,\rm dex$ as a result) and the samples fit here and by \citet{Westcott2025} which likely contains contaminant Balmer break systems, increasing the median size at a given redshift. In addition, the \citet{Westcott2025} results are fit using F444W imaging which probes the rest-frame optical rather than the UV done in this work. An analysis of wavelength dependent sizes and its link to star formation will be explored later in this paper series.

In addition to the spectroscopic sample, we fit our $13.5<z_{\rm phot}<16.5$ ``gold'' candidates using the same fitting methodology explained above, plotting the results on \autoref{fig:sizes}. Similarly to \citet{Hainline2026}, we identify a number of resolved sources with $r_{\rm c}\gtrsim0.5-1\,\rm pkpc$ suggestive of spatially extended star formation produced by extreme gas recycling induced by stochastic star formation driven outflows \citep{McClymont2025b}.

The sizes of these very early galaxies are in the same range as previous \jwst\ analyses \citep[][]{Finkelstein2023, Robertson2023}. These sources may be genuine spatially extended stellar systems, indicating that galaxy structures, and potentially disks, can be formed very early \citep[][]{Ferreira2022, Ferreira2023, Xu2024}. However these (relatively) large sizes could also result from clumpy star-formation, with neighbouring star-formation regions distributed over a $\sim0.5-1\,\rm kpc$ wide area without a dynamically settled disk \citep[see e.g.][]{Harikane2025}. Multi-wavelength sizes are required to gain more insight in the stage of star-formation in which these galaxies are observed, to distinguish evolved galaxy-scale structure from spatially distributed episodes of star formation \citep{Roberts-Borsani2026}.

It is expected that $z\sim4$ massive quiescent galaxies and dusty Balmer break interlopers are intrinsically larger than $z_{\rm phot}>13.5$ star-forming systems. Results from the \jwst\ NIRSpec Cycle~1 program ``How Many Quiescent Galaxies Are There at $3<z<4$ Really?'' (PI Glazebrook; PID: GO 2565) by \citet{Kawinwanichakij2026} suggests $r_{\rm e, F277W} = 0.72 \pm 0.08\,\rm pkpc$ for $M_{\star} = 5\times10^{10}\,\mathrm{M}_{\odot}$ systems, $\sim0.3-0.4\,\rm dex$ above the \citet{Westcott2025} relation. The four quiescent sources from \citet{Carnall2024} quoted with $r_{\rm e,F277W}\simeq\{0.61,0.31,0.73,0.91\}\,\rm pkpc$ are approximately consistent with the \citet{Kawinwanichakij2026} sizes, although $\sim0.3-1.0\,\rm dex$ larger than the three spectroscopically confirmed sources at $z_{\rm spec}>13.5$. The turnover in angular diameter distance at $z\sim1.5$ \citep{Hogg1999}, however, suggests that contaminant sources of the same intrinsic size appear a factor $D_{\rm A}(z=4)/D_{\rm A}(z=15)\simeq2.2$ ($\simeq0.35\,\rm dex$) larger at high-z, somewhat converging the two size/redshift regimes. Our extended $z_{\rm phot}>13.5$ sources which are approximately the same size are thus plausible contaminants, however one spectroscopically confirmed source (CEERS2\_588 at $z=11.635$ from \citealt{ArrabalHaro2023b}) identified with $r_{\rm c}=1.27\pm 0.02\,\rm pkpc$ highlights the possibility that these are in fact real high-redshift sources. We conclude that our size measurements cannot adequately distinguish single sources from contaminant galaxies due to the intrinsic variation in galaxy UV size across the sample.

\subsection{Are our \texorpdfstring{$z>13.5$}{z>13.5} candidates in fact lower-redshift interlopers?}
\label{sec:ultra_highz_candidates}

\begin{figure}
    \centering
    \includegraphics[width=\linewidth]{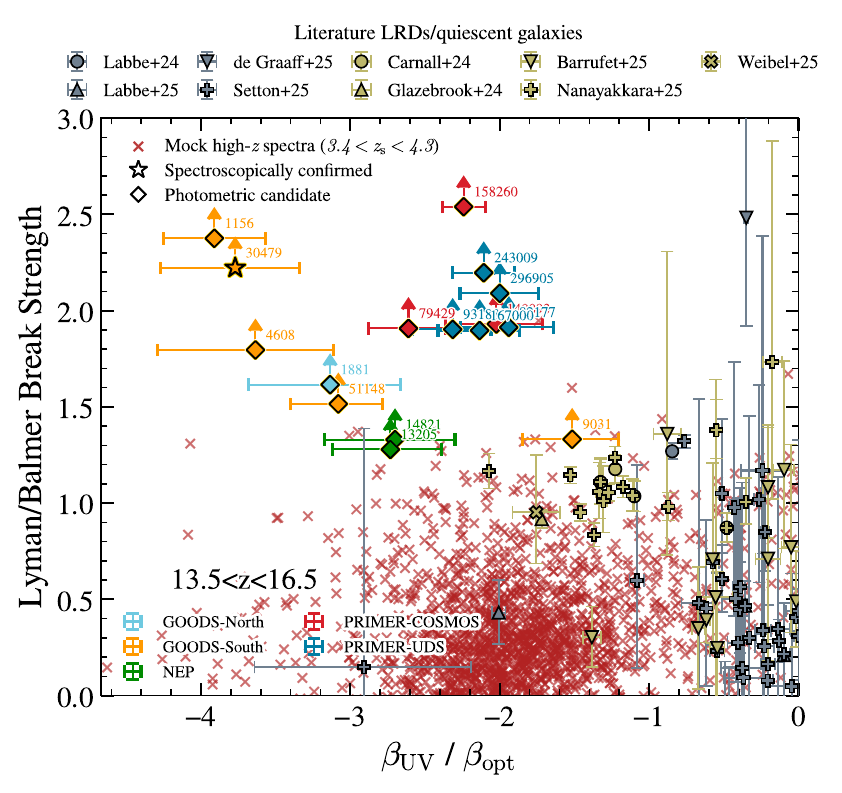}
    \caption{Lyman/Balmer break strength as a function of $\beta_{\rm UV/opt}$ for our ``gold'' sample of $16$ candidate SFGs at $13.5<z_{\rm phot}<16.5$ without NIRSpec coverage. We compare our high-redshift sources against samples of LRDs \citep[][in grey]{Labbe2024, Labbe2025, deGraaff2025, Setton2025} and massive quiescent galaxies \citep[][in khaki]{Carnall2024, Glazebrook2024, Barrufet2025, Nanayakkara2025, Weibel2025} compiled from the literature, which are likely contaminants at $z\sim4$. Red crosses highlight ``high-redshift'' properties from mock photometry of $3.4<z_{\rm spec}<4.3$ grade 3 PRISM sources shifted to $z'=(1+z_{\rm s})\times(4000\,\rm\AA/\lambda_{\rm Ly\alpha}) - 1$.}
	\label{fig:break_strength_vs_beta}
\end{figure}

It is natural to question whether our large sample of $13.5<z_{\rm phot}<16.5$ sources are in fact real high-redshift systems, especially given the computed $\eta=75_{-37}^{+21}\,\%$ contamination estimate from \autoref{fig:specz_vs_photoz_comparison}. Perhaps they are interlopers with large Balmer breaks or dusty sources with strong optical emission lines mimicking a large break and flat/blue continuum in broadband filters, as was the case with CEERS\_93316 \citep{Naidu2022b, Donnan2023, ArrabalHaro2023b}. We attempt to disentangle this redshift bimodality by comparing our candidates in ``\lya\ break strength--\uvbeta'' parameter space to lower-redshift contaminants in their equivalent Balmer break strength--$\beta_{\rm opt}$ space, with results shown in the left-hand panel of \autoref{fig:break_strength_vs_beta}. 

Samples of spectroscopically confirmed massive quiescent galaxies \citep{Carnall2024, Glazebrook2024, Barrufet2025, Nanayakkara2025, Weibel2025} and LRDs \citep{Labbe2024, Labbe2025, deGraaff2025, Setton2025} are compiled from the literature for the comparison. NIRSpec PRISM spectra are taken from the DJA v4.4, from which we compute Balmer break flux ratios, $D_{\rm Balmer}$, in $3620<\lambda_{\rm rest}\,/\,\rm \AA<3720$ and $4000<\lambda_{\rm rest}\,/\,\rm \AA<4100$ windows following the procedure in \citet{Wang2025} and derive break strengths (i.e. colors) as $2.5\log_{10}(D_{\rm Balmer})$. We note that this Balmer break (at $3646\,\rm \AA$) traces A-stars, and is different to the ``$4000\,\rm \AA$ break'' (at $3968\,\rm \AA$) produced by metal absorption lines strongest in older G, K, and M stars. $\beta_{\rm opt}$ is calculated in wavelength windows closely resembling the rest-frame UV coverage used to measure \uvbeta, albeit shifted redwards by $\Delta\lambda_{\rm rest}=4000\rm\AA/\lambda_{\rm Ly\alpha}$, and with $200\,\rm \AA$ width avoidance zones surrounding the Balmer series recombination/absorption lines and the \fion{O}{3} $\lambda\lambda 4959,5007$ doublet. While intriguing, this is strictly not a direct comparison to our high-redshift candidates since there is no guarantee our broadband photometry actually traces the rest optical continuum.

To determine whether our high-redshift sources are situated in a unique portion of the parameter space, we compare to mock photometry in the F200W, F277W, F356W, F410M, and F444W NIRCam filters from all grade 3 PRISM spectra from DJA v4.4 at $3.4<z_{\rm spec}<4.3$ shifted to $z'=(1+z_{\rm spec})\times(4000\rm\AA/\lambda_{\rm Ly\alpha}) - 1$ (i.e. $13.5<z'<16.5$). From this mock photometry we compute the ``$\rm Ly\alpha$ break strength'' and ``\uvbeta'' properties for each spectrum using the $3620<\lambda_{\rm rest}\,/\,\rm \AA<3720$ tophat flux (measured at redshift $z_{\rm spec}$ as opposed to $z'$) for the break strength anchor, which are additionally displayed in \autoref{fig:break_strength_vs_beta} as red crosses.

None of our candidates appear red enough in the UV to be considered contaminant LRDs, which are predominantly selected with $\beta_{\rm opt}\gtrsim-1$, however GS-S-9031 and potentially the two NEP candidates (13205 and 14821) sit within the realm of massive, quiescent galaxies with flat or red $\beta_{\rm UV/opt}\gtrsim-2$ and Balmer break strengths $\sim1.0-1.2\,\rm mag$. Increasing the number density of these objects at $z\sim4$ could potentially worsen tensions with hydrodynamical simulations which don't always reproduce the abundance of massive quiescent galaxies observed at this epoch \citep[e.g.][]{Lovell2023b, Carnall2024, Russell2025}.

The bundle of PRIMER-COSMOS/UDS sources at $\beta_{\rm UV/opt}\sim-2$ have $\gtrsim2.0\,\rm mag$ break strengths approximately $\sim0.8-1.0\,\rm mag$ larger than the most massive quiescent galaxies. They perhaps have Balmer breaks comparable only to the most extreme LRDs, such as ``The Cliff'' from \citet{deGraaff2025}, and could in principle host extreme Balmer breaks of AGN origin \citep{Inayoshi2025}. Since one selected source ({\tt capers-cos04-v4 39810}; removed from \autoref{fig:break_strength_vs_beta}) has already been confirmed as a strong emission line galaxy at $z_{\rm spec}=3.656$, it is perhaps more plausible that these are lower redshift interlopers; this is especially likely given their reduced depths and filter coverage compared to the other \epochstwo\ fields.

The most exciting quadrant of this plot is the upper right, where the spectroscopically confirmed GS-z14-1 at $z_{\rm spec}=14.080$ resides \citep{Carniani2024}. Our two most promising candidates, GS-W-4608 and GN-Med-1881 are also located in this $\beta_{\rm UV}\lesssim-3$, $>1.5\,\rm mag$ break strength region, as well as NGD-1156 \citep[NGD-z15a from][]{Austin2023} which is detected in just $4$ NIRCam broadband filters. Strong \fion{O}{3}+\hbeta\ could feasibly produce the F277W flux boost leading to an ultra-blue observed $\beta_{\rm UV}=-3.91\pm 0.34$ in NGD-1156 which is otherwise $\sim3\sigma$ discrepant with the youngest, burstiest, and most metal-poor Pop.~II dominated systems \citep[e.g.][]{BC03, Conroy2010_software, Eldridge2017-BPASS, Stanway2018-BPASS} predicted by models. The inclusion of the F335M and F410M medium bands in the JADES footprint containing GS-W-4608 and GN-Med-1881 somewhat breaks this degeneracy and promotes these high-$z$ solutions. NIRSpec/PRISM confirmation of these two candidates would increase the number of known sources at $z>13.5$ by $50\%$, transforming future galaxy evolution studies at the earliest times.

\section{Conclusions}
\label{sec:conclusions}

In this paper we presented the \epochstwo\ sample of $2452$ galaxy candidates at $6.5<z<16.5$ produced from PSF homogenized deep blank field \jwst/NIRCam imaging across $6$ independent sightlines, GOODS-North, GOODS-South, NEP, EGS, COSMOS, and UDS. As a consequence of our masking criteria, the area covered ranges $\sim650\,\rm arcmin^2$ ($6.50<z<7.27$) to $\sim740\,\rm arcmin^2$ ($7.27<z<16.50$). We identify $107$ compact sources with better-fitting brown dwarf than galaxy SEDs ($69$ L--type, $36$ T--type, and $2$ Y--type) computed with \bdfinder, which are removed from the initial sample, as well as $34$ LRDs with ``V--shaped'' SEDs.

Our photometric catalog is cross-matched with the DJA v4.4 spectroscopic catalog, finding $673$ unmasked ``grade 3'' NIRSpec PRISM spectra at $z_{\rm spec}>6.5$, of which $566$ are unmasked, and perform the most comprehensive completeness/contamination analysis at $z>6.5$ with \jwst\ spectra to date. A photometric--spectroscopic redshift comparison is performed for $4$ \epochstwo\ selected samples produced from \eazypy\ SED fits using the ``fsps\_larson'', ``fsps\_jades'', ``sfhz'', and ``sfh\_blue\_agn'' templates; we find that the latter produces the tightest $\sigma_{\rm NMAD}=0.017$, fewest extreme outliers, $\eta=4.3\%$, and smallest median redshift offset, $\langle\Delta z\rangle=-0.022$. $293$ NIRSpec/PRISM spectra are successfully selected ($52\%$) with this fiducial ``sfh\_blue\_agn'' template set, including $14/22$ $z_{\rm spec}>10$ objects from the literature. We find that the lowest redshift bin, $6.5<z<7.5$, has the smallest selection success rate ($47\%$), predominantly caused by the $\mathrm{SNR}_{\rm red\,Ly\alpha}>\{5.0, 5.0\}$ (\#2) and $\Delta\chi^2_{\rm low-z}>4$ (\#5) criteria, which remove $34\%$ and $17\%$ real high-$z$ sources respectively.

Power law fitting results to the $\beta_{\rm UV}-M_{\rm UV}$ relation show $\beta_{\rm UV}(M_{\rm UV}=-19)=-2.32^{+0.08}_{-0.07}$ at $z\simeq12.5$ which evolves slowly towards lower redshift and an increasing scatter $\sigma_{\beta_{\rm UV}}$ at the highest redshifts. This implies \textit{moderately} dust-polluted stellar populations with increasing diversity possibly due to rapid type II SNe dust production and stochastic star formation driving outflows.
We provide \bagpipes\ SED fitting results for the sample; measurements of the KL-divergence for each free parameter/hyperparameter highlight several parameters that are not well constrained by the ACS/WFC+NIRCam data, including $\Phi_{\rm SF}$, $\rm age_{\rm MW}$, $Z_{\rm \star}$, $\log_{10}U$, and $f_{\rm esc}^{\rm LyC}$ at $z>6.5$ and Balmer break strengths ($D_{\rm Balmer}$), $\rm sSFR_{10/100\,Myr}$, and $\xi_{\rm ion,0}$ at $z\gtrsim9$.

We provide a ``gold'' sample of $45$ $10.5<z_{\rm phot}<13.5$ and $17$ $13.5<z_{\rm phot}<16.5$ candidates with an increased $\rm SNR_{\rm red\,Ly\alpha}$ threshold, $38$ and $15$ of which have no associated NIRSpec spectra. The $13.5<z<16.5$ sample shows widely varied galaxy sizes and UV colors, a number of which are extended beyond $r_{\rm c}>1\,\rm pkpc$. Even after accounting for $60-75\%$ of the ultra-high redshift sample expected to be strong emission line or quiescent galaxy interlopers, we expect $5-7$ of these to be real, in mild tension with number density estimates from \citet{McLeod2026} and \citet{Weibel2026}. Two robust $\beta_{\rm UV}<-3$ sources with \lya\ break colors $>1.6\,\rm mag$ and $\gtrsim3\sigma$ F335M and F410M detections, which merit spectroscopic follow-up, and may somewhat relieve this tension.

This \epochstwo\ sample forms the basis of the upcoming paper series; we will compute UVLFs and stellar mass functions to constrain the galaxy--halo connection, and measure the ionizing properties of these sources to better understand the nature of reionization. Reproducible imaging, data products, and catalogs made using the \galfind\ python tool are released publicly and provide vital \jwst\ legacy imaging datasets for use in a variety of other science cases.

\begin{acknowledgments}
We acknowledge support from the ERC Advanced Investigator Grant EPOCHS (788113), as well as two studentships from the STFC. This work is based on observations made with the NASA/ESA \textit{Hubble Space Telescope} (HST) and NASA/ESA/CSA \textit{James Webb Space Telescope} (JWST) obtained from the \texttt{Mikulski Archive for Space Telescopes} (\texttt{MAST}) at the \textit{Space Telescope Science Institute} (STScI), which is operated by the Association of Universities for Research in Astronomy, Inc., under NASA contract NAS 5-03127 for JWST, and NAS 5–26555 for HST. The authors thank all involved with the construction and operation of JWST, without whom this work would not be possible. Some of the data products presented herein were retrieved from the Dawn JWST Archive (DJA). DJA is an initiative of the Cosmic Dawn Center (DAWN), which is funded by the Danish National Research Foundation under grant DNRF140.

RAW acknowledges support from NASA JWST Interdisciplinary Scientist grants
NAG5-12460, NNX14AN10G and 80NSSC18K0200 from GSFC.

\end{acknowledgments}

\vspace{1.5em}

\section*{Data Availability}

The imaging, catalogs, and all downstream products associated with this data release will be made available for public download after the final revisions to this paper have been made. Catalogs are available via reasonable request to the corresponding author. These are all produced by, and hence compatible with, the \galfind\ software package\footnote{\doi{10.5281/zenodo.18613231}} written in python. \\

\begin{contribution}

We list here the roles and contributions of the authors according to the Contributor Roles Taxonomy (CRediT)\footnote{\url{https://credit.niso.org}}. \\
\textbf{Duncan A. Austin}: Conceptualization, Data curation, Formal analysis, Investigation, Methodology, Project Administration, Resources, Software, Validation, Visualization, Writing - original draft, Writing - review \& editing \\
\textbf{Thomas Harvey}: Software, Writing - review \& editing \\
\textbf{Christopher J. Conselice}: Funding Acquisition, Project Administration, Supervision, Writing - review \& editing \\
\textbf{Nathan J. Adams}: Data curation, Resources \\
\textbf{Louis Quilley, Jordan C. J. D'Silva, Rogier A. Windhorst:} Writing - review \& editing \\
\textbf{William J. Roper:} Software \\

\end{contribution}

\facilities{\hst(ACS/WFC), \jwst(NIRCam), \jwst(NIRSpec)}


\appendix




\section{\epochstwo\ depths and areas}
\label{sec:areas_depths}

In this section, we outline the procedure within \galfind\ to calculate areas and depths for this \epochstwo\ data release. Due to the depth inhomogeneity in our surveys, it is important to divide areas of differing exposure time into separate subregions. This, of course, is challenging when considering the differing coverage across a number of filters, and thus we simplify to a maximum of two regions per ``.fits'' mosaic. These subregions are determined using a K-means clustering algorithm in each filter to compute the local depths (see \texttt{galfind.utils.Depths}), and unless otherwise stated they are based on the F277W+F356W+F444W stacked LW detection band. 

The median and $1\sigma$ local depths in each subregion is presented in \autoref{tab:EPOCHS_v2_ACS+NIRCam_SW_depths} for the ACS/WFC and NIRCam SW filters and \autoref{tab:EPOCHS_v2_NIRCam_LW_depths} for the NIRCam LW filters; the subfields from these tables are summarised below. Imaging in GOODS-South is split into $5$ mosaics: NGDEEP (Deep and Shallow); JADES-DR3-GS-North; JADES-DR3-GS-South (JOF and not JOF); JADES-DR3-GS-East (Deep and Shallow); and JADES-DR3-GS-West (Deep and Shallow, based on F335M coverage). The GOODS-North imaging is split into $3$ mosaics: JADES-DR3-GN-Deep (Deep and Shallow); JADES-DR3-GN-Medium; and JADES-DR3-GN-Parallel. NEP-TDF data is split into the $4$ spokes (labelled $1-4$), and CEERS data is computed on a pointing-by-pointing basis. PRIMER-UDS and PRIMER-COSMOS moasics are divided into Deep and Shallow regions, where the UDS subregions are computed from F444W.

While an accurate area calculation is not critical to the results presented in this overview paper, it will become important when measuring UVLFs on a sightline-by-sightline basis (Austin et al. in prep.) to reduce UVLF systematics and provide the best constraints on the build up of structure over cosmic time. Since our selection criteria in \autoref{sec:EPOCHS_v2_sample_selection} requires different filters to be detected/non-detected as a function of redshift, we compute the total unmasked area using boolean logic with the masks for each band satisfying masking criteria \#8. The \epochstwo\ areas for each depth subregion are presented as a function of redshift in \autoref{tab:EPOCHS_v2_areas}, where we show only the redshift bins where there is $\gtrsim0.5\,\rm arcmin^2$ change in total area. Note that these areas are slightly lower than quoted values from the literature due to the removal of areas covered by our relatively strict masking criteria surrounding stellar diffraction spikes and image edges.


\begin{sidewaystable*}[]
   \centering
   \caption{Median ACS/WFC and NIRCam SW depths (with $1\sigma$ error) for the NGDEEP, JADES (GOODS-South + GOODS-North), NEP, CEERS, and PRIMER (UDS + COSMOS) \epochstwo\ surveys. The division of these into subfields are explained in \autoref{sec:areas_depths}.}
   \setlength{\tabcolsep}{2pt}
   \begin{tabular}{c|ccccccccccccc}
   \toprule
Survey & F435W & F606W & F070W & F775W & F814W & F850LP & F090W & F115W & F150W & F162M & F182M & F200W & F210M \\
\hline
NGDEEP (Deep) & $28.6^{+0.2}_{-1.4}$ & $29.8^{+0.3}_{-1.3}$ & $-$ & $29.4^{+0.3}_{-0.4}$ & $29.9^{+0.3}_{-0.7}$ & $28.8^{+0.4}_{-1.5}$ & $-$ & $30.4^{+0.2}_{-0.2}$ & $30.4^{+0.2}_{-0.2}$ & $-$ & $-$ & $30.4^{+0.2}_{-0.2}$ & $-$ \\
NGDEEP (Shallow) & $28.5^{+0.2}_{-1.4}$ & $29.8^{+0.3}_{-0.5}$ & $-$ & $29.3^{+0.3}_{-0.3}$ & $29.6^{+0.5}_{-1.6}$ & $28.9^{+0.2}_{-0.4}$ & $-$ & $30.1^{+0.2}_{-0.2}$ & $30.1^{+0.2}_{-0.2}$ & $-$ & $-$ & $30.0^{+0.3}_{-0.2}$ & $-$ \\
JADES-DR3-GS-North & $28.7^{+0.3}_{-0.2}$ & $28.9^{+0.2}_{-0.2}$ & $28.8^{+0.3}_{-0.2}$ & $28.3^{+0.2}_{-0.2}$ & $28.6^{+0.3}_{-0.3}$ & $28.0^{+0.2}_{-0.2}$ & $29.3^{+0.4}_{-0.3}$ & $29.3^{+0.3}_{-0.2}$ & $29.4^{+0.3}_{-0.3}$ & $-$ & $-$ & $29.4^{+0.3}_{-0.3}$ & $-$ \\
JADES-DR3-GS-South (JOF) & $28.6^{+0.4}_{-0.2}$ & $28.9^{+0.3}_{-0.5}$ & $28.7^{+0.1}_{-0.1}$ & $28.3^{+0.2}_{-0.4}$ & $29.0^{+0.3}_{-0.2}$ & $27.9^{+0.3}_{-0.4}$ & $30.1^{+0.1}_{-0.2}$ & $30.3^{+0.1}_{-0.2}$ & $30.3^{+0.1}_{-0.2}$ & $30.0^{+0.1}_{-0.1}$ & $30.4^{+0.2}_{-0.2}$ & $30.4^{+0.1}_{-0.2}$ & $30.2^{+0.2}_{-0.2}$ \\
JADES-DR3-GS-South & $28.7^{+0.3}_{-0.2}$ & $29.1^{+0.2}_{-0.3}$ & $28.8^{+0.1}_{-0.1}$ & $28.3^{+0.2}_{-0.2}$ & $29.0^{+0.3}_{-0.6}$ & $28.0^{+0.2}_{-0.3}$ & $29.2^{+0.3}_{-0.2}$ & $29.4^{+0.3}_{-0.2}$ & $29.5^{+0.3}_{-0.2}$ & $30.0^{+0.1}_{-0.1}$ & $30.4^{+0.1}_{-0.1}$ & $29.6^{+0.3}_{-0.2}$ & $30.2^{+0.1}_{-0.2}$ \\
JADES-DR3-GS-East (Deep) & $29.2^{+1.1}_{-0.6}$ & $29.4^{+1.2}_{-0.5}$ & $-$ & $29.0^{+1.3}_{-0.6}$ & $28.9^{+0.2}_{-0.3}$ & $28.4^{+1.2}_{-0.4}$ & $30.2^{+0.2}_{-0.2}$ & $30.4^{+0.2}_{-0.2}$ & $30.4^{+0.2}_{-0.2}$ & $-$ & $29.7^{+0.2}_{-0.2}$ & $30.4^{+0.2}_{-0.2}$ & $29.5^{+0.2}_{-0.2}$ \\
JADES-DR3-GS-East (Shallow) & $28.7^{+0.2}_{-0.2}$ & $29.1^{+0.2}_{-0.3}$ & $-$ & $28.3^{+0.2}_{-0.2}$ & $28.9^{+0.2}_{-0.3}$ & $28.0^{+0.1}_{-0.2}$ & $29.3^{+0.3}_{-0.3}$ & $29.6^{+0.3}_{-0.3}$ & $29.5^{+0.3}_{-0.3}$ & $-$ & $29.7^{+0.2}_{-0.2}$ & $29.6^{+0.3}_{-0.3}$ & $29.5^{+0.1}_{-0.1}$ \\
JADES-DR3-GS-West (Deep) & $28.9^{+0.3}_{-0.5}$ & $29.2^{+0.3}_{-0.6}$ & $28.9^{+0.2}_{-0.2}$ & $28.8^{+0.3}_{-0.5}$ & $28.8^{+0.3}_{-0.4}$ & $28.3^{+0.3}_{-0.6}$ & $29.7^{+0.2}_{-0.3}$ & $29.7^{+0.2}_{-0.3}$ & $29.8^{+0.2}_{-0.2}$ & $29.2^{+0.1}_{-0.2}$ & $29.4^{+0.2}_{-0.2}$ & $29.7^{+0.2}_{-0.3}$ & $29.3^{+0.2}_{-0.2}$ \\
JADES-DR3-GS-West (Shallow) & $28.6^{+0.4}_{-0.5}$ & $28.7^{+0.5}_{-0.3}$ & $28.8^{+0.3}_{-0.1}$ & $28.1^{+0.8}_{-1.3}$ & $28.6^{+0.4}_{-0.5}$ & $27.7^{+0.6}_{-0.3}$ & $29.3^{+0.2}_{-0.2}$ & $29.3^{+0.2}_{-0.2}$ & $29.4^{+0.2}_{-0.2}$ & $29.2^{+0.1}_{-0.2}$ & $29.4^{+0.2}_{-0.3}$ & $29.4^{+0.2}_{-0.2}$ & $29.3^{+0.3}_{-0.2}$ \\
\hline
JADES-DR3-GN-Deep (Deep) & $29.1^{+0.2}_{-0.3}$ & $28.9^{+0.2}_{-0.1}$ & $-$ & $28.6^{+0.3}_{-0.2}$ & $29.1^{+0.2}_{-0.3}$ & $28.3^{+0.2}_{-0.1}$ & $29.2^{+0.1}_{-0.2}$ & $29.5^{+0.1}_{-0.2}$ & $29.4^{+0.2}_{-0.2}$ & $-$ & $-$ & $29.6^{+0.2}_{-0.2}$ & $-$ \\
JADES-DR3-GN-Deep (Shallow) & $29.0^{+0.3}_{-0.2}$ & $28.9^{+0.2}_{-0.3}$ & $-$ & $28.5^{+0.3}_{-0.3}$ & $29.1^{+0.2}_{-0.3}$ & $28.3^{+0.2}_{-0.4}$ & $28.9^{+0.1}_{-0.2}$ & $29.2^{+0.2}_{-0.2}$ & $29.1^{+0.1}_{-0.2}$ & $-$ & $-$ & $29.3^{+0.2}_{-0.2}$ & $-$ \\
JADES-DR3-GN-Medium & $28.8^{+0.2}_{-0.4}$ & $29.0^{+0.2}_{-0.3}$ & $-$ & $28.9^{+0.2}_{-0.4}$ & $29.2^{+0.2}_{-0.3}$ & $28.2^{+0.2}_{-0.3}$ & $28.4^{+0.2}_{-0.2}$ & $28.8^{+0.2}_{-0.2}$ & $28.7^{+0.2}_{-0.2}$ & $-$ & $-$ & $28.9^{+0.2}_{-0.2}$ & $-$ \\
JADES-DR3-GN-Parallel & $28.8^{+0.3}_{-0.3}$ & $28.9^{+0.2}_{-0.2}$ & $28.7^{+0.1}_{-0.1}$ & $28.7^{+0.2}_{-0.2}$ & $28.7^{+0.3}_{-0.2}$ & $28.2^{+0.2}_{-0.2}$ & $29.2^{+0.2}_{-0.2}$ & $29.3^{+0.2}_{-0.2}$ & $29.3^{+0.2}_{-0.2}$ & $-$ & $-$ & $29.4^{+0.2}_{-0.2}$ & $-$ \\
\hline
NEP-1 & $28.3^{+0.3}_{-0.3}$ & $28.9^{+0.3}_{-0.3}$ & $-$ & $-$ & $-$ & $-$ & $28.6^{+0.3}_{-0.2}$ & $28.6^{+0.3}_{-0.3}$ & $28.7^{+0.2}_{-0.3}$ & $-$ & $-$ & $28.9^{+0.3}_{-0.3}$ & $-$ \\
NEP-2 & $28.2^{+0.4}_{-0.3}$ & $28.8^{+0.4}_{-0.2}$ & $-$ & $-$ & $-$ & $-$ & $28.6^{+0.2}_{-0.2}$ & $28.6^{+0.2}_{-0.2}$ & $28.7^{+0.2}_{-0.2}$ & $-$ & $-$ & $28.9^{+0.2}_{-0.2}$ & $-$ \\
NEP-3 & $28.2^{+0.4}_{-0.3}$ & $28.8^{+0.3}_{-0.3}$ & $-$ & $-$ & $-$ & $-$ & $28.6^{+0.2}_{-0.2}$ & $28.6^{+0.2}_{-0.2}$ & $28.7^{+0.2}_{-0.2}$ & $-$ & $-$ & $28.8^{+0.2}_{-0.2}$ & $-$ \\
NEP-4 & $28.2^{+0.3}_{-0.4}$ & $28.9^{+0.3}_{-0.3}$ & $-$ & $-$ & $-$ & $-$ & $28.6^{+0.2}_{-0.2}$ & $28.7^{+0.2}_{-0.2}$ & $28.7^{+0.2}_{-0.2}$ & $-$ & $-$ & $28.9^{+0.2}_{-0.2}$ & $-$ \\
\hline
CEERSP1 & $-$ & $28.8^{+0.1}_{-0.2}$ & $-$ & $-$ & $28.4^{+0.2}_{-0.3}$ & $-$ & $-$ & $28.7^{+0.2}_{-0.3}$ & $28.7^{+0.1}_{-0.4}$ & $-$ & $-$ & $29.0^{+0.1}_{-0.2}$ & $-$ \\
CEERSP2 & $-$ & $28.8^{+0.1}_{-0.2}$ & $-$ & $-$ & $28.4^{+0.2}_{-0.3}$ & $-$ & $-$ & $28.8^{+0.2}_{-0.4}$ & $28.7^{+0.2}_{-0.3}$ & $-$ & $-$ & $29.0^{+0.1}_{-0.2}$ & $-$ \\
CEERSP3 & $-$ & $28.8^{+0.2}_{-0.2}$ & $-$ & $-$ & $28.5^{+0.1}_{-0.2}$ & $-$ & $-$ & $29.0^{+0.1}_{-0.1}$ & $28.9^{+0.1}_{-0.1}$ & $-$ & $-$ & $29.0^{+0.1}_{-0.1}$ & $-$ \\
CEERSP4 & $-$ & $28.5^{+0.2}_{-0.2}$ & $-$ & $-$ & $28.2^{+0.3}_{-0.3}$ & $-$ & $-$ & $29.0^{+0.1}_{-0.2}$ & $28.9^{+0.1}_{-0.1}$ & $-$ & $-$ & $29.0^{+0.1}_{-0.2}$ & $-$ \\
CEERSP5 & $-$ & $28.8^{+0.1}_{-0.1}$ & $-$ & $-$ & $28.6^{+0.1}_{-0.2}$ & $-$ & $-$ & $29.0^{+0.1}_{-0.1}$ & $28.9^{+0.1}_{-0.1}$ & $-$ & $-$ & $29.1^{+0.1}_{-0.1}$ & $-$ \\
CEERSP6 & $-$ & $28.7^{+0.1}_{-0.1}$ & $-$ & $-$ & $28.6^{+0.1}_{-0.1}$ & $-$ & $-$ & $29.0^{+0.1}_{-0.2}$ & $28.9^{+0.1}_{-0.1}$ & $-$ & $-$ & $29.0^{+0.1}_{-0.1}$ & $-$ \\
CEERSP7 & $-$ & $28.6^{+0.2}_{-0.2}$ & $-$ & $-$ & $28.5^{+0.1}_{-0.2}$ & $-$ & $-$ & $29.0^{+0.1}_{-0.1}$ & $28.9^{+0.1}_{-0.1}$ & $-$ & $-$ & $29.1^{+0.1}_{-0.1}$ & $-$ \\
CEERSP8 & $-$ & $28.6^{+0.2}_{-0.2}$ & $-$ & $-$ & $28.5^{+0.1}_{-0.2}$ & $-$ & $-$ & $29.0^{+0.1}_{-0.1}$ & $28.9^{+0.1}_{-0.1}$ & $-$ & $-$ & $29.1^{+0.1}_{-0.1}$ & $-$ \\
CEERSP9 & $-$ & $28.6^{+0.2}_{-0.2}$ & $-$ & $-$ & $28.5^{+0.1}_{-0.2}$ & $-$ & $-$ & $29.1^{+0.1}_{-0.2}$ & $28.8^{+0.1}_{-0.1}$ & $-$ & $-$ & $29.1^{+0.1}_{-0.1}$ & $-$ \\
CEERSP10 & $-$ & $28.5^{+0.2}_{-0.2}$ & $-$ & $-$ & $28.3^{+0.3}_{-0.4}$ & $-$ & $-$ & $29.0^{+0.1}_{-0.1}$ & $28.9^{+0.1}_{-0.1}$ & $-$ & $-$ & $29.1^{+0.1}_{-0.1}$ & $-$ \\
\hline
PRIMER-UDS (Deep) & $-$ & $28.3^{+0.2}_{-0.3}$ & $-$ & $-$ & $28.2^{+0.2}_{-0.4}$ & $-$ & $28.5^{+0.4}_{-0.3}$ & $28.6^{+0.4}_{-0.3}$ & $28.8^{+0.4}_{-0.3}$ & $-$ & $-$ & $28.9^{+0.3}_{-0.3}$ & $-$ \\
PRIMER-UDS (Shallow) & $-$ & $28.3^{+0.2}_{-0.4}$ & $-$ & $-$ & $28.2^{+0.2}_{-0.4}$ & $-$ & $28.0^{+0.3}_{-0.3}$ & $28.0^{+0.3}_{-0.3}$ & $28.2^{+0.3}_{-0.3}$ & $-$ & $-$ & $28.4^{+0.3}_{-0.3}$ & $-$ \\
\hline
PRIMER-COSMOS (Deep) & $-$ & $28.3^{+0.2}_{-0.4}$ & $-$ & $-$ & $28.1^{+0.2}_{-0.3}$ & $-$ & $28.6^{+1.0}_{-0.3}$ & $28.7^{+0.8}_{-0.3}$ & $28.9^{+4.1}_{-0.3}$ & $-$ & $-$ & $29.1^{+4.2}_{-0.3}$ & $-$ \\
PRIMER-COSMOS (Shallow) & $-$ & $28.2^{+0.2}_{-0.3}$ & $-$ & $-$ & $28.0^{+0.2}_{-0.3}$ & $-$ & $28.0^{+0.3}_{-0.4}$ & $28.1^{+0.3}_{-0.4}$ & $28.3^{+0.3}_{-0.4}$ & $-$ & $-$ & $28.4^{+0.4}_{-0.4}$ & $-$ \\
   \botrule
   \end{tabular}
   \label{tab:EPOCHS_v2_ACS+NIRCam_SW_depths}
\end{sidewaystable*}

\begin{sidewaystable*}[]
   \centering
   \caption{Median NIRCam LW depths (with $1\sigma$ error) for the NGDEEP, JADES (GOODS-South + GOODS-North), NEP, CEERS, and PRIMER (UDS + COSMOS) \epochstwo\ surveys. The division of these into subfields are explained in \autoref{sec:areas_depths}.}
   \setlength{\tabcolsep}{2pt}
   \begin{tabular}{c|cccccccccc}
   \toprule
Survey & F250M & F277W & F300M & F335M & F356W & F410M & F430M & F444W & F460M & F480M \\
\hline
NGDEEP (Deep) & $-$ & $31.0^{+0.1}_{-0.1}$ & $-$ & $-$ & $30.9^{+0.1}_{-0.1}$ & $-$ & $-$ & $30.9^{+0.1}_{-0.2}$ & $-$ & $-$ \\
NGDEEP (Shallow) & $-$ & $30.6^{+0.2}_{-0.2}$ & $-$ & $-$ & $30.4^{+0.3}_{-0.2}$ & $-$ & $-$ & $30.6^{+0.1}_{-0.2}$ & $-$ & $-$ \\
JADES-DR3-GS-North & $-$ & $29.9^{+0.4}_{-0.3}$ & $-$ & $29.3^{+0.4}_{-0.2}$ & $29.8^{+0.4}_{-0.2}$ & $29.4^{+0.4}_{-0.3}$ & $-$ & $29.6^{+0.3}_{-0.3}$ & $-$ & $-$ \\
JADES-DR3-GS-South (JOF) & $30.1^{+0.1}_{-0.1}$ & $30.7^{+0.1}_{-0.1}$ & $30.5^{+0.1}_{-0.1}$ & $30.6^{+0.1}_{-0.1}$ & $30.7^{+0.1}_{-0.1}$ & $30.1^{+0.1}_{-0.1}$ & $-$ & $30.4^{+0.1}_{-0.1}$ & $-$ & $-$ \\
JADES-DR3-GS-South & $30.1^{+0.1}_{-0.1}$ & $29.9^{+0.3}_{-0.3}$ & $30.5^{+0.1}_{-0.1}$ & $29.3^{+0.2}_{-0.1}$ & $30.0^{+0.2}_{-0.3}$ & $29.3^{+0.3}_{-0.3}$ & $-$ & $29.6^{+0.3}_{-0.3}$ & $-$ & $-$ \\
JADES-DR3-GS-East (Deep) & $-$ & $30.9^{+0.2}_{-0.2}$ & $-$ & $30.3^{+0.2}_{-0.2}$ & $30.7^{+0.2}_{-0.2}$ & $30.2^{+0.2}_{-0.2}$ & $28.9^{+0.1}_{-0.2}$ & $30.5^{+0.2}_{-0.2}$ & $28.4^{+0.1}_{-0.1}$ & $28.4^{+0.2}_{-0.2}$ \\
JADES-DR3-GS-East (Shallow) & $-$ & $29.8^{+0.3}_{-0.3}$ & $-$ & $29.5^{+0.3}_{-0.3}$ & $29.9^{+0.3}_{-0.3}$ & $29.3^{+0.3}_{-0.3}$ & $28.9^{+0.1}_{-0.1}$ & $29.5^{+0.3}_{-0.3}$ & $28.4^{+0.1}_{-0.2}$ & $28.5^{+0.2}_{-0.3}$ \\
JADES-DR3-GS-West (Deep) & $29.1^{+0.1}_{-0.1}$ & $30.1^{+0.2}_{-0.2}$ & $29.6^{+0.1}_{-0.1}$ & $29.7^{+0.1}_{-0.1}$ & $30.1^{+0.1}_{-0.1}$ & $29.6^{+0.2}_{-0.2}$ & $-$ & $29.9^{+0.2}_{-0.2}$ & $-$ & $-$ \\
JADES-DR3-GS-West (Shallow) & $29.2^{+0.1}_{-0.1}$ & $29.9^{+0.2}_{-0.2}$ & $29.6^{+0.1}_{-0.1}$ & $29.3^{+0.3}_{-0.1}$ & $29.8^{+0.3}_{-0.1}$ & $29.4^{+0.2}_{-0.1}$ & $-$ & $29.7^{+0.3}_{-0.1}$ & $-$ & $-$ \\
\hline
JADES-DR3-GN-Deep (Deep) & $-$ & $30.0^{+0.1}_{-0.1}$ & $-$ & $29.5^{+0.1}_{-0.2}$ & $30.0^{+0.1}_{-0.1}$ & $29.3^{+0.1}_{-0.1}$ & $-$ & $29.6^{+0.1}_{-0.1}$ & $-$ & $-$ \\
JADES-DR3-GN-Deep (Shallow) & $-$ & $29.7^{+0.1}_{-0.1}$ & $-$ & $29.2^{+0.2}_{-0.2}$ & $29.7^{+0.1}_{-0.1}$ & $29.0^{+0.1}_{-0.2}$ & $-$ & $29.3^{+0.1}_{-0.1}$ & $-$ & $-$ \\
JADES-DR3-GN-Medium & $-$ & $29.3^{+0.2}_{-0.2}$ & $-$ & $28.8^{+0.1}_{-0.2}$ & $29.4^{+0.1}_{-0.2}$ & $28.7^{+0.2}_{-0.2}$ & $-$ & $28.9^{+0.1}_{-0.2}$ & $-$ & $-$ \\
JADES-DR3-GN-Parallel & $-$ & $29.8^{+0.1}_{-0.1}$ & $-$ & $29.2^{+0.1}_{-0.1}$ & $29.7^{+0.1}_{-0.1}$ & $29.3^{+0.1}_{-0.1}$ & $-$ & $29.5^{+0.1}_{-0.1}$ & $-$ & $-$ \\
\hline
NEP-1 & $-$ & $29.4^{+0.3}_{-0.2}$ & $-$ & $-$ & $29.4^{+0.3}_{-0.2}$ & $28.7^{+0.3}_{-0.2}$ & $-$ & $29.0^{+0.3}_{-0.2}$ & $-$ & $-$ \\
NEP-2 & $-$ & $29.4^{+0.3}_{-0.2}$ & $-$ & $-$ & $29.4^{+0.3}_{-0.2}$ & $28.6^{+0.3}_{-0.1}$ & $-$ & $28.9^{+0.3}_{-0.1}$ & $-$ & $-$ \\
NEP-3 & $-$ & $29.4^{+0.2}_{-0.2}$ & $-$ & $-$ & $29.4^{+0.2}_{-0.1}$ & $28.6^{+0.2}_{-0.1}$ & $-$ & $28.9^{+0.3}_{-0.1}$ & $-$ & $-$ \\
NEP-4 & $-$ & $29.4^{+0.2}_{-0.2}$ & $-$ & $-$ & $29.4^{+0.2}_{-0.1}$ & $28.7^{+0.3}_{-0.1}$ & $-$ & $29.0^{+0.3}_{-0.1}$ & $-$ & $-$ \\
\hline
CEERSP1 & $-$ & $29.4^{+0.1}_{-0.1}$ & $-$ & $-$ & $29.4^{+0.1}_{-0.1}$ & $28.6^{+0.1}_{-0.2}$ & $-$ & $28.9^{+0.1}_{-0.1}$ & $-$ & $-$ \\
CEERSP2 & $-$ & $29.4^{+0.1}_{-0.1}$ & $-$ & $-$ & $29.4^{+0.1}_{-0.1}$ & $28.6^{+0.1}_{-0.1}$ & $-$ & $28.9^{+0.1}_{-0.1}$ & $-$ & $-$ \\
CEERSP3 & $-$ & $29.4^{+0.1}_{-0.1}$ & $-$ & $-$ & $29.4^{+0.1}_{-0.1}$ & $28.6^{+0.1}_{-0.1}$ & $-$ & $28.9^{+0.1}_{-0.1}$ & $-$ & $-$ \\
CEERSP4 & $-$ & $29.4^{+0.1}_{-0.1}$ & $-$ & $-$ & $29.4^{+0.1}_{-0.1}$ & $28.6^{+0.1}_{-0.1}$ & $-$ & $28.8^{+0.1}_{-0.1}$ & $-$ & $-$ \\
CEERSP5 & $-$ & $29.4^{+0.1}_{-0.1}$ & $-$ & $-$ & $29.4^{+0.1}_{-0.1}$ & $28.6^{+0.1}_{-0.1}$ & $-$ & $28.8^{+0.1}_{-0.1}$ & $-$ & $-$ \\
CEERSP6 & $-$ & $29.4^{+0.1}_{-0.1}$ & $-$ & $-$ & $29.4^{+0.1}_{-0.1}$ & $28.6^{+0.1}_{-0.1}$ & $-$ & $28.9^{+0.1}_{-0.1}$ & $-$ & $-$ \\
CEERSP7 & $-$ & $29.4^{+0.1}_{-0.1}$ & $-$ & $-$ & $29.4^{+0.1}_{-0.1}$ & $28.6^{+0.1}_{-0.1}$ & $-$ & $28.8^{+0.1}_{-0.1}$ & $-$ & $-$ \\
CEERSP8 & $-$ & $29.4^{+0.1}_{-0.1}$ & $-$ & $-$ & $29.4^{+0.1}_{-0.1}$ & $28.6^{+0.1}_{-0.1}$ & $-$ & $28.8^{+0.1}_{-0.1}$ & $-$ & $-$ \\
CEERSP9 & $-$ & $29.4^{+0.1}_{-0.1}$ & $-$ & $-$ & $29.3^{+0.1}_{-0.1}$ & $28.6^{+0.1}_{-0.1}$ & $-$ & $29.2^{+0.1}_{-0.1}$ & $-$ & $-$ \\
CEERSP10 & $-$ & $29.3^{+0.1}_{-0.1}$ & $-$ & $-$ & $29.4^{+0.1}_{-0.1}$ & $28.6^{+0.1}_{-0.1}$ & $-$ & $28.8^{+0.1}_{-0.1}$ & $-$ & $-$ \\
\hline
PRIMER-UDS (Deep) & $-$ & $29.2^{+0.2}_{-0.3}$ & $-$ & $-$ & $29.2^{+0.3}_{-0.3}$ & $28.4^{+0.2}_{-0.3}$ & $-$ & $28.8^{+0.2}_{-0.2}$ & $-$ & $-$ \\
PRIMER-UDS (Shallow) & $-$ & $28.6^{+0.3}_{-0.3}$ & $-$ & $-$ & $28.7^{+0.3}_{-0.2}$ & $27.9^{+0.3}_{-0.2}$ & $-$ & $28.2^{+0.2}_{-0.2}$ & $-$ & $-$ \\
\hline
PRIMER-COSMOS (Deep) & $-$ & $29.4^{+0.3}_{-0.2}$ & $-$ & $-$ & $29.5^{+0.3}_{-0.3}$ & $28.7^{+0.2}_{-0.2}$ & $-$ & $29.0^{+0.2}_{-0.2}$ & $-$ & $-$ \\
PRIMER-COSMOS (Shallow) & $-$ & $28.8^{+0.4}_{-0.4}$ & $-$ & $-$ & $28.8^{+0.3}_{-0.4}$ & $28.1^{+0.3}_{-0.3}$ & $-$ & $28.4^{+0.3}_{-0.4}$ & $-$ & $-$ \\
   \botrule
   \end{tabular}
   \label{tab:EPOCHS_v2_NIRCam_LW_depths}
\end{sidewaystable*}

\begin{sidewaystable*}[]
   \centering
   \caption{Unmasked areas for the NGDEEP, JADES (GOODS-South + GOODS-North), NEP, CEERS, and PRIMER (UDS + COSMOS) \epochstwo\ surveys as a function of redshift. The total unmasked area is given in bold.}
   \setlength{\tabcolsep}{2pt}
   \begin{tabular}{c|c|c|c|c|c|c}
   \toprule
\multirow{2}{*}{\textbf{Survey}} & \multicolumn{6}{c}{Area within redshift range / $\mathrm{arcmin}^2$} \\
& $6.50<z<6.53$ & $6.53<z<7.27$ & $7.27<z<9.97$ & $9.97<z<13.09$ & $13.09<z<15.19$ & $15.19<z<16.50$ \\
\hline
NGDEEP (Deep) & $7.7$ & $7.7$ & $7.7$ & $7.7$ & $7.7$ & $7.7$ \\
NGDEEP (Shallow) & $2.9$ & $2.9$ & $2.9$ & $2.9$ & $2.9$ & $2.9$ \\
JADES-DR3-GS-North & $22.9$ & $22.9$ & $22.9$ & $23.8$ & $23.8$ & $23.8$ \\
JADES-DR3-GS-South (JOF) & $8.8$ & $8.8$ & $8.8$ & $8.9$ & $8.9$ & $8.9$ \\
JADES-DR3-GS-South & $30.5$ & $30.5$ & $30.7$ & $31.4$ & $30.4$ & $30.4$ \\
JADES-DR3-GS-East (Deep) & $24.0$ & $24.1$ & $24.1$ & $24.1$ & $24.1$ & $23.9$ \\
JADES-DR3-GS-East (Shallow) & $9.3$ & $9.3$ & $9.4$ & $9.5$ & $9.5$ & $9.1$ \\
JADES-DR3-GS-West (Deep) & $13.1$ & $15.3$ & $16.4$ & $16.9$ & $16.5$ & $16.5$ \\
JADES-DR3-GS-West (Shallow) & $9.5$ & $7.8$ & $11.9$ & $12.3$ & $8.6$ & $8.6$ \\
\hline
JADES-DR3-GN-Deep (Deep) & $10.3$ & $10.3$ & $10.3$ & $10.3$ & $10.3$ & $10.3$ \\
JADES-DR3-GN-Deep (Shallow) & $16.3$ & $16.3$ & $16.3$ & $16.3$ & $16.3$ & $16.3$ \\
JADES-DR3-GN-Medium & $28.2$ & $28.2$ & $28.2$ & $28.3$ & $28.3$ & $28.3$ \\
JADES-DR3-GN-Parallel & $25.9$ & $25.9$ & $25.9$ & $26.0$ & $26.0$ & $26.0$ \\
\hline
NEP-1 & $13.4$ & $13.4$ & $13.9$ & $14.2$ & $14.2$ & $14.2$ \\
NEP-2 & $13.8$ & $13.8$ & $14.7$ & $14.7$ & $14.7$ & $14.7$ \\
NEP-3 & $13.3$ & $13.3$ & $14.2$ & $14.2$ & $14.2$ & $14.2$ \\
NEP-4 & $14.0$ & $14.0$ & $14.5$ & $14.5$ & $14.5$ & $14.5$ \\
\hline
CEERSP1 & $8.7$ & $8.7$ & $8.6$ & $8.9$ & $8.9$ & $8.9$ \\
CEERSP2 & $8.7$ & $8.7$ & $8.6$ & $8.9$ & $8.9$ & $8.9$ \\
CEERSP3 & $8.3$ & $8.3$ & $8.3$ & $9.0$ & $9.0$ & $9.0$ \\
CEERSP4 & $8.0$ & $8.0$ & $8.1$ & $8.8$ & $8.8$ & $8.8$ \\
CEERSP5 & $8.1$ & $8.1$ & $8.1$ & $8.8$ & $8.8$ & $8.8$ \\
CEERSP6 & $8.2$ & $8.2$ & $8.2$ & $8.9$ & $8.9$ & $8.9$ \\
CEERSP7 & $8.2$ & $8.2$ & $8.2$ & $9.0$ & $9.0$ & $9.0$ \\
CEERSP8 & $8.0$ & $8.0$ & $8.0$ & $8.8$ & $8.8$ & $8.8$ \\
CEERSP9 & $8.1$ & $8.1$ & $8.1$ & $8.9$ & $8.9$ & $8.9$ \\
CEERSP10 & $8.0$ & $8.0$ & $8.0$ & $8.7$ & $8.7$ & $8.7$ \\
\hline
PRIMER-UDS (Deep) & $101.4$ & $101.5$ & $113.2$ & $113.3$ & $113.2$ & $113.2$ \\
PRIMER-UDS (Shallow) & $70.6$ & $70.7$ & $125.2$ & $125.2$ & $125.2$ & $125.2$ \\
\hline
PRIMER-COSMOS (Deep) & $68.8$ & $68.8$ & $68.9$ & $68.9$ & $68.9$ & $68.9$ \\
PRIMER-COSMOS (Shallow) & $64.3$ & $64.5$ & $71.4$ & $71.4$ & $71.4$ & $71.4$ \\
\botrule
\textbf{Total} & $\mathbf{651.3}$ & $\mathbf{652.3}$ & $\mathbf{733.7}$ & $\mathbf{743.5}$ & $\mathbf{738.3}$ & $\mathbf{737.7}$ \\
\botrule
   \end{tabular}
   \label{tab:EPOCHS_v2_areas}
\end{sidewaystable*}

\section{Photometric candidates at $z>10.5$}
\label{sec:gold_properties}

Photometric properties for our \epochstwo\ ``gold'' sample of $z_{\rm phot}>10.5$ galaxies are presented in \autoref{tab:gold_sample}, which includes sources without spectroscopic confirmation, as well as those confirmed at high-redshift (shown in \autoref{tab:highz_spec}) and as low-redshift contaminants. We cross-match this sample to a number of photometric NIRCam candidates compiled from the literature \citep{Bouwens2023,Donnan2023,Donnan2023b,Austin2023,Leung2023,Hainline2024a,Finkelstein2024,Conselice2025,Whitler2025,Hainline2026,McLeod2026} using a $1.0''$ radius, identifying a number of new high-redshift candidates. We additionally search radio and X-ray source catalogs, finding $3$ sources with detections in the MeerKAT International Gigahertz Tiered Extragalactic Explorations \citep[MIGHTEE;][]{Jarvis2016} data \citep{Heywood2022,Malefahlo2026} in the COSMOS and UDS PRIMER NIRCam footprints. In addition, GS-S-5497 at $z_{\rm phot}=10.47^{+0.50}_{-0.15}$ is X-ray detected in the Chandra Deep Field South \citep[CDFS;][]{Weigel2015, Cappelluti2016, Luo2017, Li2019, Yang2022, Evans2024}, which may highlight the AGN nature of the source \citep[as in][]{Bogdan2024}, however a more thorough investigation into the nature of this source is required to confirm this scenario.


\begin{deluxetable*}{l|ccccccccc}
    \tablewidth{\textwidth}
    \setlength{\tabcolsep}{3pt}
    \tablecaption{Properties of the \epochstwo\ ``gold'' sample of $30$ $10.5<z<11.5$, $15$ $11.5<z<13.5$, and $17$ $13.5<z<16.5$ galaxy candidates. The redshift bins were produced using the best-fit ``zbest'' from \eazypy\ using the ``sfhz\_blue\_agn'' template set, and the columns show the median and $1\sigma$ errors of the full posterior distribution. The reference column shows cross-matched sources to the NIR photometric studies of: [1]: \citet{Donnan2023}; [2]: \citet{Bouwens2023}; [3]: \citet{Donnan2023b} [4]: \citet{Austin2023}; [5]: \citet{Leung2023}; [6]: \citet{Hainline2024a}; [7]: \citet{Finkelstein2024}; [8]: \citet{Conselice2025}; [9]: \citet{Whitler2025}; [10]: \citet{Hainline2026}; [11]: \citet{McLeod2026}.    \label{tab:gold_sample}}

    \tablehead{
        \colhead{Source name} & \colhead{RA\,/\,deg} & \colhead{Dec\,/\,deg} & \colhead{$z_{\rm EaZy}$} & \colhead{$\chi^2_{\rm red}$} & \colhead{$M_{\rm UV}$} & \colhead{$\beta_{\rm UV}$} & \colhead{$\mathrm{SNR}^{\rm stack}_{\rm blue}$} & \colhead{$\mathrm{SNR}_{\rm red\,Ly\alpha}$} & \colhead{Ref.}
    }
    \startdata
\multicolumn{10}{c}{$\mathbf{10.5<z<11.5}$} \\
\hline
GS-N-4826 & $53.10618$ & $-27.74776$ & $11.18^{+0.17}_{-0.10}$ & $2.3$ & $-19.86^{+0.07}_{-0.07}$ & $-2.54^{+0.19}_{-0.19}$ & $-0.7$ & $\{13.5,13.1\}$ & [6,10] \\
GS-S-2719 & $53.12161$ & $-27.90814$ & $11.30^{+0.23}_{-0.05}$ & $0.9$ & $-19.43^{+0.07}_{-0.07}$ & $-2.24^{+0.21}_{-0.21}$ & $-0.6$ & $\{16.6,16.8\}$ & [6,9,10] \\
GS-S-5497$^{\S}$ & $53.14636$ & $-27.87095$ & $10.47^{+0.50}_{-0.15}$ & $1.7$ & $-18.60^{+0.13}_{-0.12}$ & $-0.95^{+0.29}_{-0.28}$ & $0.6$ & $\{8.5,17.1\}$ & -- \\
GS-S-10008$^{\dagger}$ & $53.11764$ & $-27.88817$ & $11.06^{+0.20}_{-0.13}$ & $1.6$ & $-19.68^{+0.08}_{-0.07}$ & $-3.25^{+0.23}_{-0.24}$ & $0.9$ & $\{20.7,11.6\}$ & [6,10] \\
GS-S-18899 & $53.07173$ & $-27.91422$ & $11.18^{+0.29}_{-0.21}$ & $0.8$ & $-19.01^{+0.07}_{-0.07}$ & $-2.31^{+0.21}_{-0.23}$ & $-0.5$ & $\{13.6,13.0\}$ & [6,9,10] \\
GS-S-43988$^{\dagger}$ & $53.07076$ & $-27.86543$ & $10.59^{+0.27}_{-0.01}$ & $1.7$ & $-19.35^{+0.08}_{-0.08}$ & $-2.55^{+0.18}_{-0.17}$ & $-0.3$ & $\{11.8,21.1\}$ & [6,10] \\
GS-E-19485 & $53.18993$ & $-27.77149$ & $11.30^{+0.23}_{-0.17}$ & $2.2$ & $-18.58^{+0.08}_{-0.08}$ & $-2.97^{+0.21}_{-0.22}$ & $0.6$ & $\{10.3,14.1\}$ & [6,8,9,10] \\
GS-E-23918 & $53.16863$ & $-27.79275$ & $10.82^{+0.39}_{-0.22}$ & $2.1$ & $-18.83^{+0.07}_{-0.07}$ & $-2.85^{+0.18}_{-0.19}$ & $0.1$ & $\{8.5,11.9\}$ & [2,3,6,8,9,10] \\
GS-E-26233 & $53.14528$ & $-27.82359$ & $10.59^{+0.21}_{-0.05}$ & $1.9$ & $-18.90^{+0.05}_{-0.05}$ & $-2.21^{+0.17}_{-0.17}$ & $-1.6$ & $\{17.6,22.6\}$ & [6,8,9,10] \\
GS-W-12273 & $52.94429$ & $-27.82560$ & $10.70^{+0.26}_{-0.08}$ & $1.7$ & $-18.80^{+0.08}_{-0.07}$ & $-2.28^{+0.19}_{-0.18}$ & $-0.8$ & $\{9.3,13.6\}$ & [10] \\
GS-W-48120 & $52.99738$ & $-27.76973$ & $10.59^{+0.17}_{-0.16}$ & $0.7$ & $-19.16^{+0.06}_{-0.06}$ & $-2.42^{+0.17}_{-0.16}$ & $-0.2$ & $\{10.8,12.8\}$ & [10] \\
NGD-228 & $53.24948$ & $-27.88332$ & $10.82^{+0.29}_{-0.23}$ & $1.9$ & $-18.63^{+0.11}_{-0.11}$ & $-2.98^{+0.31}_{-0.32}$ & $0.5$ & $\{9.2,9.1\}$ & [5,10] \\
NGD-2009 & $53.24209$ & $-27.85508$ & $10.70^{+0.16}_{-0.17}$ & $1.5$ & $-18.74^{+0.08}_{-0.07}$ & $-2.21^{+0.22}_{-0.21}$ & $0.2$ & $\{16.4,21.0\}$ & [4,5,8,10] \\
GN-Par-9815 & $189.42822$ & $62.33057$ & $11.18^{+0.30}_{-0.23}$ & $0.7$ & $-19.14^{+0.10}_{-0.10}$ & $-2.43^{+0.26}_{-0.27}$ & $0.2$ & $\{8.9,11.8\}$ & [10] \\
GN-Par-13474 & $189.39712$ & $62.30648$ & $10.70^{+0.21}_{-0.09}$ & $1.8$ & $-19.44^{+0.07}_{-0.07}$ & $-2.30^{+0.20}_{-0.21}$ & $-1.7$ & $\{29.5,16.8\}$ & [9] \\
CEERS-2-9698 & $214.86093$ & $52.88116$ & $11.06^{+0.40}_{-0.32}$ & $0.2$ & $-19.67^{+0.10}_{-0.10}$ & $-1.94^{+0.26}_{-0.27}$ & $0.6$ & $\{9.3,14.4\}$ & [8] \\
CEERS-2-12686 & $214.92739$ & $52.91169$ & $11.18^{+0.25}_{-0.33}$ & $1.3$ & $-19.66^{+0.10}_{-0.10}$ & $-2.68^{+0.32}_{-0.31}$ & $1.1$ & $\{9.2,9.0\}$ & -- \\
NEP-2-811 & $260.86760$ & $65.81360$ & $11.43^{+0.22}_{-0.18}$ & $1.8$ & $-20.16^{+0.09}_{-0.08}$ & $-2.75^{+0.26}_{-0.24}$ & $0.7$ & $\{10.5,13.3\}$ & -- \\
NEP-2-10330 & $260.81916$ & $65.84473$ & $11.30^{+0.31}_{-0.19}$ & $2.3$ & $-20.13^{+0.10}_{-0.09}$ & $-2.03^{+0.26}_{-0.26}$ & $-1.1$ & $\{9.3,20.8\}$ & -- \\
NEP-2-14859 & $260.85786$ & $65.84657$ & $10.59^{+0.31}_{-0.10}$ & $1.1$ & $-19.39^{+0.12}_{-0.11}$ & $-2.28^{+0.29}_{-0.28}$ & $-1.6$ & $\{9.3,11.4\}$ & [8] \\
NEP-3-4068 & $260.68065$ & $65.85158$ & $10.82^{+0.34}_{-0.14}$ & $1.0$ & $-19.87^{+0.10}_{-0.09}$ & $-2.46^{+0.27}_{-0.25}$ & $-1.6$ & $\{10.8,11.6\}$ & [8] \\
UDS-67087$^{\P}$ & $34.41042$ & $-5.27233$ & $10.59^{+0.22}_{-0.26}$ & $1.2$ & $-20.43^{+0.10}_{-0.09}$ & $-1.90^{+0.24}_{-0.24}$ & $-0.8$ & $\{10.5,13.3\}$ & -- \\
UDS-146725 & $34.27549$ & $-5.20046$ & $10.94^{+0.27}_{-0.24}$ & $0.9$ & $-20.55^{+0.13}_{-0.12}$ & $-1.84^{+0.30}_{-0.28}$ & $0.5$ & $\{8.2,14.6\}$ & -- \\
UDS-295313 & $34.35505$ & $-5.09664$ & $10.82^{+0.29}_{-0.25}$ & $0.7$ & $-20.40^{+0.10}_{-0.09}$ & $-2.00^{+0.28}_{-0.28}$ & $0.5$ & $\{10.8,9.9\}$ & -- \\
UDS-302436$^{\ddagger}$ & $34.22773$ & $-5.09102$ & $11.30^{+0.32}_{-0.37}$ & $1.8$ & $-20.49^{+0.11}_{-0.10}$ & $-1.70^{+0.30}_{-0.29}$ & $-0.2$ & $\{8.6,11.3\}$ & -- \\
COSMOS-5948 & $150.10786$ & $2.17861$ & $11.18^{+0.29}_{-0.60}$ & $1.1$ & $-20.93^{+0.10}_{-0.09}$ & $-1.92^{+0.26}_{-0.25}$ & $-0.8$ & $\{9.8,13.2\}$ & -- \\
COSMOS-103852$^{\P}$ & $150.11872$ & $2.31469$ & $11.43^{+0.28}_{-0.20}$ & $2.6$ & $-19.74^{+0.08}_{-0.08}$ & $-1.25^{+0.25}_{-0.24}$ & $0.7$ & $\{10.8,18.8\}$ & -- \\
COSMOS-171869$^{\ddagger}$ & $150.08630$ & $2.41955$ & $10.59^{+0.17}_{-2.01}$ & $1.3$ & $-20.56^{+0.12}_{-0.12}$ & $-2.60^{+0.30}_{-0.30}$ & $1.2$ & $\{8.7,10.9\}$ & -- \\
COSMOS-173946$^{\ddagger,\P}$ & $150.10463$ & $2.41731$ & $11.43^{+0.30}_{-0.23}$ & $1.8$ & $-19.95^{+0.10}_{-0.09}$ & $-1.60^{+0.27}_{-0.26}$ & $0.4$ & $\{9.3,20.9\}$ & -- \\
COSMOS-174516 & $150.17667$ & $2.39501$ & $10.47^{+0.24}_{-0.23}$ & $1.1$ & $-19.86^{+0.12}_{-0.12}$ & $-2.77^{+0.28}_{-0.28}$ & $1.7$ & $\{9.2,11.4\}$ & -- \\
\midrule
\multicolumn{10}{c}{$\mathbf{11.5<z<13.5}$} \\
\hline
GS-S-5056 & $53.14476$ & $-27.87419$ & $11.43^{+0.44}_{-0.23}$ & $1.0$ & $-18.78^{+0.11}_{-0.10}$ & $-2.49^{+0.36}_{-0.35}$ & $-0.6$ & $\{8.0,8.1\}$ & -- \\
GS-S-34290 & $53.08468$ & $-27.86666$ & $13.44^{+0.41}_{-0.49}$ & $1.0$ & $-18.17^{+0.10}_{-0.09}$ & $-1.51^{+0.23}_{-0.21}$ & $-0.4$ & $\{10.8,10.9\}$ & [6,9,11] \\
GS-E-15655$^{\dagger}$ & $53.16635$ & $-27.82155$ & $12.33^{+1.06}_{-0.14}$ & $1.1$ & $-18.76^{+0.06}_{-0.05}$ & $-2.06^{+0.12}_{-0.12}$ & $-0.1$ & $\{12.4,21.0\}$ & [6,8,9,10,11] \\
GS-E-26267 & $53.19050$ & $-27.74982$ & $11.55^{+0.22}_{-0.01}$ & $2.6$ & $-19.24^{+0.06}_{-0.06}$ & $-2.97^{+0.17}_{-0.17}$ & $1.1$ & $\{17.2,18.3\}$ & [6,8,9,10] \\
NGD-7528 & $53.26661$ & $-27.87658$ & $11.81^{+0.17}_{-0.13}$ & $1.5$ & $-18.87^{+0.06}_{-0.06}$ & $-2.32^{+0.22}_{-0.21}$ & $0.4$ & $\{14.7,20.5\}$ & [4,5,8,10] \\
GN-Par-19105 & $189.43359$ & $62.27893$ & $11.81^{+0.29}_{-0.18}$ & $2.9$ & $-19.70^{+0.08}_{-0.08}$ & $-1.85^{+0.19}_{-0.18}$ & $0.7$ & $\{9.5,20.7\}$ & [9,10] \\
CEERS-2-9216$^{\dagger}$ & $214.94314$ & $52.94245$ & $11.43^{+0.32}_{-0.14}$ & $2.9$ & $-20.06^{+0.07}_{-0.06}$ & $-3.01^{+0.21}_{-0.21}$ & $-0.4$ & $\{13.7,14.1\}$ & [1,7,8] \\
NEP-4-3626 & $260.69763$ & $65.80557$ & $13.01^{+0.75}_{-0.67}$ & $2.9$ & $-19.72^{+0.14}_{-0.13}$ & $-1.76^{+0.31}_{-0.31}$ & $0.3$ & $\{8.1,12.2\}$ & -- \\
NEP-4-10688 & $260.62468$ & $65.79584$ & $11.68^{+0.37}_{-0.22}$ & $3.1$ & $-20.06^{+0.09}_{-0.08}$ & $-3.43^{+0.28}_{-0.29}$ & $0.8$ & $\{9.5,10.7\}$ & [8] \\
UDS-6611 & $34.26094$ & $-5.31818$ & $12.07^{+0.75}_{-0.16}$ & $0.9$ & $-20.44^{+0.06}_{-0.06}$ & $-1.56^{+0.23}_{-0.24}$ & $0.4$ & $\{22.7,11.5\}$ & -- \\
UDS-34215 & $34.36529$ & $-5.29496$ & $12.33^{+0.49}_{-0.30}$ & $1.9$ & $-20.16^{+0.08}_{-0.07}$ & $-1.83^{+0.17}_{-0.16}$ & $1.2$ & $\{8.6,11.3\}$ & -- \\
UDS-118600 & $34.31914$ & $-5.23014$ & $11.55^{+0.27}_{-0.16}$ & $2.2$ & $-20.57^{+0.08}_{-0.07}$ & $-2.00^{+0.23}_{-0.22}$ & $1.4$ & $\{11.6,16.2\}$ & -- \\
UDS-200517 & $34.47869$ & $-5.16110$ & $12.33^{+0.78}_{-0.04}$ & $2.1$ & $-19.90^{+0.05}_{-0.05}$ & $-2.80^{+0.18}_{-0.19}$ & $-0.6$ & $\{21.4,12.1\}$ & -- \\
UDS-310943 & $34.41549$ & $-5.08117$ & $11.55^{+0.33}_{-0.28}$ & $2.7$ & $-20.70^{+0.10}_{-0.09}$ & $-2.94^{+0.33}_{-0.34}$ & $1.1$ & $\{8.7,8.5\}$ & -- \\
COSMOS-156491 & $150.16167$ & $2.36890$ & $12.07^{+0.52}_{-0.28}$ & $0.9$ & $-19.51^{+0.09}_{-0.08}$ & $-2.74^{+0.32}_{-0.31}$ & $0.8$ & $\{9.1,9.3\}$ & -- \\
\bottomrule
\enddata
\end{deluxetable*}

\addtocounter{table}{-1}
\begin{deluxetable*}{l|ccccccccc}
    \tablewidth{\textwidth}
    \setlength{\tabcolsep}{3pt}
    \tablecaption{Continued...}
    \tablehead{
        \colhead{Source name} & \colhead{RA\,/\,deg} & \colhead{Dec\,/\,deg} & \colhead{$z_{\rm EaZy}$} & \colhead{$\chi^2_{\rm red}$} & \colhead{$M_{\rm UV}$} & \colhead{$\beta_{\rm UV}$} & \colhead{$\mathrm{SNR}^{\rm stack}_{\rm blue}$} & \colhead{$\mathrm{SNR}_{\rm red\,Ly\alpha}$} & \colhead{Ref.}
    }
    \startdata
\multicolumn{10}{c}{$\mathbf{13.5<z<16.5}$} \\
\hline
GS-S-9031 & $53.13974$ & $-27.86616$ & $13.59^{+0.47}_{-1.00}$ & $1.5$ & $-18.86^{+0.14}_{-0.14}$ & $-1.52^{+0.31}_{-0.33}$ & $-1.2$ & $\{10.0,8.9\}$ & [9,10] \\
GS-S-30479$^{\dagger}$ & $53.07427$ & $-27.88592$ & $14.03^{+0.28}_{-0.25}$ & $3.2$ & $-19.36^{+0.21}_{-0.21}$ & $-3.77^{+0.43}_{-0.50}$ & $-0.7$ & $\{13.9,10.8\}$ & [6,10] \\
GS-E-51148 & $53.11160$ & $-27.79905$ & $16.46^{+1.76}_{-0.36}$ & $2.0$ & $-19.30^{+0.10}_{-0.09}$ & $-3.08^{+0.30}_{-0.32}$ & $1.3$ & $\{11.3,10.0\}$ & -- \\ 
GS-W-4608 & $52.93064$ & $-27.83888$ & $14.03^{+0.46}_{-1.26}$ & $1.9$ & $-19.41^{+0.19}_{-0.22}$ & $-3.64^{+0.52}_{-0.66}$ & $0.5$ & $\{12.7,8.7\}$ & -- \\
NGD-1156 & $53.24949$ & $-27.87572$ & $14.96^{+0.28}_{-0.12}$ & $2.0$ & $-19.43^{+0.10}_{-0.09}$ & $-3.91^{+0.34}_{-0.34}$ & $1.6$ & $\{16.7,9.8\}$ & [4,5] \\
GN-Med-1881 & $189.16732$ & $62.31028$ & $14.96^{+0.40}_{-0.34}$ & $1.9$ & $-20.39^{+0.15}_{-0.16}$ & $-3.14^{+0.47}_{-0.55}$ & $0.7$ & $\{10.1,10.6\}$ & [6,9,10] \\
NEP-1-14821 & $260.74818$ & $65.74352$ & $15.28^{+0.62}_{-0.43}$ & $2.0$ & $-20.22^{+0.11}_{-0.12}$ & $-2.70^{+0.40}_{-0.48}$ & $-1.1$ & $\{11.1,10.3\}$ & -- \\
NEP-4-13205 & $260.64767$ & $65.78252$ & $13.73^{+0.53}_{-1.00}$ & $1.5$ & $-20.53^{+0.13}_{-0.14}$ & $-2.73^{+0.34}_{-0.39}$ & $0.9$ & $\{11.8,9.8\}$ & [8] \\
UDS-82177 & $34.51114$ & $-5.26080$ & $15.12^{+0.50}_{-0.58}$ & $0.3$ & $-20.24^{+0.10}_{-0.09}$ & $-1.94^{+0.30}_{-0.32}$ & $0.5$ & $\{12.2,11.4\}$ & -- \\
UDS-93188 & $34.46519$ & $-5.25142$ & $15.61^{+0.46}_{-0.55}$ & $0.7$ & $-20.41^{+0.09}_{-0.08}$ & $-2.32^{+0.26}_{-0.27}$ & $0.4$ & $\{10.5,10.4\}$ & -- \\
UDS-167000 & $34.51762$ & $-5.17913$ & $14.64^{+0.28}_{-0.33}$ & $2.7$ & $-20.93^{+0.10}_{-0.10}$ & $-2.13^{+0.27}_{-0.28}$ & $1.9$ & $\{15.8,9.0\}$ & -- \\
UDS-243009 & $34.33360$ & $-5.13209$ & $14.03^{+0.28}_{-0.44}$ & $1.3$ & $-20.88^{+0.09}_{-0.09}$ & $-2.11^{+0.21}_{-0.21}$ & $1.9$ & $\{18.3,15.1\}$ & -- \\
UDS-296905 & $34.38833$ & $-5.09499$ & $14.49^{+0.32}_{-0.23}$ & $3.1$ & $-20.90^{+0.10}_{-0.10}$ & $-2.00^{+0.26}_{-0.27}$ & $1.7$ & $\{12.7,8.7\}$ & -- \\
COSMOS-74973$^{\ddagger}$ & $150.13357$ & $2.27101$ & $16.46^{+0.44}_{-0.30}$ & $1.0$ & $-20.14^{+0.06}_{-0.06}$ & $-2.24^{+0.19}_{-0.21}$ & $1.4$ & $\{13.2,10.8\}$ & -- \\
COSMOS-79429 & $150.08652$ & $2.29339$ & $16.12^{+0.59}_{-0.19}$ & $2.0$ & $-20.21^{+0.07}_{-0.07}$ & $-2.61^{+0.25}_{-0.27}$ & $1.6$ & $\{13.4,9.6\}$ & -- \\
COSMOS-140093 & $150.13713$ & $2.35898$ & $14.96^{+0.47}_{-0.51}$ & $0.9$ & $-19.56^{+0.11}_{-0.11}$ & $-2.02^{+0.31}_{-0.34}$ & $1.1$ & $\{11.0,8.4\}$ & -- \\
COSMOS-158260 & $150.14694$ & $2.37687$ & $16.12^{+0.39}_{-0.15}$ & $1.3$ & $-21.05^{+0.06}_{-0.06}$ & $-2.24^{+0.15}_{-0.14}$ & $1.0$ & $\{19.2,21.1\}$ & -- \\
\bottomrule
\enddata
\tablecomments{``$\dagger$'': Spectroscopically confirmed within $\Delta z<\lvert0.15\times(1+z_{\rm spec})\rvert$ of the photometric redshift.\\
``$\ddagger$'': Spectroscopically confirmed at lower redshift $z_{\rm spec}<z_{\rm phot}-0.15\times(1+z_{\rm spec})$.\\``$\S$'': X-ray detected in CDFS \citep[][]{Weigel2015, Cappelluti2016, Luo2017, Li2019, Yang2022, Evans2024}.\\``$\P$: Radio detected in MIGHTEE data \citep[][]{Heywood2022, Malefahlo2026}.}
\end{deluxetable*}

\bibliography{bib/brown_dwarfs, bib/surveys, bib/uvlfs, bib/uv_beta, bib/misc, bib/evs, bib/spec_sources, bib/highz_candidates, bib/halo_modelling, bib/instruments, bib/stellar_mass_functions, bib/fundamentals, bib/sed_fitting, bib/dust, bib/lrds, bib/cold_gas, bib/sps, bib/gal_clustering_and_abundance_matching, bib/simulations, bib/reionization, bib/morph, bib/quiescent_gals, bib/imf, bib/star_formation, bib/cross_matched}{}
\bibliographystyle{aasjournalv7}

\end{document}